\documentclass[5p,authoryear]{elsarticle}
\journal{Icarus}
\usepackage[latin1]{inputenc}
\usepackage[T1]{fontenc}
\usepackage[english]{babel}
\usepackage{graphicx}
\usepackage{epstopdf} 
\usepackage{amssymb}
\usepackage{amsmath}
\usepackage{mathabx}
\usepackage{tabularx}
\usepackage{rotating}
\usepackage{threeparttable}

\usepackage{lineno}

\biboptions{authoryear}

\newcommand{\ud}{\mathrm{d}}

\renewcommand{\deg}{^\circ}

\newcommand{\au}{\,\mathrm{au}}
\newcommand{\km}{\,\mathrm{km}}
\newcommand{\meter}{\,\mathrm{m}}

\renewcommand{\second}{\,\mathrm{s}}

\begin{document}

\begin{frontmatter}

\title{Identification of meteorite source regions in the Solar System}
\author[uh,ltu]{Mikael Granvik\corref{cor1}}
\author[uwo,cpsx]{Peter Brown}
\address[uh]{Department of Physics, P.O. Box 64, 00014 University of 
Helsinki, Finland}
\address[ltu]{Department of Computer Science, Electrical and Space Engineering, Lule\aa{} University of Technology, Box 848, S-98128 Kiruna, Sweden}
\address[uwo]{Department of Physics and Astronomy, University of Western Ontario, N6A 3K7, London, Canada}
\address[cpsx]{Centre for Planetary Science and Exploration, University of Western Ontario, N6A 5B7, London, Canada}
\cortext[cor1]{Corresponding author, mgranvik@iki.fi}

\begin{abstract}
Over the past decade there has been a large increase in the number of
automated camera networks that monitor the sky for fireballs. One of
the goals of these networks is to provide the necessary information
for linking meteorites to their pre-impact, heliocentric orbits and
ultimately to their source regions in the solar system. We re-compute
heliocentric orbits for the 25 meteorite falls published to date from
original data sources. Using these orbits, we constrain their most
likely escape routes from the main asteroid belt and the cometary
region by utilizing a state-of-the-art orbit model of the
near-Earth-object population, which includes a size-dependence in
delivery efficiency. While we find that our general results for escape
routes are comparable to previous work, the role of trajectory
measurement uncertainty in escape-route identification is explored for
the first time. Moreover, our improved size-dependent delivery model
substantially changes likely escape routes for several meteorite
falls, most notably Tagish Lake which seems unlikely to have
originated in the outer main belt as previously suggested.  We find
that reducing the uncertainty of fireball velocity measurements below
$\sim0.1$~km/s does not lead to reduced uncertainties in the
identification of their escape routes from the asteroid belt and,
further, their ultimate source regions. This analysis suggests that
camera networks should be optimized for the largest possible number of
meteorite recoveries with measured speed precisions of order 0.1~km/s.
\end{abstract}


\end{frontmatter}


\section{Introduction}

Understanding meteorite source regions in the solar system is one of
the central problems in contemporary planetary science. Association of
certain meteorite classes with families of asteroids or specific
asteroids based on commonalities in their reflectance spectra is one
approach to bridging the asteroid-meteorite divide
\citep{Binzel2015}. As meteorites are delivered to the Earth through a
long process of stochastic gravitational and radiative perturbations
\citep{Morbidelli1998}, the orbit immediately prior to impact loses
memory of its original, specific parent body. However, delivery from
the main asteroid belt is sufficiently rapid and well-understood, that
statistical inferences as to the source region for meteorite classes
may be possible if pre-atmospheric orbits for enough meteorites are
measured, a process which has been understood for some time
\citep{Wetherill1985, Wisdom2017}. To date roughly two dozen meteorite
falls have had orbits measured \citep{Borovicka2015a}.

The importance of measuring large numbers of meteorite orbits has
motivated development of a new generation of all-sky fireball networks
whose primary goal is the instrumental recording of meteorite
producing fireballs 
\citep{2017ExA....43..237H,Spurny2007b,2016eMetN...1...67C}. An 
outstanding question for such fireball networks is the precision needed 
in measurement to meaningfully associate a given meteorite orbit with
potential source regions. Clearly the smaller the measurement
uncertainty the better, but at what point do reduced uncertainties no
longer improve the accuracy of statistical associations with meteorite
source regions?  Also, what are the limitations on this inversion
process imposed by the assumptions of the delivery model?

In this work we have two goals. First, we aim to examine the currently
known two dozen meteorite orbits and apply a new near-Earth object
(NEO) model which has a size dependence \citep{2016Natur.530..303G} to
estimate the probable escape/entrance routes/regions (ER) from,
primarily, the main asteroid belt for different meteorite
classes. This is the necessary first step in trying to ultimately
associate meteorites to original parent bodies, generally presumed to
be associated with asteroid collisional families. Secondly, we use the
database of two dozen meteorite orbits to examine the role measurement
errors, particularly in speed, play in the orbital uncertainty and
propagate this through the ER model of \citet{2016Natur.530..303G}. We
hope to gain quantitative insight into the corresponding source region
uncertainties. This in turn will provide operational guidance to those
building new fireball networks as to the required velocity measurement
precision for meteorite dropping fireballs needed to make meaningful
source region identifications.

\section{Methods}

\subsection{Computation of meteorite orbits}

Our computation of the pre-atmospheric heliocentric orbit and
associated uncertainties for meteorite falls follows well-established
algorithms used to convert apparent meteor trajectories to orbits
\citep{Ceplecha1987}. The method has been validated through comparison
with independent numerical approaches \citep{Clark2011}. The essence
of the algorithm is to use an in-atmosphere local trajectory for a
fireball and then translate this to an equivalent heliocentric
orbit. Here we take the trajectories reported in various literature
sources and transform the measured Right Ascension (RA) and
Declination (Dec) of an apparent radiant and the measured speed
$v_\infty$ to the initial state vector. We assume that the average
speed $\overline v=v_\infty-1.0\km\second^{-1}$. The nominal
uncertainties on all measured quantities are also taken from published
sources when available.  The Meteor Toolkit
\citep{2015P&SS..117..223D}, the only open-source implementation of
meteor orbit computation software as far as we know, would have been
an alternative to our in-house orbit computation software.  As it
produces very similar results to the technique based on
\citet{Ceplecha1987}, which was also used in the majority of the
fireball producing meteorite fall studies published to date, we chose
to remain with that original software.

A major systematic uncertainty in all meteorite computed orbits is the
correction applied for pre-luminous flight deceleration. For most
published meteorite orbits, insufficient information is provided in
the original references to compute this correction or even determine
if it has been applied to the final orbital elements. We do not
explore this correction further, but merely emphasize that for those
studies which do not model the meteoroid entry, the apparent initial
speed may be underestimated by of order a few tens of meters per
second in some extreme cases.

\subsection{Prediction of meteorite escape routes}
\label{sec:sourcetheory}

A key ingredient for predicting a meteoroid's likely source region in
the solar system is a model describing NEO orbit and size
distributions \citep{2016Natur.530..303G}. The model also provides
probabilistic information about NEO ERs as a function of semimajor
axis $a$, eccentricity $e$, inclination $i$ and the absolute magnitude
$H$ for NEOs with $17<H<25$. The $H$ range corresponds to a diameter
$35\meter \lesssim D \lesssim 1400\meter$ when assuming an average
geometric albedo $p_V=0.14$ \citep{2012Icar..221..365P}. The model
accounts for 7 different ERs in the solar system: Hungaria asteroids,
Phocaea asteroids and Jupiter-family comets as well as asteroids
escaping through the $\nu_6$ secular resonance or the 3:1J, 5:2J and
2:1J mean-motion resonances (MMR) with Jupiter. We note that less
important yet non-negligible escape routes have been incorporated into
the adjacent main resonances.

When neglecting correlations between orbital elements derived from a
meteoroid trajectory, the probability for a meteoroid with orbital
elements and corresponding uncertainties ($a \pm \delta a$,$e \pm
\delta e$,$i \pm \delta i$) to originate in ER $s$ is
\begin{equation}
\begin{split}
p_s(a, e, i, \delta a, \delta e, \delta i) = \\
 \hspace{-10pt} \frac{\displaystyle{\int_{a-\delta a}^{a+\delta a}\int_{e-\delta e}^{e+\delta e}\int_{i-\delta i}^{i+\delta i}\int_{H=17}^{25}N_s(a,e,i,H)\,\ud a\,\ud e\,\ud i\,\ud H}}{\displaystyle{\sum_s \int_{a-\delta a}^{a+\delta a}\int_{e-\delta e}^{e+\delta e}\int_{i-\delta i}^{i+\delta i}\int_{H=17}^{25} N_s(a,e,i,H)\,\ud a\,\ud e\,\ud i\,\ud H}}\,, \label{eq:sourcepint}
\end{split}
\end{equation}
where $N_s(a,e,i,H)$ is the debiased differential number distribution
of NEOs originating in ER $s$. So in essence the probability is the
ratio of the number of asteroids on the specified orbit originating in
ER $s$ and the total number of asteroids on the specified orbit. In
practice, $N_s(a,e,i,H)$ is discretized with bin sizes ($\Delta a$,
$\Delta e$, $\Delta i$, $\Delta H$) and used as a four-dimensional
look-up table. In what follows Eq.~\ref{eq:sourcepint} is therefore
approximated as
\begin{equation}
\begin{split}
p_s(a, e, i, \delta a, \delta e, \delta i, \Delta a, \Delta e, \Delta i) = \\
 \hspace{-10pt} \frac{\displaystyle{\sum_{j=j_\mathrm{min}}^{j_\mathrm{max}}\sum_{k=k_\mathrm{min}}^{k_\mathrm{max}}\sum_{l=l_\mathrm{min}}^{l_\mathrm{max}}\sum_{m=1}^M N_s(a_j,e_k,i_l,H_m)}}{\displaystyle{\sum_{s=1}^7 \sum_{j=j_\mathrm{min}}^{j_\mathrm{max}}\sum_{k=k_\mathrm{min}}^{k_\mathrm{max}}\sum_{l=l_\mathrm{min}}^{l_\mathrm{max}}\sum_{m=1}^M N_s(a_j,e_k,i_l,H_m)}}\,, \label{eq:sourcepsum}
\end{split}
\end{equation}
where the summation over $s$ is over all 7 ERs and the summation over
$m$ is over all $M$ $H$ bins available. The limits for the summation
over $j$ are defined as
\begin{equation*}
  j_\mathrm{min} =
  \begin{cases}
    \lceil\frac{a - \delta a - a_\mathrm{min}}{\Delta a}\rceil & \text{when } a_\mathrm{min} < a - \delta a, \\
    1 & \text{when }  a_\mathrm{min} \ge a - \delta a.
  \end{cases}
\end{equation*}
\begin{equation*}
  j_\mathrm{max} =
  \begin{cases}
    0 & \text{when } a_\mathrm{min} > a + \delta a, \\
    \lceil\frac{a + \delta a - a_\mathrm{min}}{\Delta a}\rceil & \text{when } a_\mathrm{min} < a + \delta a < a_\mathrm{max}, \\
    J & \text{when }  a_\mathrm{max} \le a + \delta a.
  \end{cases}    
\end{equation*}
where $a_\mathrm{min}$ and $a_\mathrm{max}$ are the minimum and
maximum semimajor axes covered by the model, respectively, $J$ is the
number of semimajor axis bins in the discretized model, and $\lceil x
\rceil$ denotes the ceiling of $x$. The limits for $k$ and $l$ are
obtained by replacing ($j_\mathrm{min}$, $j_\mathrm{max}$, $a$,
$\delta a$, $a_\mathrm{min}$, $a_\mathrm{max}$, $\Delta a$, $J$) with
($k_\mathrm{min}$, $k_\mathrm{max}$, $e$, $\delta e$,
$e_\mathrm{min}$, $e_\mathrm{max}$, $\Delta e$, $K$) and
($l_\mathrm{min}$, $l_\mathrm{max}$, $i$, $\delta i$,
$i_\mathrm{min}$, $i_\mathrm{max}$, $\Delta i$, $L$), respectively.

The model by \citet{2016Natur.530..303G} is bounded by
$a_\mathrm{min}=0.3\au$, $a_\mathrm{max}=4.2\au$, $e_\mathrm{min}=0$,
$e_\mathrm{max}=1$, $i_\mathrm{min}=0\deg$, and
$i_\mathrm{max}=180\deg$. Note that \citet{2016Natur.530..303G} did
not force $a_\mathrm{min}=0.3\au$, but this is a naturally occuring
lower limit caused by the fact that there does not exist a mechanism
to decouple an asteroid's orbit from that of Mercury. In terms of the
resolution ($\Delta a, \Delta e, \Delta i$) we here use both low
resolution ($0.1\au, 0.04, 4\deg$) and high resolution ($0.05\au,
0.02, 2\deg$).

As with the NEO model of \citet{bot2002a}, the calculation assumes
that the meteoroid orbit is similar to the orbit of its asteroidal (or
cometary) parent body and the estimate for the source region applies,
strictly speaking, to that parent body. We emphasize that, although
large objects affect $p_s$, this value is primarily driven by the
smallest objects included in the NEO model, that is, those with
diameters of a few tens of meters or so, due to the steep NEO size
distribution.

\section{Data}
\label{sec:data}

We collected the observed trajectories as well as meteorite
classifications, densities, and cosmic-ray exposure ages (the latter
three collectively called geophysical data hereafter) for known
fireball-producing meteorite falls from the literature (Tables
\ref{table:trajectorydata} and \ref{table:geophysdata}). The list of
meteorite falls should be complete less a few recent and as of yet
unpublished events. All meteorite-producing fireballs with orbits that
we have used have trajectories and speeds computed using some
combination of instrumental records, including dedicated camera
networks, calibrated casual video recordings, seismic, infrasound,
satellite or pre-atmospheric telescopic detection. Details of the
nature of each event and quality of the resulting orbits is discussed
in \citet{Borovicka2015a}. We have extracted from the original
published reference for each event (all primary references are listed
in Tables \ref{table:trajectorydata}) the local trajectory and
corresponding quoted uncertainties, and then independently computed
orbits for consistency.

\begin{sidewaystable*}
  \begin{threeparttable}
  \caption{Equivalent trajectory data for meteorite falls having
    instrumentally measured orbits. All angular elements are
    J2000.0. Latitude and longitude for the reference positions, 
    equivalent equatorial geocentric radiant coordinates,
    speeds prior to deceleration from the Earth's atmosphere
    ($v_{\infty}$) are extracted from references, but modified to
    agree with published primary data in some cases. In cases where
    references do not provide formal errors, the last significant
    figure is taken as the uncertainty.}\label{table:trajectorydata}
  \begin{tabular}{lcrrrrrcc}
\hline
Meteorite name & \multicolumn{2}{c}{Fall date and time} & \multicolumn{1}{c}{Lat} & \multicolumn{1}{c}{Lon} & \multicolumn{1}{c}{RA} & \multicolumn{1}{c}{Dec} & \multicolumn{1}{c}{$v_\infty$}  & \multicolumn{1}{c}{Ref.} \\
& \multicolumn{2}{c}{UTC} & \multicolumn{1}{c}{[deg]} & \multicolumn{1}{c}{[deg]} & \multicolumn{1}{c}{[deg]} & \multicolumn{1}{c}{[deg]} & \multicolumn{1}{c}{[km/s]} & \\
\hline
P\v{r}\'ibram              &  1959-04-07 &  19.5033  &  49.510 & 14.830 & $192.34 \pm 0.01$ & $ 17.47  \pm 0.01$ & $ 20.93 \pm 0.01  $  &	$^{27,34,35}$    \\
Lost City                  &  1970-01-04 &  2.2333   &  36.005 & -95.090 & $315.0 \pm 0.1$ & $ 39.0  \pm 0.1$ & $ 14.235 \pm 0.002  $  &	$^{18,19,20}$    \\
Innisfree                  &  1977-02-06 &  2.2939   &  53.415 & -111.338 & $  7.43 \pm 0.3$ & $ 66.52  \pm 0.01$ & $ 14.5 \pm 0.1  $  &	$^{13,14}$       \\
Bene\v sov                 &  1991-05-07 &  23.0647  &  49.662 & 14.635 & $227.62 \pm 0.01$ & $ 39.909  \pm 0.002$ & $ 21.272 \pm 0.005  $  &	$^{4,5}$         \\
Peekskill                  &  1992-10-09 &  23.8000  &  39.663 & -78.206 & $208.9 \pm 0.2$ & $ -29.2 \pm 0.3$ & $ 14.72 \pm 0.050  $  &	$^{27,32,33}$    \\
Tagish Lake                &  2000-01-18 &  16.7283  &  60.000 & -134.200 & $ 90.4 \pm 1.9$ & $ 29.9  \pm 2.8$ & $ 15.8 \pm 0.6  $  &	$^{38,39,40}$    \\
Mor\'avka                  &  2000-05-06 &  11.8639  &  50.230 & 18.450 & $250.1 \pm 0.7$ & $ 54.95  \pm 0.25$ & $ 22.5 \pm 0.3 $  &	$^{26,27}$       \\
Neuschwanstein             &  2002-04-06 &  20.3383  &  47.304 & 11.552 & $192.33 \pm 0.03$ & $ 19.55  \pm 0.04$ & $ 20.960 \pm 0.040  $  &	$^{28,29}$       \\
Park Forest                &  2003-03-27 &  5.8406   &  41.130 & -87.900 & $171.8 \pm 1.3$ & $ 11.2  \pm 0.5$ & $ 19.5 \pm 0.3  $  &	$^{31}$       \\
Villalbeto de la Pe\~na    &  2004-01-04 &  16.7792  &  42.771 & -4.789 & $311.4 \pm 1.4$ & $ -18.0 \pm 0.7$ & $ 16.90 \pm 0.40  $  &	$^{41,42}$       \\
Bunburra Rockhole          &  2007-07-20 &  19.2311  &  -31.450 & 129.827 & $ 80.70 \pm 0.05$ & $ -14.22 \pm 0.04$ & $ 13.395 \pm 0.007  $  &	$^{6}$           \\
Almahata Sitta             &  2008-10-07 &  2.7611   &  20.858 & 31.804 & $348.1 \pm 0.1$ & $ 7.6   \pm 0.1$ & $ 12.760 \pm 0.001  $  &	$^{1,2}$         \\
Buzzard Coulee             &  2008-11-21 &  0.4453   &  53.183 & -109.875 & $290.0 \pm 0.7$ & $ 77.0  \pm 0.3$ & $ 18.05 \pm 0.40 $  &	$^{7,8}$         \\
Maribo                     &  2009-01-17 &  19.1411  &  54.585 & 13.657 & $124.7 \pm 1.0$ & $ 19.7  \pm 0.5$ & $ 28.30 \pm 0.20  $  &	$^{21,22,23,24}$ \\
Jesenice                   &  2009-04-09 &  0.9944   &  46.662 & 13.692 & $159.9 \pm 1.2$ & $ 58.7  \pm 0.5$ & $ 13.80 \pm 0.25  $  &	$^{15}$          \\
Grimsby                    &  2009-09-26 &  1.0496   &  43.534 & -80.194 & $242.60 \pm 0.26$ & $ 54.98  \pm 0.12$ & $ 20.95 \pm 0.19  $  &	$^{12}$        \\
Ko\v sice                  &  2010-02-28 &  22.4128  &  20.705 & 48.667 & $114.3 \pm 1.7$ & $ 29.0  \pm 3.0$ & $ 14.90 \pm 0.35 $  &	$^{16}$          \\
Mason Gully                &  2010-04-13 &  10.6036  &  -30.275 & 128.215 & $148.36 \pm 0.05$ & $ 9.00   \pm 0.05$ & $ 14.68 \pm 0.011  $  &	$^{25}$          \\
Kri\v zevci                &  2011-02-04 &  23.3444  &  45.733 & 16.430 & $131.22 \pm 0.05$ & $ 19.53  \pm 0.04$ & $ 18.21 \pm 0.07  $  &	$^{17}$          \\
Sutter's Mill              &  2012-04-22 &  14.6036  &  38.804 & -120.908 & $ 24.0 \pm 1.3$ & $ 12.7  \pm 1.7$ & $ 28.6 \pm 0.6  $  &	$^{36,37}$       \\
Novato                     &  2012-10-18 &  2.7417   &  36.295 & -123.463 & $268.1 \pm 0.5$ & $ -48.9 \pm 0.6$ & $ 13.75 \pm 0.12  $  &	$^{30}$          \\
Chelyabinsk                &  2013-02-15 &  3.3389   &  54.454 & 64.477 & $332.81 \pm 0.14$ & $ 0.29   \pm 0.14$ & $ 19.03 \pm 0.13  $  &	$^{9,10}$        \\
Annama                     &  2014-04-18 &  22.2358  &  68.775 & 30.787 & $213.0 \pm 0.2$ & $ 8.8   \pm 0.4$ & $ 24.20 \pm 0.50 $  &	$^{3}$           \\
\v Zd'\'ar nad S\'azavou   &  2014-12-09 &  16.2806  &  49.941 & 18.002 & $ 69.31 \pm 0.02$ & $ 26.98  \pm 0.02$ & $ 21.97 \pm 0.030  $  &	$^{43}$          \\
Ejby                       &  2016-02-06 &  21.0750  &  55.449 & 11.912 & $ 77.74 \pm 0.09$ & $ 26.89  \pm 0.38$ & $ 14.50 \pm 0.100  $  &	$^{11}$          \\
\hline
\end{tabular}
\begin{tablenotes}
\small
\item $^1$\cite{jen2009a} $^2$\cite{Welten2010}
  $^3$\cite{Trigo-Rodriguez2015} $^4$\cite{Spurny1994}
  $^5$\cite{Borovicka1998} $^6$\cite{Spurny2012a}
  $^7$\cite{Milley2010a} $^8$\cite{Milley2010}
  $^9$\cite{Borovicka2013a} $^{10}$\cite{Brown2013}
  $^{11}$\cite{Spurny2016a} $^{12}$\cite{Brown2011}
  $^{13}$\cite{Halliday1978}
  $^{14}$\cite{Halliday1981} $^{15}$\cite{Spurny2010}
  $^{16}$\cite{Borovicka2013b} $^{17}$\cite{Borovicka2015b}
  $^{18}$\cite{McCrosky1971} $^{19}$\cite{Ceplecha1996a}
  $^{20}$\cite{Ceplecha2005} $^{21}$\cite{Keurer2009}
  $^{22}$\cite{Haak2010} $^{23}$\cite{Haak2011a}
  $^{24}$\cite{SpurnyP.a2013} $^{25}$\cite{Spurny2012b}
  $^{26}$\cite{Borovicka2003a} $^{27}$\cite{Borovicka2003b}
  $^{28}$\cite{Spurny2003} $^{29}$\cite{Spurny2002}
  $^{30}$\cite{Jenniskens2014} $^{31}$\cite{Brown2004}
  $^{32}$\cite{Brown1994}
  $^{33}$\cite{Ceplecha1996} $^{34}$\cite{Ceplecha1961}
  $^{35}$\cite{Ceplecha1977} $^{36}$\cite{Jenniskens2012}
  $^{37}$\cite{Nishiizumi2014} $^{38}$\cite{Brown2001a}
  $^{39}$\cite{Hildebrand2006} $^{40}$\cite{Brown2002f}
  $^{41}$\cite{Trigo-Rodriguez2006b} $^{42}$\cite{Llorca2005}
  $^{43}$\cite{Spurny2015} 
\end{tablenotes}
\end{threeparttable}
\end{sidewaystable*}

\begin{table*}
  \begin{threeparttable}
  \caption{Geophysical data for meteorite falls having instrumentally
    measured orbits. This includes meteorite classification, bulk
    density and cosmic-ray exposure age (CRE). All data are extracted
    from the associated references.}\label{table:geophysdata}
  \begin{tabular}{lcccccc}
\hline
Meteorite name & \multicolumn{1}{c}{Classification} & \multicolumn{1}{c}{Ref.} & \multicolumn{1}{c}{Bulk density} & \multicolumn{1}{c}{Ref.} & \multicolumn{1}{c}{CRE age} & \multicolumn{1}{c}{Ref.}  \\
& & & [kg$\,$m$^{-3}$] & & [Myr] & \\
\hline
P\v{r}\'ibram                    & H5        &$^{73}$ & 3570 & $^{46}$ & 12 & $^{47}$ \\
Lost City                  & H5        &$^{74}$ & -- & & 8 & $^{48}$ \\
Innisfree                  & L5/LL5(?) &$^{75}$ & -- & & 26--28 & $^{49}$ \\
Bene\v sov                    & H5/LL3.5  &$^{76}$ & -- & & -- & \\
Peekskill                  & H6        &$^{77}$ & -- & & 32 & $^{50}$ \\
Tagish Lake                & C2        & $^{40}$ & 1640 & $^{41}$ & >3 & $^{51}$ \\
Mor\'avka                    & H5        & $^{28}$ & 3590 & $^{28}$ & 5.7--7.7 & $^{28}$ \\
Neuschwanstein             & EL6       & $^{52}$ & 3500 & $^{53}$ & 43--51 & $^{52}$ \\
Park Forest                & L5        & $^{33}$ & 3370 & $^{54}$ & 12--16 & $^{55}$ \\
Villalbeto de la Pe\~na      & L6        & $^{44}$ & 3420 & $^{44}$ & 43--53 & $^{44}$ \\
Bunburra Rockhole          & Euc-Anom  & $^{56}$ & -- & & 22 & $^{57}$ \\
Almahata Sitta             & Ure-Anom  & $^1$ & 2900--3300 & $^{58}$ & 16--22 & $^{59}$ \\
Buzzard Coulee             & H4        & $^{60}$ & 3370--3550 & $^{61}$ & -- & \\
Maribo                     & CM2       & $^{62}$ & -- & & 0.8--1.4 & $^{62}$ \\
Jesenice                   & L6        & $^{63}$ & -- & & 4 & $^{63}$ \\
Grimsby                    & H5        & $^{64}$ & 3340--3370 & $^{64}$ & 21--26 & $^{13}$ \\
Ko\v sice                     & H5        & $^{65}$ & 3320--3540 & $^{66}$ & 5--7 &$^{78}$ \\
Mason Gully                & H5        & $^{67}$ & 3280--3360 & $^{54}$ & -- & \\
Kri\v zevci                   & H6        & $^{68}$ & -- & & -- & \\
Sutter's Mill              & CM2       & $^{69}$ & 2156--2358 & $^{38}$ &0.074--0.09 &$^{39}$ \\
Novato                     & L6        & $^{31}$ & 3190--3350 & $^{31}$ & 8--10 & $^{31}$ \\
Chelyabinsk                & LL5       & $^{70}$ & 3321--3329 & $^{70}$ & 1 & $^{71}$ \\
Annama                     & H5        & $^{72}$ & 3500 & $^{72}$ & 26--34 & $^{72}$ \\
\v Zd'\'ar nad S\'azavou  & L3/L3.9   & $^{45}$ & 3050 & $^{45}$ & -- & \\
Ejby                       & H5/6      & $^{11}$ & -- & & -- & \\
\hline
\end{tabular}
\begin{tablenotes}
\small
\item $^1$\cite{jen2009a} $^{11}$\cite{Spurny2016a} 
  $^{13}$\cite{Cartwright2010} $^{28}$\cite{Borovicka2003b}
  $^{31}$\cite{Jenniskens2014} $^{33}$\cite{Simon2004} 
  $^{38}$\cite{Jenniskens2012}
  $^{39}$\cite{Nishiizumi2014} $^{40}$\cite{Brown2001a}
  $^{41}$\cite{Hildebrand2006} $^{44}$\cite{Llorca2005}
  $^{45}$\cite{Spurny2015} $^{46}$\cite{Britt2003} 
  $^{47}$\cite{Stauffer1962}
  $^{48}$\cite{Baxter1971}
  $^{49}$\cite{Goswami1978}
  $^{50}$\cite{Graf1997}
  $^{51}$\cite{Herzog2014}
  $^{52}$\cite{Zipfel2010}
  $^{53}$\cite{Kohout2010} 
  $^{54}$\cite{Macke2010}
  $^{55}$\cite{Meier2017} 
  $^{56}$\cite{Bland2009} 
  $^{57}$\cite{Welten2012}
  $^{58}$\cite{Kohout2011}
  $^{59}$\cite{Horstmann2014}  
  $^{60}$\cite{Hutson2009} 
  $^{61}$\cite{Fry2013}  
  $^{62}$\cite{Haak2012}
  $^{63}$\cite{Bischoff2011}
  $^{64}$\cite{McCausland2010a}
  $^{65}$\cite{Ozdin2015}
  $^{66}$\cite{Kohout2014b}
  $^{67}$\cite{Dyl2016}
  $^{68}$\cite{Lyon2014}
  $^{69}$\cite{Zolensky2014}
  $^{70}$\cite{Kohout2014a}
  $^{71}$\cite{Righter2015}
  $^{72}$\cite{Kohout2016}
  $^{73}$\cite{Tucek1961}
  $^{74}$\cite{Nava1971}
  $^{75}$\cite{Kallemeyn1989}
  $^{76}$\cite{Spurny2014}
  $^{77}$\cite{Wlotzka1993}
  $^{78}$\cite{Povinec2015}
\end{tablenotes}
\end{threeparttable}
\end{table*}

\section{Results and discussion} 

\subsection{Orbits of known meteorite falls}

We recomputed the orbits of known meteorite falls based on the
observational data presented in Sect.~\ref{sec:data}, using the
original information from literature sources as much as possible. In
some cases we could not independently match the published orbit with
the radiant and speed given in the same reference. In those cases we
have tried to use the original trajectory information (and stated
errors) as much as possible to recompute the orbit from fundamental
measurements and propagate the uncertainties through to a final
geocentric radiant and orbit. Where this was not possible, we match
the geocentric radiant and geocentric velocity as published to produce
the final orbital elements, even if the final orbits differ slightly
from published values. In most cases these systematic differences are
much smaller than the formal uncertainties. In some cases, quoted
errors represent only the uncertainty in the mathematical model fit in
the trajectory solution to data. This ignores other uncertainties in
the system such as centroid pick errors or differences in calibration
precision within the image as well as model assumptions of the fit.

These differences are often due to ambiguities in the original
reference, such as whether apparent radiants refer to the beginning of
the trajectory or the mid-point (or end). We assume where this is not
stated that apparent radiants are computed for the beginning of the
fireball.

In Table \ref{table:trajectorydata} we refer to an "equivalent"
radiant and/or speed which produces the literature values of the
original apparent trajectory or geocentric radiant/geocentric
velocity. The final orbits are then found using our adopted
corrections as described above.

The resulting orbits are similar to typical NEO orbits with 20 falls
having nominal semimajor axes inside the 3:1J mean-motion resonance
(MMR) with Jupiter (J) and the remaining four having nominal semimajor
axes suggesting an origin in the outer main belt exterior to the 3:1J
MMR but inside the 5:2J MMR (Table \ref{table:orbits}). The 1-$\sigma$
uncertainty on the semimajor axis is typically $\delta a < 0.05\au$
but for the C and CM meteorite falls the uncertainty is $\delta a \sim
0.3\au$ which immediately suggests that the ER for C and CM meteorites
cannot be accurately determined. The other elements are
well-constrained less two falls with essentially unconstrained
longitude of ascending node and argument of perihelion due to near
zero-inclination orbits.

Our recomputed orbits are generally in statistical agreement with
those reported in the original literature sources. However, some
meteorite orbits do show noticeable differences; for example in cases
where complete initial state vector information (like apparent
radiant) are omitted (such as Lost City and Annama). In other cases,
these differences are due to the low inclination of the orbits (e.g.,
Maribo and Villalbeto de la Pe\~na) where small differences in corrections will make large differences to the angular
elements. For Chelyabinsk, we have two independent literature
estimates of the original orbit and a low inclination which makes our
independent recomputed orbital elements differ slightly from both
literature values. We have chosen to use our recomputed values using
uniform corrections (like average speed) recognizing there may remain
small systematic uncertainties due to lack of consistency in reported
state vector quantities. This is in addition to quality differences
resulting from the techniques used to make trajectory measurements
\citep[cf.][]{Borovicka2015a}.

\begin{sidewaystable*}
    \caption{Orbital elements for meteorite-dropping
      fireballs. Semimajor axis ($a$), eccentricity $e$, inclination
      $i$, longitude of ascending node $\Omega$, argument of
      perihelion $\omega$.}\label{table:orbits}
    \begin{tabular}{lcccccc}
\hline
Meteorite name & \multicolumn{1}{c}{$a$ [au]} & \multicolumn{1}{c}{$e$} & \multicolumn{1}{c}{$i$ [deg]} & \multicolumn{1}{c}{$\Omega$ [deg]} & \multicolumn{1}{c}{$\omega$ [deg]} & \multicolumn{1}{c}{MJD UTC} \\
\hline
P\v{r}\'ibram                   & $ 2.4050 \pm 0.0043 $ & $ 0.6717 \pm 0.0006 $ & $ 10.4841 \pm 0.0086 $ & $ 17.8019 \pm 0.0000 $  & $ 241.723 \pm 0.020 $ & 36665.812639 \\
Lost City                 & $ 1.6486 \pm 0.0037 $ & $ 0.4133 \pm 0.0013 $ & $ 11.940 \pm 0.025 $   & $ 283.7568 \pm 0.0000 $ & $ 160.67 \pm 0.14 $   & 40590.093056 \\
Innisfree                 & $ 1.866 \pm 0.051 $   & $ 0.472 \pm 0.014 $   & $ 12.250 \pm 0.22 $    & $ 317.5223 \pm 0.0002 $ & $ 177.97 \pm 0.13 $   & 43180.095579 \\
Bene\v sov                   & $ 2.4826 \pm 0.0024 $ & $ 0.6273 \pm 0.0004 $ & $ 23.9807 \pm 0.0074 $ & $ 47.0004 \pm 0.0000 $  & $ 218.369 \pm 0.017 $ & 48383.961030 \\
Peekskill                 & $ 1.489 \pm 0.015 $   & $ 0.4052 \pm 0.0055 $ & $ 4.88 \pm 0.15 $      & $ 17.0289 \pm 0.0001 $  & $ 307.46 \pm 0.61 $   & 48904.991667 \\
Tagish Lake               & $ 1.98 \pm 0.19 $     & $ 0.554 \pm 0.045 $   & $ 2.06 \pm 0.97 $      & $ 297.9013 \pm 0.0007 $ & $ 224.3 \pm 2.5 $     & 51561.697014 \\
Mor\'avka                   & $ 1.851 \pm 0.079 $   & $ 0.469 \pm 0.022 $   & $ 32.28 \pm 0.42 $     & $ 46.2573 \pm 0.0000 $  & $ 203.5 \pm 1.8 $     & 51670.494329 \\
Neuschwanstein            & $ 2.398 \pm 0.017 $   & $ 0.6693 \pm 0.0024 $ & $ 11.416 \pm 0.032 $   & $ 16.8257 \pm 0.0000 $  & $ 241.185 \pm 0.061 $ & 52370.847431 \\
Park Forest               & $ 2.53 \pm 0.21 $     & $ 0.680 \pm 0.025 $   & $ 3.25 \pm 0.66 $      & $ 6.1153 \pm 0.0003 $   & $ 237.5 \pm 2.0 $     & 52725.243356 \\
Villalbeto de la Pe\~na     & $ 2.30 \pm 0.22 $     & $ 0.627 \pm 0.036 $   & $ 0.02 \pm 0.49 $      & $ 282 \pm 180 $         & $ 134 \pm 180 $       & 53008.699132 \\
Bunburra Rockhole         & $ 0.8530 \pm 0.0010 $ & $ 0.2425 \pm 0.0014 $ & $ 8.937 \pm 0.072 $    & $ 297.5945 \pm 0.0000 $ & $ 210.04 \pm 0.15 $   & 54301.801296 \\
Almahata Sitta            & $ 1.3085 \pm 0.0003 $ & $ 0.3060 \pm 0.0001 $ & $ 2.4373 \pm 0.0023 $  & $ 194.0822 \pm 0.0000 $ & $ 232.405 \pm 0.019 $ & 54746.115046 \\
Buzzard Coulee            & $ 1.246 \pm 0.029 $   & $ 0.228 \pm 0.018 $   & $ 25.07 \pm 0.78 $     & $ 238.9374 \pm 0.0001 $ & $ 211.3 \pm 1.5 $     & 54791.018553 \\
Maribo                    & $ 2.48 \pm 0.27 $     & $ 0.807 \pm 0.015 $   & $ 0.1 \pm 1.3 $        & $ 297 \pm 48 $          & $ 279 \pm 49 $        & 54848.797546 \\
Jesenice                  & $ 1.744 \pm 0.068 $   & $ 0.429 \pm 0.023 $   & $ 9.58 \pm 0.48 $      & $ 19.1950 \pm 0.0002 $  & $ 190.49 \pm 0.54 $   & 54930.041435 \\
Grimsby                   & $ 2.035 \pm 0.048 $   & $ 0.518 \pm 0.011 $   & $ 28.08 \pm 0.28 $     & $ 182.9560 \pm 0.0000 $ & $ 159.85 \pm 0.24 $   & 55100.043731 \\
Ko\v sice                    & $ 2.73 \pm 0.23 $     & $ 0.650 \pm 0.030 $   & $ 1.99 \pm 0.47 $      & $ 340.0713 \pm 0.0024 $ & $ 204.2 \pm 1.5 $     & 55255.933866 \\
Mason Gully               & $ 2.5566 \pm 0.0094 $ & $ 0.6159 \pm 0.0014 $ & $ 0.895 \pm 0.032 $    & $ 203.2134 \pm 0.0015 $ & $ 19.01 \pm 0.08 $    & 55299.441817 \\
Kri\v zevci                  & $ 1.5428 \pm 0.0087 $ & $ 0.5205 \pm 0.0034 $ & $ 0.638 \pm 0.026 $    & $ 315.5544 \pm 0.0035 $ & $ 254.383 \pm 0.081 $ & 55596.972685 \\
Sutter's Mill             & $ 2.57 \pm 0.38 $     & $ 0.823 \pm 0.022 $   & $ 2.4 \pm 1.5 $        & $ 32.716 \pm 0.015 $    & $ 77.9 \pm 3.3 $      & 56039.608484 \\
Novato                    & $ 2.091 \pm 0.079 $   & $ 0.528 \pm 0.018 $   & $ 5.52 \pm 0.11 $      & $ 24.9625 \pm 0.0002 $  & $ 347.36 \pm 0.30 $   & 56218.114236 \\
Chelyabinsk               & $ 1.770 \pm 0.022 $   & $ 0.5792 \pm 0.0061 $ & $ 4.78 \pm 0.13 $      & $ 326.4136 \pm 0.0003 $ & $ 109.22 \pm 0.18 $   & 56338.139120 \\
Annama                    & $ 2.08 \pm 0.13 $     & $ 0.690 \pm 0.021 $   & $ 14.23 \pm 0.49 $     & $ 28.5990 \pm 0.0001 $  & $ 263.08 \pm 0.74 $   & 56765.926493 \\
\v Zd'\'ar nad S\'azavou & $ 2.1020 \pm 0.0088 $ & $ 0.6808 \pm 0.0015 $ & $ 2.809 \pm 0.018 $    & $ 257.2618 \pm 0.0002 $ & $ 257.748 \pm 0.043 $ & 57000.678356 \\
Ejby                      & $ 2.805 \pm 0.094 $   & $ 0.655 \pm 0.012 $   & $ 0.958 \pm 0.098 $    & $ 317.2078 \pm 0.0080 $ & $ 197.75 \pm 0.25 $   & 57424.878125 \\
\hline
\end{tabular}
\end{sidewaystable*}

None of the meteoroid orbits have perihelion distances
$q\lesssim0.4\au$, that is, within the orbit of Mercury (Table
\ref{table:qcapqvgvh}). A lack of low-$q$ orbits has also been
reported for fireballs produced by meter-scale Earth impacting
meteoroids in general \citep{2016Icar..266...96B}. A similar lack of
objects having orbits with q < 0.4 AU is present in the collection of
57 fireballs reported by \citet{Halliday1996} having terminal masses
in excess of 100~g and therefore probable meteorite
producers. Super-catastrophic disruptions of asteroids at small $q$
was recently proposed by \citet{2016Natur.530..303G} as an explanation
for the lack of known NEOs at small $q$. As mentioned in
Sect.~\ref{sec:sourcetheory}, the model by \citet{2016Natur.530..303G}
was calibrated with the observed distribution of NEOs with $17<H<25$,
that is, diameters $35\meter \lesssim D \lesssim 1.4\km$ when assuming
a geometric albedo $p_V=0.14$. As shown in the same paper there is a
linear correlation between $H$ and the perihelion distance at which
the disruption typically happens, $q^\star$. A simple linear
extrapolation in ($H$,$q^\star$) space from the model's lower diameter
limit of about $D\approx35\meter$ down to $D\approx1\meter$
($H\approx32$) suggests that for 1-meter-diameter objects
$q^\star\approx0.38\au$. This is remarkably well in agreement with the
fireball data and suggest that the lack of meter-scale meteoroids with
$q\lesssim0.4\au$ is real and that, similar to larger asteroids, they
are destroyed at non-trivial distances from the Sun. In addition,
whatever mechanism or mechanisms are destroying these objects, the
correlation between disruption distance and size is remarkably simple
over a size range covering more than three orders of magnitude.

The aphelion distances are all well within the perihelion distance of
Jupiter indicating that these objects are detached from Jupiter's
direct influence. This suggests that they are of asteroidal rather
than cometary origin. The only border-line cases are those with
$Q\gtrsim4.5\au$, that is, Ko\v sice, Ejby, Maribo, and Sutter's
Mill. Especially for the latter two the probability for a cometary
origin is increased because of their classification as carbonaceous
chondrites. The strong anticorrelation between strong, probable
meteorite-producing fireballs, and trans-Jovian orbits is apparent in
the larger set of Prairie Network (PN) fireball data as well
\citep{wetherill1982}, indicating that very few obviously cometary
fireballs yield meteorites.

\begin{table*}
  \caption{Orbital parameters for meteorite-dropping fireballs
    computed from their apparent radiants. The epoch of osculation for
    $q$ and $Q$ is identical to that in Table \ref{table:orbits}. Here
    $v_g$ is the geocentric speed of the meteoroid prior to impact,
    that is, the speed it would have in the absence of gravitational
    acceleration, whereas $v_h$ is the heliocentic velocity of the
    meteorite at the epoch of impact.}\label{table:qcapqvgvh}
  \begin{tabular}{lcccc}
\hline
Meteorite name & \multicolumn{1}{c}{$q$ [au]} & \multicolumn{1}{c}{$Q$ [au]} & \multicolumn{1}{c}{$v_g$ [km/s]} & \multicolumn{1}{c}{$v_h$ [km/s]} \\
\hline
P\v{r}\'ibram                   & $ 0.7896 \pm 0.0001$ & $4.0204 \pm 0.0087$ & $17.437  \pm 0.012$  & $37.4600 \pm 0.0089 $ \\
Lost City                 & $ 0.9672 \pm 0.0002$ & $2.3300 \pm 0.0074$ & $ 9.1984 \pm 0.0018$ & $35.584  \pm 0.017 $ \\
Innisfree                 & $ 0.9860 \pm 0.0000$ & $2.75   \pm 0.10$   & $ 9.40   \pm 0.23$   & $36.38   \pm 0.18 $ \\
Bene\v sov                   & $ 0.9252 \pm 0.0001$ & $4.0401 \pm 0.0048$ & $18.0807 \pm 0.0059$ & $37.4271 \pm 0.0046 $ \\
Peekskill                 & $ 0.8856 \pm 0.0018$ & $2.093  \pm 0.029$  & $10.090  \pm 0.075$  & $34.369  \pm 0.087 $ \\
Tagish Lake               & $ 0.884  \pm 0.011$  & $3.08   \pm 0.39$   & $11.31   \pm 0.84$   & $36.82   \pm 0.59 $ \\
Mor\'avka                   & $ 0.9822 \pm 0.0038$ & $2.72   \pm 0.16$   & $19.60   \pm 0.23$   & $35.77   \pm 0.28 $ \\
Neuschwanstein            & $ 0.7930 \pm 0.0004$ & $4.003  \pm 0.033$  & $17.506  \pm 0.048$  & $37.453  \pm 0.034 $ \\
Park Forest               & $ 0.8107 \pm 0.0098$ & $4.251  \pm 0.419$  & $16.05   \pm 0.37$   & $37.79   \pm 0.39 $ \\
Villalbeto de la Pe\~na     & $ 0.8596 \pm 0.0074$ & $3.75   \pm 0.45$   & $12.92   \pm 0.53$   & $37.67   \pm 0.50 $ \\
Bunburra Rockhole         & $ 0.6461 \pm 0.0019$ & $1.0599 \pm 0.0003$ & $ 6.735  \pm 0.039$  & $26.571  \pm 0.022 $ \\
Almahata Sitta            & $ 0.9081 \pm 0.0000$ & $1.7088 \pm 0.0005$ & $ 6.8024 \pm 0.0019$ & $33.1258 \pm 0.0021 $ \\
Buzzard Coulee            & $ 0.9612 \pm 0.0015$ & $1.530  \pm 0.058$  & $14.19   \pm 0.51$   & $32.92   \pm 0.25 $ \\
Maribo                    & $ 0.478  \pm 0.018$  & $4.48   \pm 0.52$   & $25.80   \pm 0.22$   & $38.03   \pm 0.51 $ \\
Jesenice                  & $ 0.9965 \pm 0.0007$ & $2.49   \pm 0.14$   & $ 8.28   \pm 0.42$   & $35.54   \pm 0.28 $ \\
Grimsby                   & $ 0.9817 \pm 0.0003$ & $3.088  \pm 0.096$  & $17.89   \pm 0.22$   & $36.52   \pm 0.14 $ \\
Ko\v sice                    & $ 0.9563 \pm 0.0042$ & $4.50   \pm 0.47$   & $10.31   \pm 0.44$   & $38.29   \pm 0.36 $ \\
Mason Gully               & $ 0.9820 \pm 0.0002$ & $4.131  \pm 0.019$  & $ 9.324  \pm 0.016$  & $37.715  \pm 0.017 $ \\
Kri\v zevci                  & $ 0.7397 \pm 0.0012$ & $2.346  \pm 0.019$  & $14.456  \pm 0.089$  & $34.995  \pm 0.047 $ \\
Sutter's Mill             & $ 0.456  \pm 0.022$  & $4.69   \pm 0.74$   & $25.96   \pm 0.65$   & $37.68   \pm 0.67 $ \\
Novato                    & $ 0.9879 \pm 0.0004$ & $3.19   \pm 0.16$   & $ 8.22   \pm 0.20$   & $36.83   \pm 0.22 $ \\
Chelyabinsk               & $ 0.7447 \pm 0.0021$ & $2.795  \pm 0.045$  & $15.14   \pm 0.16$   & $35.99   \pm 0.086 $ \\
Annama                    & $ 0.6436 \pm 0.0054$ & $3.51   \pm 0.26$   & $21.45   \pm 0.56$   & $36.60   \pm 0.36 $ \\
\v Zd'\'ar nad S\'azavou & $ 0.6709 \pm 0.0003$ & $3.533  \pm 0.018$  & $18.607  \pm 0.035$  & $37.141  \pm 0.024 $ \\
Ejby                      & $ 0.9677 \pm 0.0006$ & $4.64   \pm 0.19$   & $ 9.44   \pm 0.15$   & $38.51   \pm 0.14 $ \\
\hline
\end{tabular}
\end{table*}

\subsection{Escape routes and source regions}

Based on the orbits derived above we first estimate the likely ERs for
these meteorites. Table \ref{table:sourceregionslores} provides
estimates of each meteorite-producing fireball's probable ER based on
its measured orbit using a debiased low-resolution model for NEO ERs
(see Sect.~\ref{sec:sourcetheory}). The differences are negligible
when using an 8 times higher resolution for the orbital distribution
as shown in Table \ref{table:sourceregionshires}. The insensitivity to
the resolution of the orbital distribution suggests that even the low
resolution is able to reproduce the typical dynamical features in the
steady-state orbit distributions for each ER. It thus implies that
estimates of the most likely ERs will not become more accurate by
increasing the resolution of the orbit model (see discussion in
Sect.~\ref{sec:velsens}).

\begin{sidewaystable*}
\begin{center}
  \caption{Probabilities of different escape regions for meteorite
    falls based on the low-resolution NEO orbit model. Meteorite
    classification has been included to help
    interpretation.}\label{table:sourceregionslores}
  \begin{tabular}{lcccccccc}
    \hline
    Meteorite name & Classification & \multicolumn{1}{c}{Hungaria} & \multicolumn{1}{c}{$\nu_6$} & \multicolumn{1}{c}{Phocaea} & \multicolumn{1}{c}{3:1J} & \multicolumn{1}{c}{5:2J} & \multicolumn{1}{c}{2:1J} & \multicolumn{1}{c}{JFC} \\
& & [\%] & [\%] & [\%] & [\%] & [\%] & [\%] & [\%] \\
    \hline
P\v{r}\'ibram & H5 & $0.8 \pm 0.4$ & $15 \pm 7$ & $0.0 \pm 0.3$ & $70 \pm 20$ & $10 \pm 9$ & $0.0 \pm 0.2$ & $2 \pm 2$ \\
Lost City & H5 & $24 \pm 7$ & $69 \pm 8$ & $0.0 \pm 0.2$ & $7 \pm 2$ & $0.3 \pm 0.2$ & $0.02 \pm 0.06$ & $0 \pm 0$ \\
Innisfree & L5/LL5? & $28 \pm 4$ & $60 \pm 4$ & $0.0 \pm 0.5$ & $11 \pm 2$ & $1.1 \pm 0.5$ & $0.0 \pm 0.1$ & $0 \pm 0$ \\
Bene\v sov & H5/LL3.5 & $0.8 \pm 0.4$ & $5 \pm 2$ & $0 \pm 7$ & $80 \pm 30$ & $9 \pm 9$ & $0.1 \pm 0.3$ & $2 \pm 3$ \\
Peekskill & H6 & $20 \pm 3$ & $72 \pm 4$ & $0.00 \pm 0.07$ & $7 \pm 1$ & $0.17 \pm 0.08$ & $0.0 \pm 0.1$ & $0 \pm 0$ \\
Tagish Lake & C2 & $6.7 \pm 0.6$ & $82 \pm 2$ & $0.00 \pm 0.01$ & $11 \pm 1$ & $0.28 \pm 0.09$ & $0.004 \pm 0.006$ & $0.2 \pm 0.1$ \\
Mor\'avka & H5 & $61 \pm 5$ & $27 \pm 3$ & $0 \pm 2$ & $10 \pm 1$ & $1.2 \pm 0.4$ & $0.04 \pm 0.05$ & $0 \pm 0$ \\
Neuschwanstein & EL6 & $1.4 \pm 0.4$ & $26 \pm 6$ & $0.0 \pm 0.2$ & $63 \pm 13$ & $9 \pm 6$ & $0.0 \pm 0.1$ & $1 \pm 1$ \\
Park Forest & L5 & $1.6 \pm 0.2$ & $25 \pm 3$ & $0.001 \pm 0.006$ & $48 \pm 5$ & $11 \pm 4$ & $0.1 \pm 0.1$ & $15 \pm 5$ \\
Villalbeto de la Pe\~na & L6 & $3.4 \pm 0.3$ & $50 \pm 2$ & $0.001 \pm 0.006$ & $41 \pm 3$ & $2.7 \pm 0.8$ & $0.03 \pm 0.04$ & $3 \pm 1$ \\
Bunburra Rockhole & Euc- Anom & $24 \pm 7$ & $68 \pm 9$ & $0.00 \pm 0.04$ & $7 \pm 3$ & $0 \pm 0$ & $0.1 \pm 0.5$ & $0 \pm 0$ \\
Almahata Sitta & Ure - Anom & $25 \pm 7$ & $70 \pm 9$ & $0.0 \pm 0.2$ & $5 \pm 2$ & $0 \pm 0$ & $0 \pm 1$ & $0 \pm 0$ \\
Buzzard Coulee & H4 & $38 \pm 7$ & $55 \pm 7$ & $0.0 \pm 0.7$ & $5 \pm 1$ & $2 \pm 2$ & $0 \pm 0$ & $0 \pm 0$ \\
Maribo & CM2  & $2.3 \pm 0.3$ & $33 \pm 3$ & $0.000 \pm 0.006$ & $36 \pm 4$ & $6 \pm 2$ & $0.07 \pm 0.09$ & $22 \pm 8$ \\
Jesenice & L6  & $21 \pm 31$ & $70 \pm 3$ & $0.00 \pm 0.06$ & $9 \pm 1$ & $0.5 \pm 0.2$ & $0.01 \pm 0.03$ & $0 \pm 0$ \\
Grimsby & H5 & $45 \pm 6$ & $23 \pm 3$ & $0 \pm 5$ & $25 \pm 4$ & $7 \pm 2$ & $0.1 \pm 0.1$ & $0.0010 \pm 0.0008$ \\
Ko\v sice & H5 & $0.7 \pm 0.1$ & $13 \pm 2$ & $0.000 \pm 0.003$ & $37 \pm 4$ & $16 \pm 5$ & $0.08 \pm 0.09$ & $33 \pm 11$ \\
Mason Gully & H5 & $1.1 \pm 0.5$ & $21 \pm 8$ & $0 \pm 0$ & $64 \pm 18$ & $7 \pm 7$ & $0.0 \pm 0.2$ & $7 \pm 9$ \\
Kri\v zevci & H6 & $10 \pm 2$ & $73 \pm 6$ & $0.00 \pm 0.05$ & $17 \pm 4$ & $0.6 \pm 0.4$ & $0 \pm 0$ & $0 \pm 0$ \\
Sutter's Mill & CM2 & $2.6 \pm 0.3$ & $36 \pm 3$ & $0.001 \pm 0.006$ & $31 \pm 3$ & $5 \pm 1$ & $0.07 \pm 0.08$ & $26 \pm 11$ \\
Novato & L6 & $4.6 \pm 0.9$ & $89 \pm 5$ & $0.000 \pm 0.006$ & $6 \pm 1$ & $0.3 \pm 0.2$ & $0.0004 \pm 0.0006$ & $0.08 \pm 0.08$ \\
Chelyabinsk & LL5 & $8 \pm 3$ & $80 \pm 8$ & $0.0 \pm 0.1$ & $11 \pm 4$ & $0.7 \pm 0.7$ & $0.01 \pm 0.02$ & $0 \pm 0$ \\
Annama & H5 & $6.8 \pm 0.9$ & $62 \pm 3$ & $0.0 \pm 0.3$ & $25 \pm 3$ & $5 \pm 2$ & $0.04 \pm 0.07$ & $0.3 \pm 0.2$ \\
\v Zd'\'ar nad S\'azavou & L3/L3.9 & $6 \pm 1$ & $72 \pm 5$ & $0.00 \pm 0.02$ & $19 \pm 3$ & $1.0 \pm 0.5$ & $0.002 \pm 0.007$ & $1 \pm 1$ \\
Ejby & H5/6 & $0.3 \pm 0.1$ & $5 \pm 2$ & $0 \pm 0$ & $15 \pm 7$ & $32 \pm 24$ & $0.1 \pm 0.3$ & $50 \pm 30$ \\
\hline
\end{tabular}
\end{center}
\end{sidewaystable*}

\begin{sidewaystable*}
\begin{center}
  \caption{Probabilities of different escape regions for meteorite
    falls based on the high-resolution NEO orbit model. Meteorite
    classification has been included to help
    interpretation.}\label{table:sourceregionshires}
  \begin{tabular}{lcccccccc}
    \hline
    Meteorite name & Classification & \multicolumn{1}{c}{Hungaria} & \multicolumn{1}{c}{$\nu_6$} & \multicolumn{1}{c}{Phocaea} & \multicolumn{1}{c}{3:1J} & \multicolumn{1}{c}{5:2J} & \multicolumn{1}{c}{2:1J} & \multicolumn{1}{c}{JFC} \\
& & [\%] & [\%] & [\%] & [\%] & [\%] & [\%] & [\%] \\
    \hline
P\v{r}\'ibram  & H5                 & $1.0 \pm 0.5$   & $18 \pm 8$     & $0.0 \pm 0.3$     & $70 \pm 20$    & $10 \pm 10$     & $0.0 \pm 0.2$       & $2 \pm 3$ \\
Lost City & H5                & $27 \pm 5$      & $65 \pm 6$     & $0.0 \pm 0.1$     & $7 \pm 2$      & $0.1 \pm 0.1$   & $0.01 \pm 0.04$     & $0 \pm 0$ \\
Innisfree  & L5/LL5?               & $26 \pm 2$      & $61 \pm 3$     & $0.0 \pm 0.2$     & $12 \pm 1$     & $0.9 \pm 0.3$   & $0.03 \pm 0.05$     & $0 \pm 0$ \\
Bene\v sov & H5/LL3.5              & $0.5 \pm 0.3$   & $3 \pm 2$      & $0 \pm 6$         & $90 \pm 30$    & $9 \pm 9$       & $0.1 \pm 0.3$       & $2 \pm 3$ \\
Peekskill & H6                & $20 \pm 3$      & $72 \pm 4$     & $0.00 \pm 0.06$   & $8 \pm 1$      & $0.2 \pm 0.1$   & $0.02 \pm 0.06$     & $0 \pm 0$ \\
Tagish Lake  & C2             & $6.7 \pm 0.3$   & $81 \pm 1$     & $0.001 \pm 0.005$ & $11.7 \pm 0.5$ & $0.28 \pm 0.04$ & $0.004 \pm 0.005$   & $0.22 \pm 0.06$ \\
Mor\'avka  & H5                 & $62 \pm 3$      & $28 \pm 2$     & $0 \pm 1$         & $9.5 \pm 0.9$  & $0.9 \pm 0.2$   & $0.03 \pm 0.03$     & $0 \pm 0$ \\
Neuschwanstein  & EL6          & $1.4 \pm 0.4$   & $24 \pm 6$     & $0.0 \pm 0.2$     & $60 \pm 10$    & $10 \pm 7$      & $0.0 \pm 0.1$       & $1 \pm 2$ \\
Park Forest & L5              & $1.6 \pm 0.1$   & $27 \pm 2$     & $0.001 \pm 0.004$ & $53 \pm 3$     & $9 \pm 2$       & $0.08 \pm 0.06$     & $9 \pm 2$ \\
Villalbeto de la Pe\~na   & L6  & $3.4 \pm 0.2$   & $46 \pm 1$     & $0.001 \pm 0.004$ & $42 \pm 2$     & $2.6 \pm 0.4$   & $0.05 \pm 0.04$     & $6 \pm 1$ \\
Bunburra Rockhole   & Euc - Anom      & $24 \pm 7$      & $71 \pm 9$     & $0.00 \pm 0.03$   & $5 \pm 2$      & $0 \pm 0$       & $0.2 \pm 0.9$       & $0 \pm 0$ \\
Almahata Sitta  & Ure - Anom          & $24 \pm 7$      & $71 \pm 9$     & $0.0 \pm 0.3$     & $5 \pm 2$      & $0 \pm 0$       & $0 \pm 2$           & $0 \pm 0$ \\
Buzzard Coulee  & H4          & $41 \pm 4$      & $54 \pm 4$     & $0.0 \pm 0.4$     & $3.8 \pm 0.6$  & $1.0 \pm 0.6$   & $0 \pm 0$           & $0 \pm 0$ \\
Maribo   & CM2                 & $2.2 \pm 0.2$   & $29 \pm 2$     & $0.000 \pm 0.003$ & $29 \pm 3$     & $5 \pm 1$       & $0.07 \pm 0.06$     & $35 \pm 6$ \\
Jesenice  & L6                & $23 \pm 1$      & $68 \pm 2$     & $0.00 \pm 0.03$   & $8.0 \pm 0.6$  & $0.6 \pm 0.1$   & $0.01 \pm 0.02$     & $0 \pm 0$ \\
Grimsby   & H5                & $41 \pm 5$      & $20 \pm 2$     & $1 \pm 4$         & $29 \pm 4$     & $9 \pm 3$       & $0.1 \pm 0.1$       & $0 \pm 0$ \\
Ko\v sice   & H5                 & $0.62 \pm 0.04$ & $11.5 \pm 0.7$ & $0.000 \pm 0.001$ & $35 \pm 2$     & $18 \pm 3$      & $0.09 \pm 0.05$     & $35 \pm 5$ \\
Mason Gully  & H5             & $1.2 \pm 0.4$   & $20 \pm 5$     & $0.00 \pm 0.01$   & $60 \pm 10$    & $7 \pm 5$       & $0.0 \pm 0.2$       & $13 \pm 9$ \\
Kri\v zevci & H6                 & $9 \pm 2$       & $71 \pm 5$     & $0.00 \pm 0.02$   & $20 \pm 3$     & $0.05 \pm 0.05$ & $0 \pm 0$           & $0 \pm 0$ \\
Sutter's Mill  & CM2           & $2.4 \pm 0.1$   & $34 \pm 1$     & $0.001 \pm 0.003$ & $32 \pm 2$     & $5.4 \pm 0.8$   & $0.07 \pm 0.04$     & $26 \pm 5$ \\
Novato     & L6               & $5.1 \pm 0.6$   & $87 \pm 3$     & $0.000 \pm 0.003$ & $7.5 \pm 0.8$  & $0.4 \pm 0.1$   & $0.0001 \pm 0.0003$ & $0.11 \pm 0.06$ \\
Chelyabinsk  & LL5             & $8 \pm 1$       & $79 \pm 4$     & $0.00 \pm 0.05$   & $12 \pm 2$     & $0.5 \pm 0.2$   & $0.0003 \pm 0.0008$ & $0 \pm 0$ \\
Annama     & H5               & $7.0 \pm 0.4$   & $64 \pm 1$     & $0.0 \pm 0.1$     & $24 \pm 1$     & $4.9 \pm 0.7$   & $0.05 \pm 0.04$     & $0.27 \pm 0.07$ \\
\v Zd'\'ar nad S\'azavou & L3/L3.9 & $6 \pm 1$       & $74 \pm 5$     & $0.00 \pm 0.02$   & $18 \pm 3$     & $0.9 \pm 0.4$   & $0 \pm 0$           & $0.9 \pm 0.7$ \\
Ejby    & H5/6                  & $0.21 \pm 0.06$ & $4 \pm 1$      & $0 \pm 0$         & $11 \pm 4$     & $21 \pm 9$      & $0.1 \pm 0.1$       & $60 \pm 30$ \\
\hline
\end{tabular}
\end{center}
\end{sidewaystable*}

\begin{figure}
  \centering
  \includegraphics[width=1.0\columnwidth]{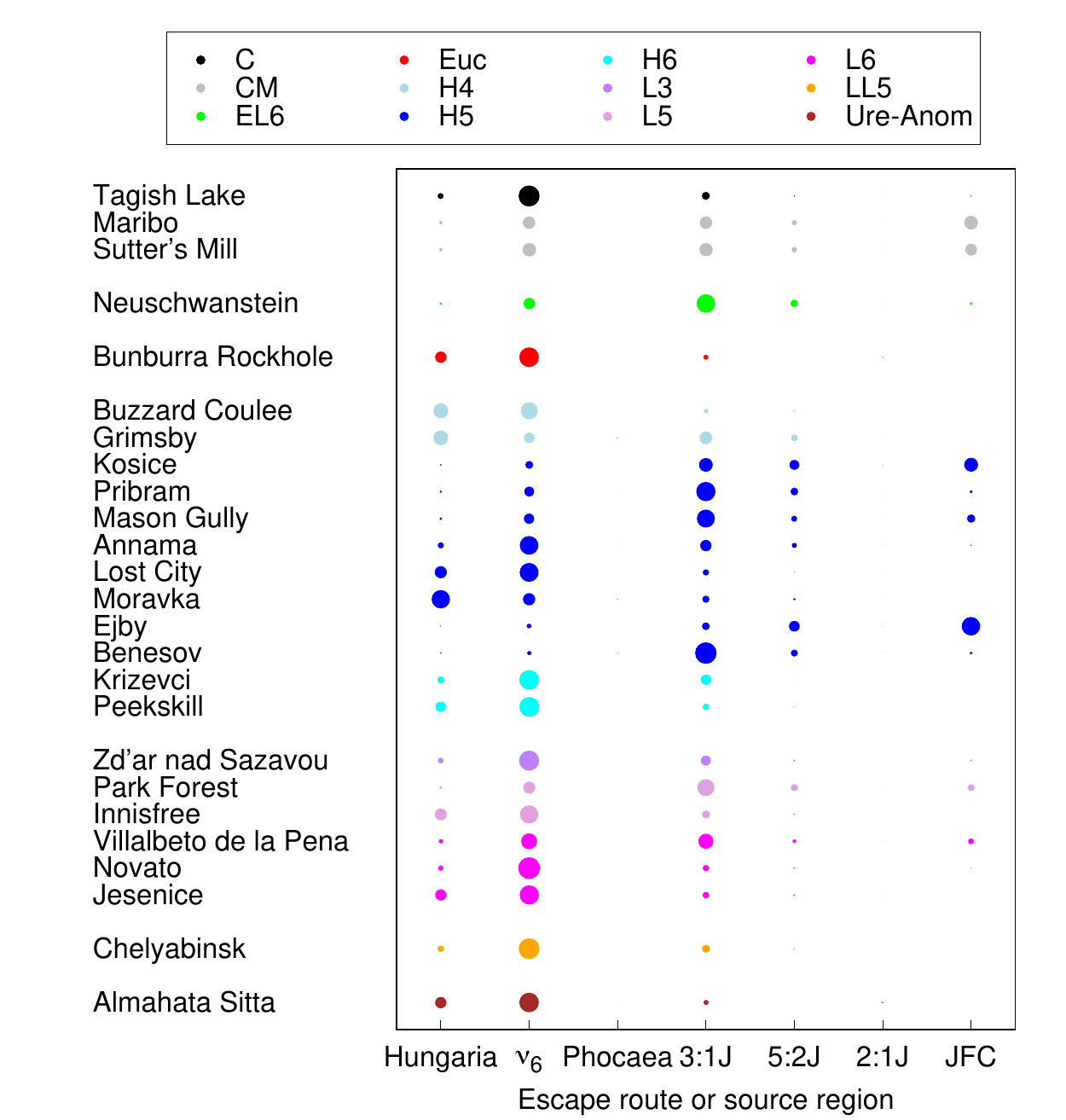}
  \caption{Escape regions for different meteorite groups. The area of
    the filled circle is proportional to the probability that the
    meteorite originates in that source region or escapes the main
    asteroid belt through that escape route. Note that Bene\v sov and
    Innisfree have mixed classifications as shown in Table
    \ref{table:geophysdata} of LL3 and LL5, respectively.}
  \label{fig:sourceprobmet}
\end{figure}

When correlating meteorite types with their likely ERs it becomes
clear that the inner main belt and the Hungaria group are the likely
ERs for almost all known meteorite falls
(Fig.~\ref{fig:sourceprobmet}). However, for meteorites (or their
parent bodies) escaping the main belt through the 3:1J MMR, we cannot
rule out an origin in the middle belt combined with subsequent orbital
evolution to a smaller semimajor axis, such as through Yarkovsky
drift. There is, nevertheless, substantial evidence from current
meteorite orbits and the present delivery model that few meteorite
falls directly originate in the outer main belt or from the immediate
JFC population. The bias is largely explained by orbital dynamics as
asteroids from the outer main belt have a small contribution to the
part of NEO orbital-element phase space that harbor the most likely
Earth-impacting orbits, that is, those with $a\sim1\au$, $e\sim0$, and
$i\sim0\deg$.

The material strength of meteoroids can be expected to also cause (a
part of) this bias, particularly assuming outer belt asteroids and
JFCs are composed of weaker material which is much more easily
destroyed upon impact with the Earth's atmosphere leaving no
macroscopic meteorites to be found. This notion is consistent with the
fireball record \citep{Flynn2017} where most fireballs having
comet-like orbits also have lower penetrating ability on average,
indicative of more fragile structure, than fireballs on more
asteroid-like orbits, a long established result
\citep{wetherill1982, Borovicka2007}.

Asteroid types with weaker material strength are also expected to be
fairly common in the inner belt in the size range of meteorite parent
bodies \citep{2014Natur.505..629D}. While there are 3 carbonaceous
chondrites among the known meteorite producing falls with orbits, this
still amounts to only about 13\% of the entire sample, almost an order
of magnitude higher than the relative abundance of carbonaceous
chondrites among all recovered meteorites (3.5\%).

In contrast, approximately 1/3 of asteroids with diameters
$5\km<D<20\km$ in the inner main-belt are C types, while the fraction
of NEOs which are C-types is in the 20-35\% range
\citep{Binzel2015}. It is notable, that these three meteorite
producing fireballs are also extraordinary in that they were all
produced by comparatively large impactors, ranging from 1.5 tonnes for
Maribo to 56 tonnes for Tagish Lake \citep{Borovicka2015a}. Taken
together, this suggests a material strength bias is present.

The essentially complete lack of falls connected to the Phocaea group
and to asteroids escaping the belt through the 2:1 mean-motion
resonance with Jupiter can be explained by both dynamical and
material-strength arguments. In terms of dynamical arguments these
groups have a factor of several smaller impact probabilities with the
Earth compared to other asteroidal sources \citep{granviketal2018a}, 
and hence a collision with the Earth is less likely. As
for material strength, the outer asteroid belt is well-known to be
mass dominated in carbonaceous asteroids that are assumed to produce
weak meteoroids\citep{2014Natur.505..629D}.  Recently a low albedo
(and hence likely carbonaceous) asteroid family was found in the
Phocaea region, likely dominating the size distribution of Phocaeas at
scales akin to meteoroid parent bodies \citep{2017AJ....153..266N}.

CM meteorites have been linked to B-type asteroids
\citep{2012Icar..218..196D} and Ch/Cgh-type asteroids
\citep{2016AJ....152...54V}. Ch-like asteroids with a 0.7-$\mu$m
hydration band are found across the asteroid belt with a minor
preference for the middle belt ($2.5\au < a < 2.8\au$)
\citep{2012Icar..221..744R}. B-type asteroids (SMASS II) are also
found across the belt with a slight preference for the middle
belt. C-type asteroids that lack the 0.7-$\mu$m hydration band are
expected to typically be found closer to the Sun where temperatures
are high enough to drive out volatiles. Our results show that Tagish
Lake, an ungrouped C meteorite, is very likely a sample of such a
C-type asteroid as it likely originates in the inner part of the main
belt and does not show signs of water. This is in contrast to earlier
work \citep{Hildebrand2006} which suggested a comparatively large
probability (1/4) that Tagish Lake originated directly from the outer
part of the main belt.

Sutter's Mill and Maribo, the two CM2 meteorites in our sample, have
very similar orbits and ER source probabilities. Both have relatively
high ER affinities with the JFC population but, even within the upper
bounds of uncertainty, are more likely to be from the inner to middle
main belt. It is most likely these are from middle belt Ch/Cgh-type or
B-type asteroids.

The likely source region for Chelyabinsk, an LL5 meteorite, is in the
inner main belt with an escape through $\nu_6$ as has also previously
been shown \citep{2013Sci...342.1069P}. There is some controversy over
the classification of Innisfree as either an L5 \citep[generally
  accepted;][]{1990GeCoA..54.1217R} or LL5 \citep{Smith1980}.  If
Innisfree belongs to the LL5 group it implies that we would have two
LL5 meteorites with associated orbits and both indicate a likely
origin in the inner main belt with an escape through $\nu_6$. This
could indicate that these meteorites were originally part of a common
parent body. The spectral similarity of LL chondrites and (8) Flora as
well as the proximity of the Flora family to $\nu_6$ has been
suggested as evidence for Flora being the parent body of LL chondrites
\citep{2008Natur.454..858V,2013Icar..222..273D}. A modern re-analysis
could reduce the uncertainty on the classification of Innisfree, but
to the best of our knowledge such a re-analysis has not been
performed. We note that Stubenberg, which we do not include in our
study as the lack of complete published primary trajectory data does
not permit us to do an independent computation consistent with other
published events, is an LL6 chondrite. The currently published orbital
elements \citep{Spurny2016b} suggest a high ER-likelihood for $\nu_6$
(70\%) and lesser likelihood for 3:1J and Phocaea (23\% and 6\%,
respectively). Interestingly, Stubenberg's orbit has the largest
probability of originating from the Phocaea group among all our
examined cases.

H chondrites comprise almost half our total sample. In most cases an
ER from either the 3:1J MMR or the $\nu_6$ is indicated. This is
consistent with some of the H-chondrite's being from in the inner main
belt \citep{Trigo-Rodriguez2015}, potentially linked to (6) Hebe
\citep{Gaffey1998} or smaller parents located in this zone
\citep{Binzel2015}. However, two of the H5 chondrites with measured
orbits (Ejby and Ko\v sice) show comparatively strong probabilities
(with large error) as being from JFCs, a surprising result. This may
simply reflect the mixing between main belt asteroids and JFCs which
occurs near the $T_j \approx$ 3 region as noted by
\citet{Tancredi2014}.

The five measured L chondrite orbits are strongly associated with ERs
in the inner main belt, in particular the $\nu_6$ and to a lesser
degree the 3:1J MMR. These remain small number statistics, but this is
somewhat at odds with an origin from the Gefion family
\citep{Nesvorny2009,Jenniskens2014} in the outer belt which would
require an escape predominantly from the 5:2J or, less likely, the
3:1J MMR.

Our predictions for the ERs of Almahata Sitta and Bunburra Rockhole,
both primarily $\nu_6$, are in agreement with previous studies
\citep{2012MNRAS.424..508G,Bland2009}.

\begin{figure}
  \centering
  \includegraphics[width=1.0\columnwidth]{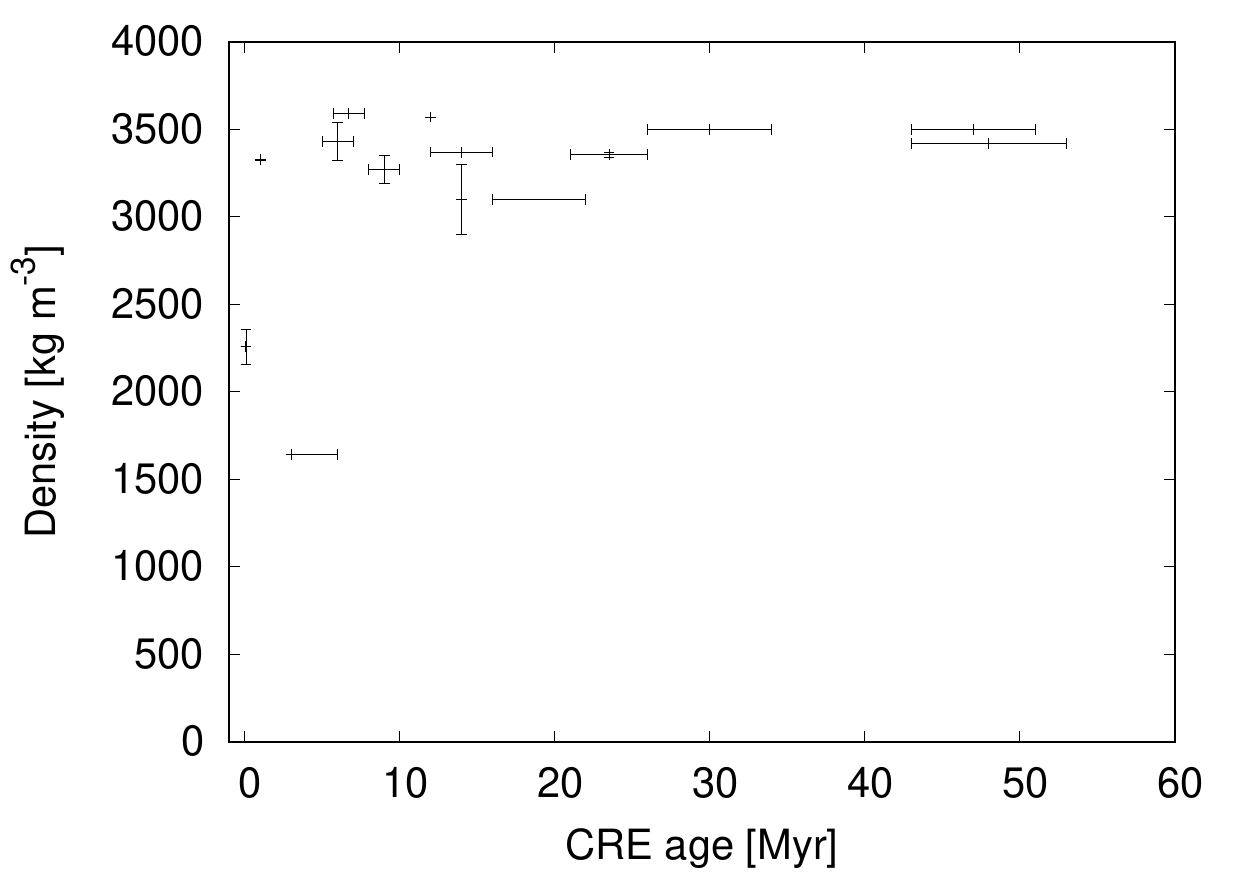}
  \caption{CRE age and bulk density for meteorites for which both have
    been measured. The error bars show the range of possible values
    with the nominal value equal to the midpoint of the range. Note
    that only the nominal value has been reported for many
    measurements (Table \protect \ref{table:geophysdata}). The
    meteorite with the lowest measured density (Tagish Lake) only has
    a measured lower limit for the CRE age, but for plotting reasons
    we have here used an (artificial) upper limit of 6~Myr.}
  \label{fig:credensity}
\end{figure}

The CRE age and bulk density are correlated in the sense that low
densities are only measured for meteorites with short CRE ages
(Fig.~\ref{fig:credensity}). In principle, the correlation could be
due to either material properties (low-density materials are more
fragile and therefore CRE ages are typically short) or dynamical
reasons (low-density materials are more common in the outer asteroid
belt and dynamical lifetimes for objects originating in the outer belt
are shorter than for objects originating in the inner belt). The
latter mechanism is likely to be less important, because $\nu_6$ is
the most likely ER for the meteorites with the lowest densities and
short CRE ages (Tagish Lake and Sutter's Mill). The interpretation is
currently based on the measurements of only 13 meteorites, but it can
be tested by measuring CRE age and/or bulk density for the remaining
12 meteorites.

Additional insight to the origins of meteorites can be obtained by
comparing the CRE age to the dynamical age typical for objects
originating in a specific ER. A particular question is whether
meteorites separated from their parent bodies while still in the
asteroid belt or if they separated later from NEOs. We estimated the
dynamical age of the meteorites by calculating the time it takes for
test asteroids in orbital integrations carried out by
\citet{2016Natur.530..303G} to evolve from a $q=1.3\au$ orbit to the
orbit of the meteorite immediately prior to the impact with the Earth
(Table~\ref{table:dynage}). For orbital similarity we required the
test asteroid's orbital elements to reproduce the meteorite orbit to
within the latter orbit's uncertainty. The relatively small
uncertainties implied that only a limited number of test asteroids,
out of the more than 90 thousand integrated, ever reached orbits
similar to those of the meteorites considered here, and hence the
uncertainties are substantial.

\begin{sidewaystable*}
\begin{center}
  \caption{Dynamical age of meteorites assuming an origin in the
    asteroid belt. The dynamical age is calculated as the time it
    takes to evolve from an asteroidal ER (at the NEO-MBO border) to
    the orbit just before impact with the Earth's atmosphere. We only
    consider test asteroids whose orbits ($a$,$e$,$i$) reproduce the
    meteorite orbits to better than 1-$\sigma$ as reported in Table
    \ref{table:orbits}. The estimates are based on the orbital
    integration of more than 90 thousand test asteroids as reported by
    \protect \citet{2016Natur.530..303G}. The numbers given for each
    meteorite and ER are median dynamical age in Myr, range of
    dynamical ages in Myr, and the number of test asteroids that
    reproduce the meteorite orbit. Dynamical ages corresponding to the
    most likely ERs are highlighted in boldface font. Note that Maribo
    and Ejby are likely to originate from the JFC population. Note
    also that the orbit of Almahata Sitta is known so accurately that
    none of the test asteroids available to us reproduce its orbital
    elements to within the orbital uncertainty.}\label{table:dynage}
  \begin{tabular}{lcccccc}
    \hline
    Meteorite name & \multicolumn{1}{c}{Hungaria} & \multicolumn{1}{c}{$\nu_6$} & \multicolumn{1}{c}{Phocaea} & \multicolumn{1}{c}{3:1J} & \multicolumn{1}{c}{5:2J} & \multicolumn{1}{c}{2:1J} \\
    \hline
P\v{r}\'ibram             &  --                    &  5.9/0.9--11.0/2    &  1.5/1.5--1.5/1        &  {\bf 0.1/0.1--0.1/1}     &  --                 &  --                 \\
Lost City            &  2.3/2.3--2.3/1        &  {\bf 3.7/1.9--11.4/6}    &  --                    &  13.2/13.2--13.2/1  &  --                 &  --                 \\
Innisfree           &  47.1/47.1--47.1/1     &  {\bf 3.6/0.8--22.8/7}    &  53.8/53.8--53.8/1     &  16.6/9.5--23.2/5   &  6.3/1.6--8.3/3     &  7.1/3.6--10.6/2    \\
Bene\v sov             &  --                    &  --                 &  --                    &  {\bf 0.3/0.3--0.3/1}     &  --                 &  --                 \\
Peekskill           &  60.7/60.7--60.7/1     &  {\bf 8.1/3.7--16.8/7}   &  167.0/167.0--167.0/1  &  19.1/3.4--24.3/3   &  --                 &  --                 \\
Tagish Lake          &  74.7/74.7--74.7/1     &  {\bf 1.8/0.0--16.4/7}    &  115.8/115.8--115.8/1  &  2.3/0.8--7.2/4     &  11.2/1.6--20.9/2   &  0.4/0.4--0.4/1     \\
Mor\'avka             &  {\bf 15.8/15.8--15.8/1}     &  22.4/4.4--167.5/7  &  79.1/79.1--79.1/1     &  10.1/2.1--17.6/4   &  8.7/5.0--12.4/2    &  12.6/12.6--12.6/1  \\
Neuschwanstein      &  553.7/553.7--553.7/1  &  3.0/0.8--39.5/7    &  1.0/1.0--1.0/1        &  {\bf 0.8/0.2--4.3/5}     &  7.6/1.8--13.4/2    &  --                 \\
Park Forest          &  74.9/74.9--74.9/1     &  7.4/1.4--22.0/7    &  55.4/55.4--55.4/1     &  {\bf 1.2/0.0--6.0/5}     &  1.0/0.0--5.8/3     &  0.6/0.1--0.9/3     \\
Villalbeto de la Pe\~na  &  74.8/74.8--74.8/1     &  {\bf 3.8/1.1--31.6/7}    &  185.6/185.6--185.6/1  &  0.9/0.0--1.7/4     &  0.7/0.1--1.5/3     &  0.1/0.1--0.1/1     \\
Bunburra Rockhole    &  99.3/99.3--99.3/1     &  {\bf 10.0/4.6--18.3/3}   &  --                    &  --                 &  --                 &  --                 \\
Almahata Sitta       &  --                    &  {\bf --}                 &  --                    &  --                 &  --                 &  --                 \\
Buzzard Coulee       &  28.6/28.6--28.6/1     &  {\bf 13.8/5.8--145.2/7}  &  122.3/122.3--122.3/1  &  26.3/17.8--46.7/3  &  45.6/21.8--69.3/2  &  --                 \\
Maribo              &  74.9/74.9--74.9/1     &  3.7/1.8--24.1/7    &  185.8/185.8--185.8/1  &  1.0/0.0--1.9/4     &  0.4/0.2--1.0/3     &  1.5/0.7--6.3/3     \\
Jesenice            &  296.9/296.9--296.9/1  &  {\bf 4.9/0.4--38.3/7}    &  53.0/53.0--53.0/1     &  16.6/6.3--24.4/5   &  11.3/9.5--13.0/2   &  10.7/10.7--10.7/1  \\
Grimsby             &  {\bf 43.3/43.3--43.3/1}     &  56.2/4.0--172.4/7  &  12.8/12.8--12.8/1     &  3.5/1.6--26.6/5    &  4.7/2.9--6.6/2     &  12.4/1.5--23.3/2   \\
Ko\v sice              &  74.9/74.9--74.9/1     &  3.5/0.8--22.6/7    &  55.5/55.5--55.5/1     &  {\bf 1.0/0.0--6.5/5}     &  1.0/0.0--5.8/3     &  0.3/0.2--0.8/3     \\
Mason Gully          &  0.4/0.4--0.4/1        &  15.4/2.7--20.7/4   &  --                    &  {\bf 0.2/0.1--3.6/3}     &  --                 &  --                 \\
Kri\v zevci            &  1.1/1.1--1.1/1        &  {\bf 10.4/1.7--18.2/4}   &  --                    &  --                 &  --                 &  --                 \\
Sutter's Mill         &  74.9/74.9--74.9/1     &  {\bf 2.8/0.4--21.9/7}    &  50.7/50.7--50.7/1     &  0.3/0.0--23.8/5    &  0.3/0.0--6.2/3     &  0.4/0.2--1.0/4     \\
Novato              &  950.3/950.3--950.3/1  &  {\bf 5.5/0.0--19.1/7}    &  115.8/115.8--115.8/1  &  18.0/0.8--35.2/4   &  8.0/0.5--15.5/2    &  --                 \\
Chelyabinsk         &  60.9/60.9--60.9/1     &  {\bf 5.6/2.2--27.8/7}    &  1.9/1.9--1.9/1        &  2.1/1.3--3.3/3     &  7.3/1.9--12.6/2    &  --                 \\
Annama              &  54.9/54.9--54.9/1     &  {\bf 4.3/0.3--58.1/7}    &  24.3/24.3--24.3/1     &  11.0/0.5--129.5/5  &  1.3/0.6--1.6/3     &  8.0/1.7--14.3/2    \\
Zd'nar nad Sazavou     &  1.0/1.0--1.0/1        &  {\bf 8.2/2.0--48.9/7}    &  --                    &  1.5/0.5--3.4/3     &  --                 &  --                 \\
Ejby                &  0.7/0.7--0.7/1        &  21.0/0.7--30.4/7   &  53.5/53.5--53.5/1     &  0.8/0.4--5.1/4     &  0.1/0.1--0.3/3     &  0.3/0.3--0.3/1     \\
\hline
\end{tabular}
\end{center}
\end{sidewaystable*}

\begin{figure}
  \centering
  \includegraphics[width=1.0\columnwidth]{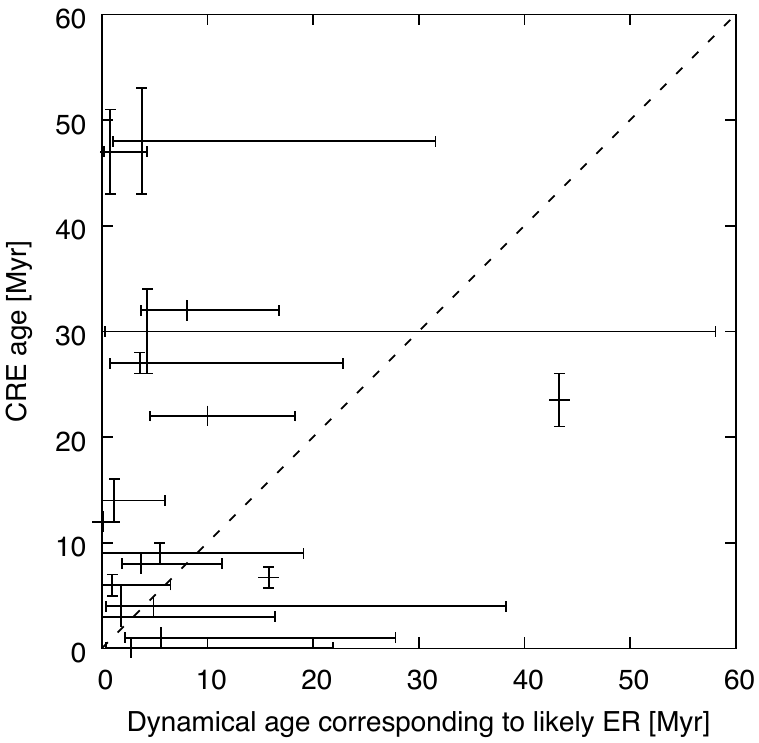}
  \caption{Comparison of dynamical age corresponding to the most
    likely ER (Table~\ref{table:dynage}) and the reported CRE age
    (Table~\ref{table:geophysdata}). The dashed line corresponds to an
    equal dynamical and CRE age with points above (below) the line
    implying shorter (longer) dynamical ages than CRE age.}
  \label{fig:dynagevscreage}
\end{figure}

By comparing the reported CRE ages to the dynamical ages we find that
most meteorites considered here were released from their parent bodies
prior to entering their most likely ERs
(Fig.~\ref{fig:dynagevscreage}). This is particularly evident for
meteorites with the largest CRE ages. The two apparent outliers that
are below the dashed line (Mor\'avka and Grimsby) suffer from small
number statistics and the range of dynamical ages cannot be
estimated. Hence it is possible that improved statistics in the
dynamical analysis would shorten their likely dynamical ages. Given
the relatively slow dynamical evolution prior to entering a resonance
we find it reasonable to assume that the CRE ages are typically reset
at the time of the catastrophic impacts in the main asteroid belt that
sends fragments towards planet-crossing orbits. This picture is
consistent with the apparent lack of "meteorite-streams"
\citep{Pauls2005} and is further supported by the apparent lack of
near-Earth asteroid streams \citep{Schunova2012}. It is broadly
consistent with the notion that most of the parent meteoroid's time
spent as a meter-sized object is in the main belt undergoing Yarkovsky
drift to one of the ERs \citep{Vok2000}.

\subsection{Sensitivity of ER prediction to accuracy of fireball velocity}
\label{sec:velsens}

In this section we will quantitatively assess the impact that
improving the determination of fireball velocity would have on our
knowledge of meteorite ERs. We start with the observed atmospheric
trajectories for the 25 fireballs discussed above, but rather than
using the nominal published values for the velocity uncertainty, we
then calculate their heliocentric orbits using 8 different assumptions
for the pre-atmosphere velocity uncertainty: $0.001\km\second^{-1}$,
$0.003\km\second^{-1}$, $0.01\km\second^{-1}$, $0.03\km\second^{-1}$, 
$0.1\km\second^{-1}$, \newline $0.3\km\second^{-1}$, $1\km\second^{-1}$, 
and $3\km\second^{-1}$. Note that the two last velocity uncertainties are 
larger than reported for any of these events.

\begin{figure*}
  \centering
  \includegraphics[width=0.8\columnwidth]{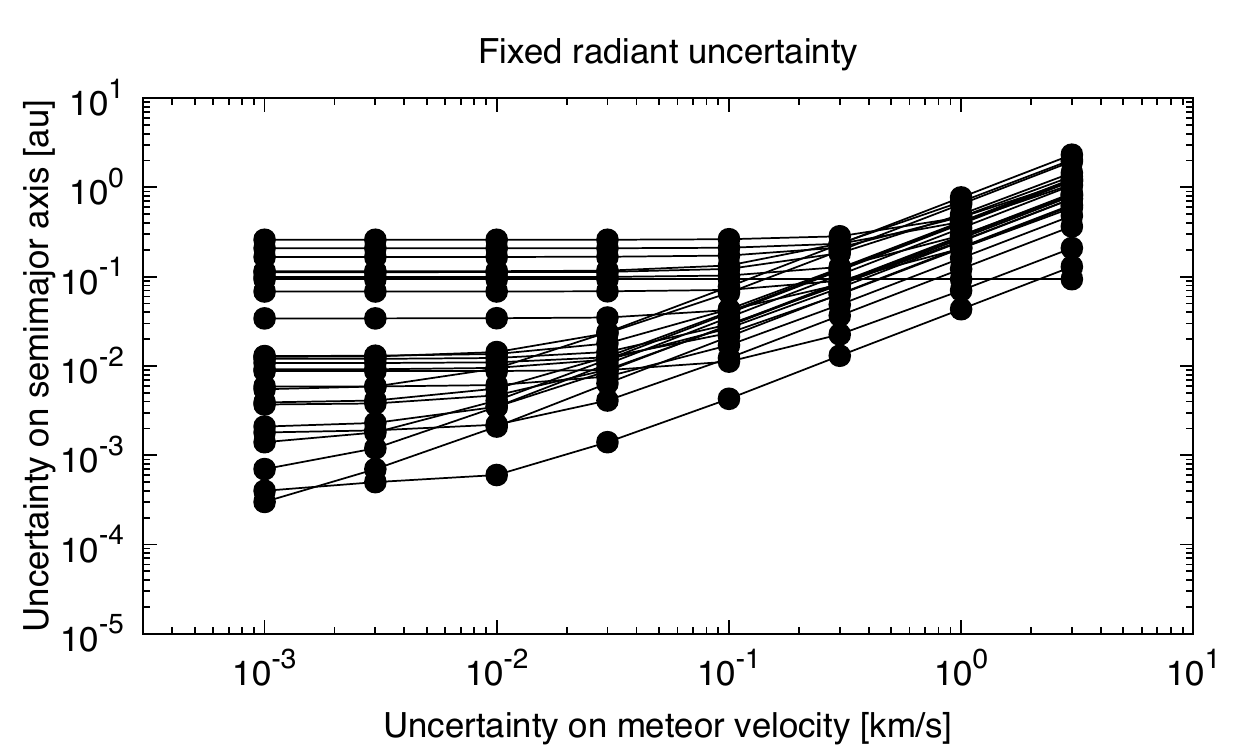}
  \includegraphics[width=0.8\columnwidth]{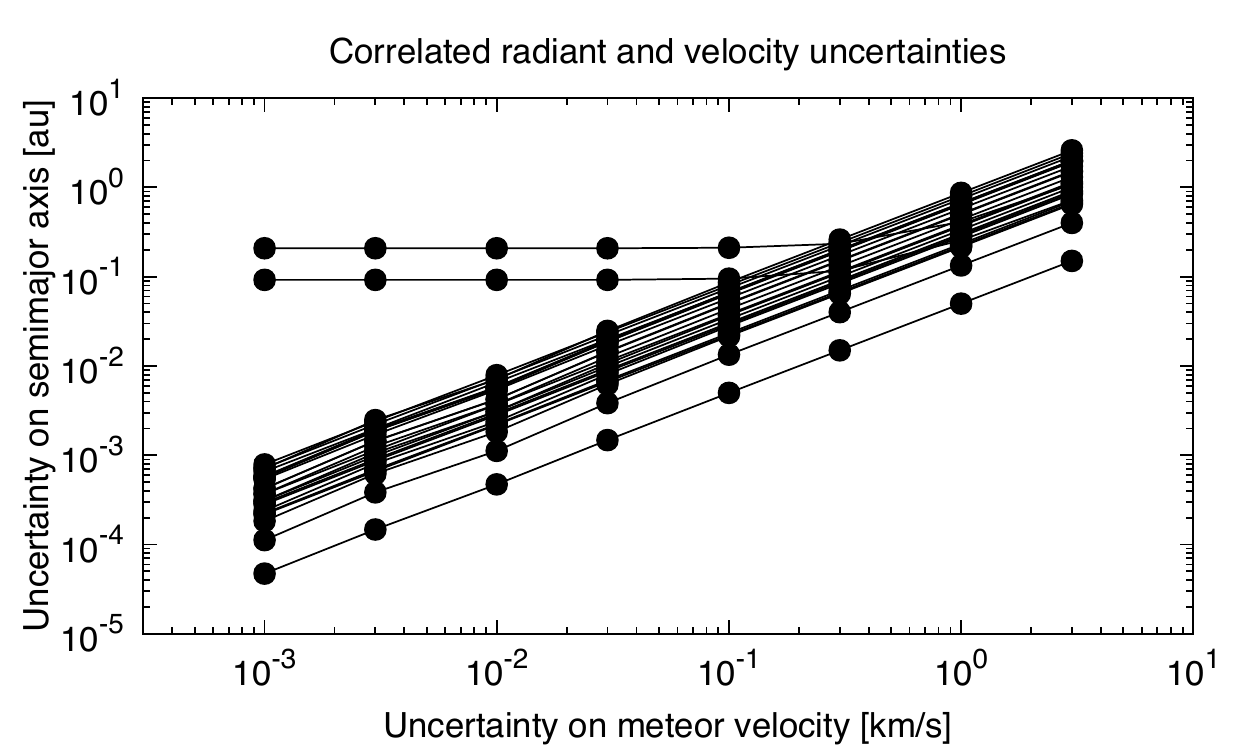}
  \includegraphics[width=0.8\columnwidth]{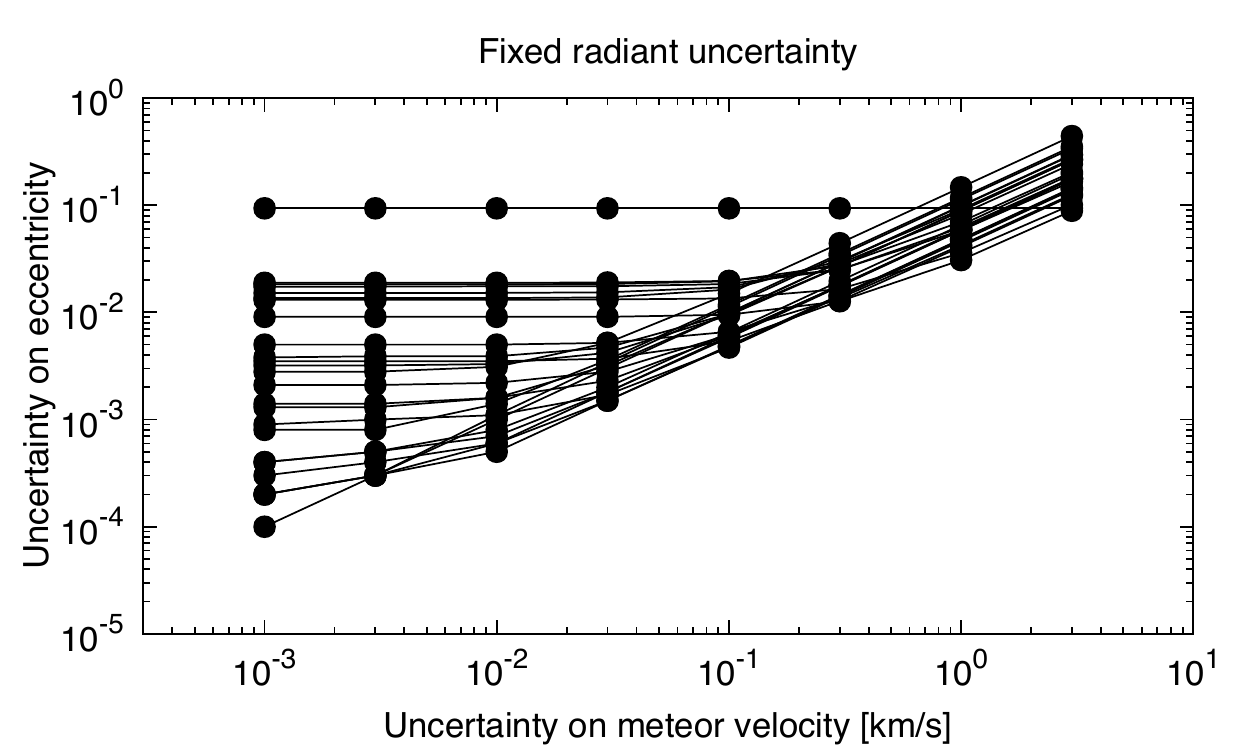}
  \includegraphics[width=0.8\columnwidth]{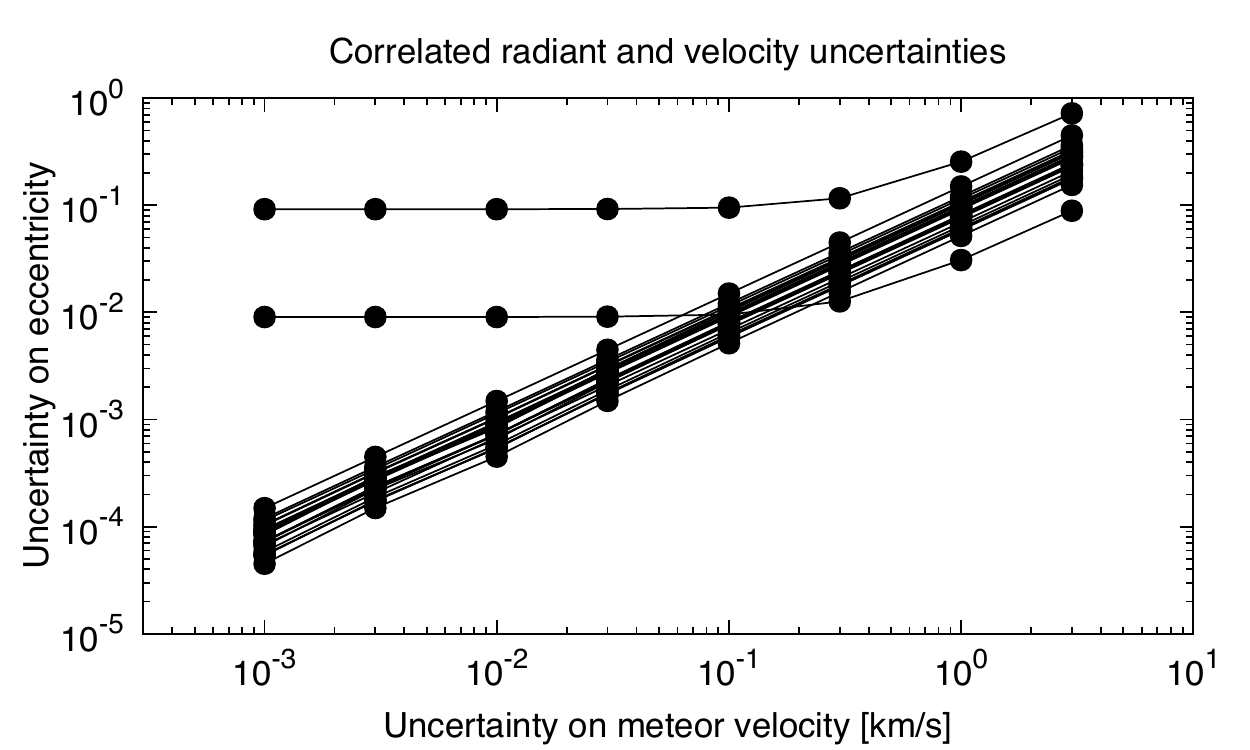}
  \includegraphics[width=0.8\columnwidth]{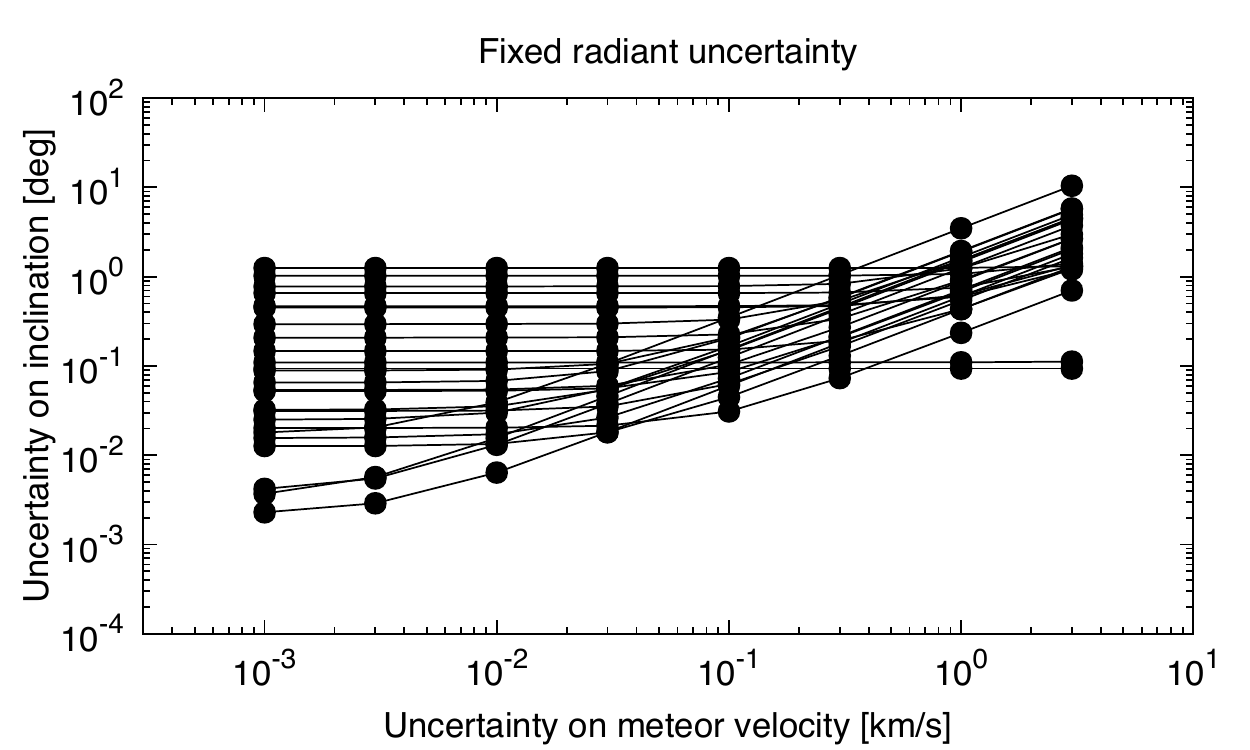}
  \includegraphics[width=0.8\columnwidth]{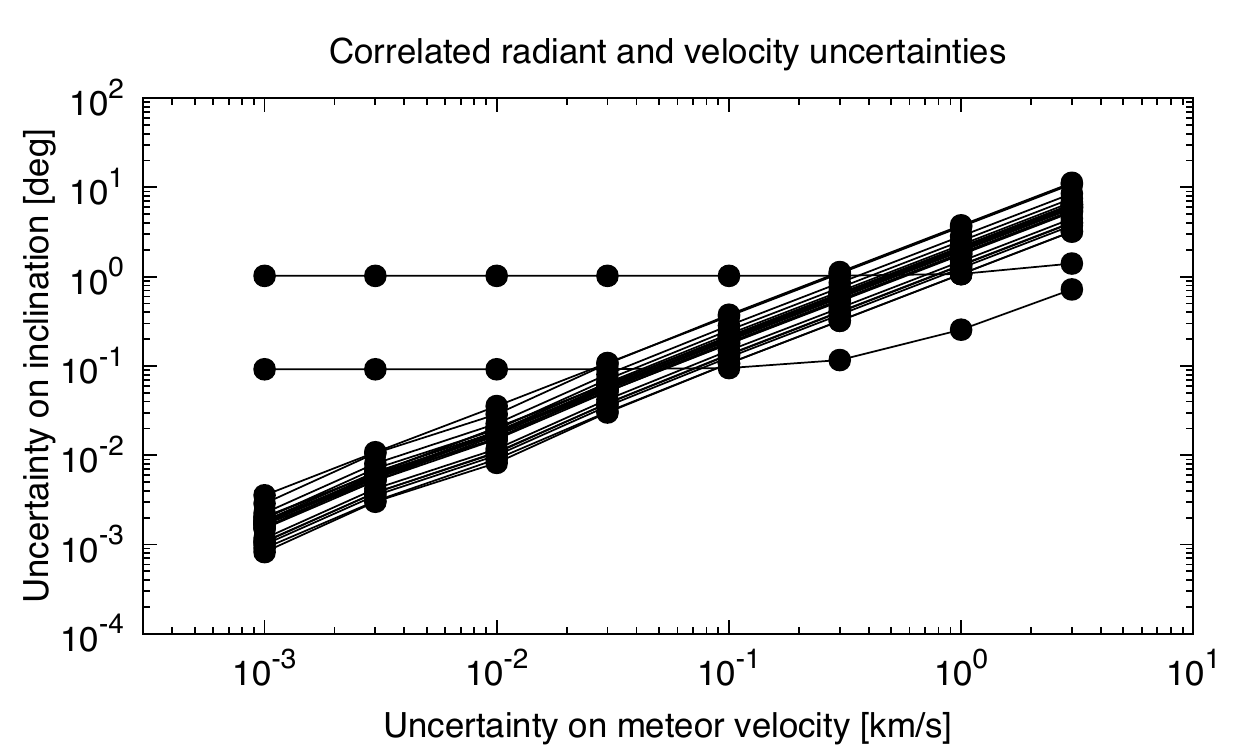}
  \caption{Resulting orbital-element uncertainty as a function of
    assumed velocity uncertainty when using a fixed radiant
    uncertainty (left) and a radiant uncertainty correlated with the
    velocity uncertainty (right).}
  \label{fig:velunc_vs_orb}
\end{figure*}

As expected, the uncertainty on the orbital elements is reduced when
the velocity uncertainty becomes smaller and the radiant uncertainty
remains constant (left column in Fig.~\ref{fig:velunc_vs_orb}). The
apparent insensitivity to velocity uncertainties below about
$0.1\km\second^{-1}$ is explained as a consequence of a relatively
large radiant uncertainty, which is kept at the nominal observed
value. At some point the velocity uncertainty becomes negligible
compared to the radiant uncertainty which now dominates the total
uncertainty budget. An improvement in the orbital uncertainty can thus
no longer be obtained by reducing the velocity uncertainty. For
example, all but one of the cases that show $\Delta a > 0.01\au$ for a
velocity uncertainty of $0.001\km\second^{-1}$ have radiant
uncertainties $\geq0.3\deg$ whereas the ones that show $\Delta a <
0.01\au$ have radiant uncertainties $\leq0.3\deg$. The only exception
to this rule is P\v{r}\'ibram which has a radiant uncertainty of only
$0.01\deg$ and for which a velocity uncertainty of
$0.001\km\second^{-1}$ results in $\Delta a \approx 0.013\au$.

\begin{figure*}
  \centering
  \includegraphics[width=0.95\textwidth]{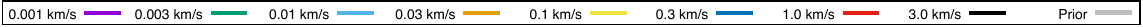}
  \includegraphics[width=0.66\columnwidth]{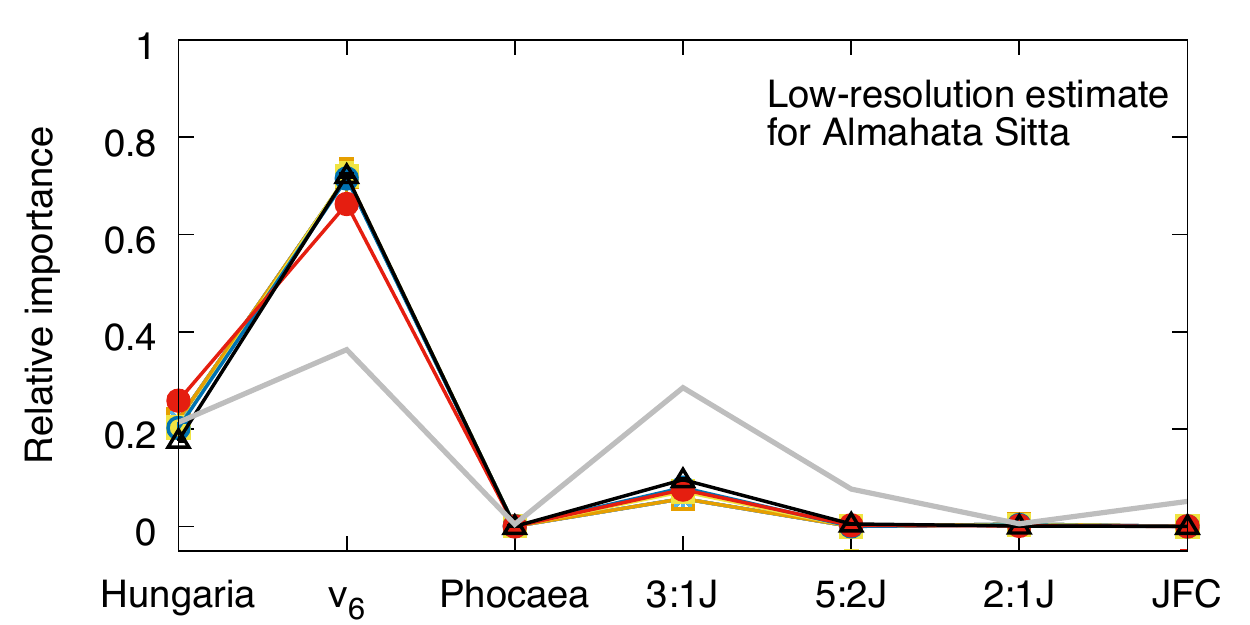}
  \includegraphics[width=0.66\columnwidth]{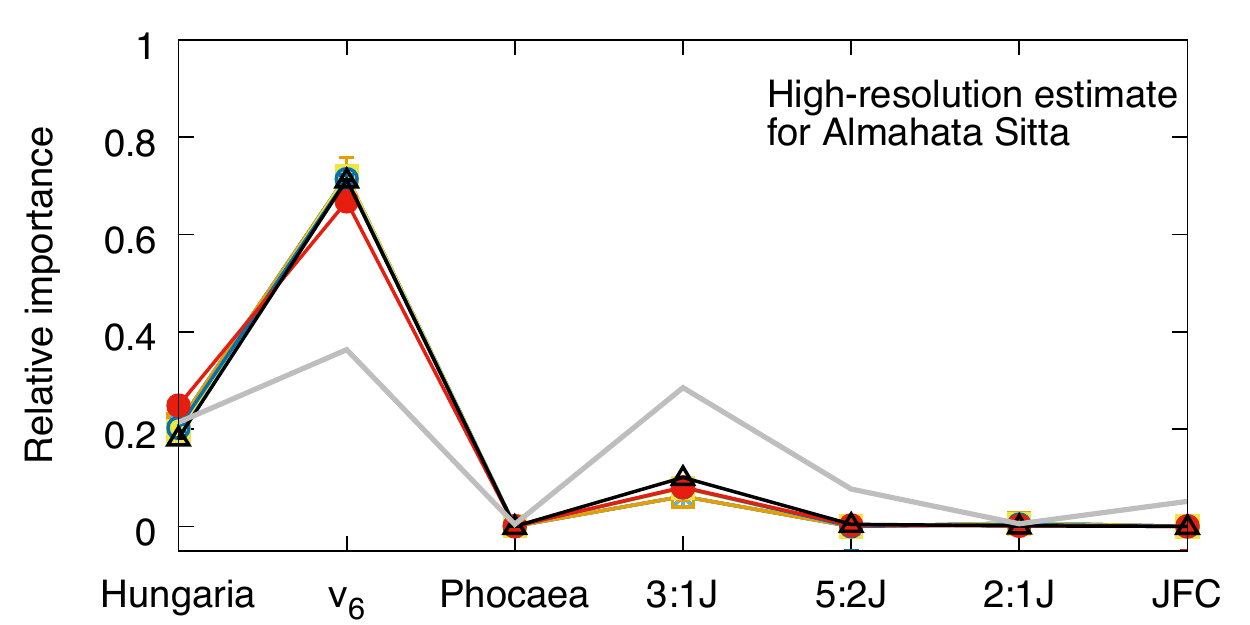}
  \includegraphics[width=0.66\columnwidth]{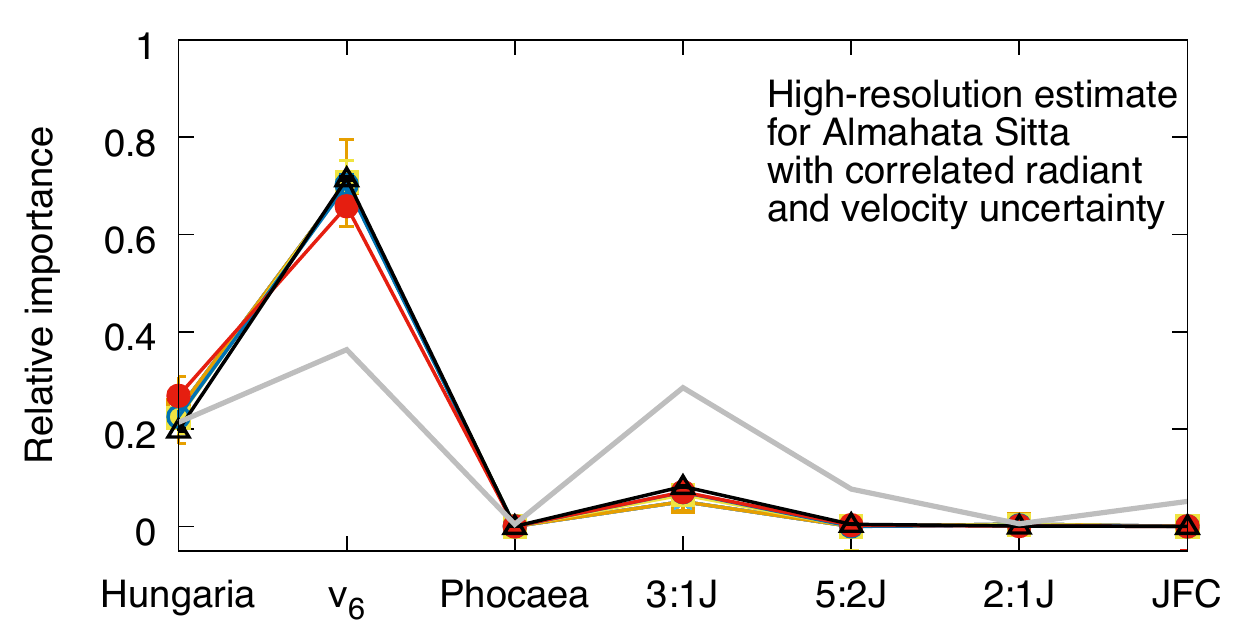}
  \includegraphics[width=0.66\columnwidth]{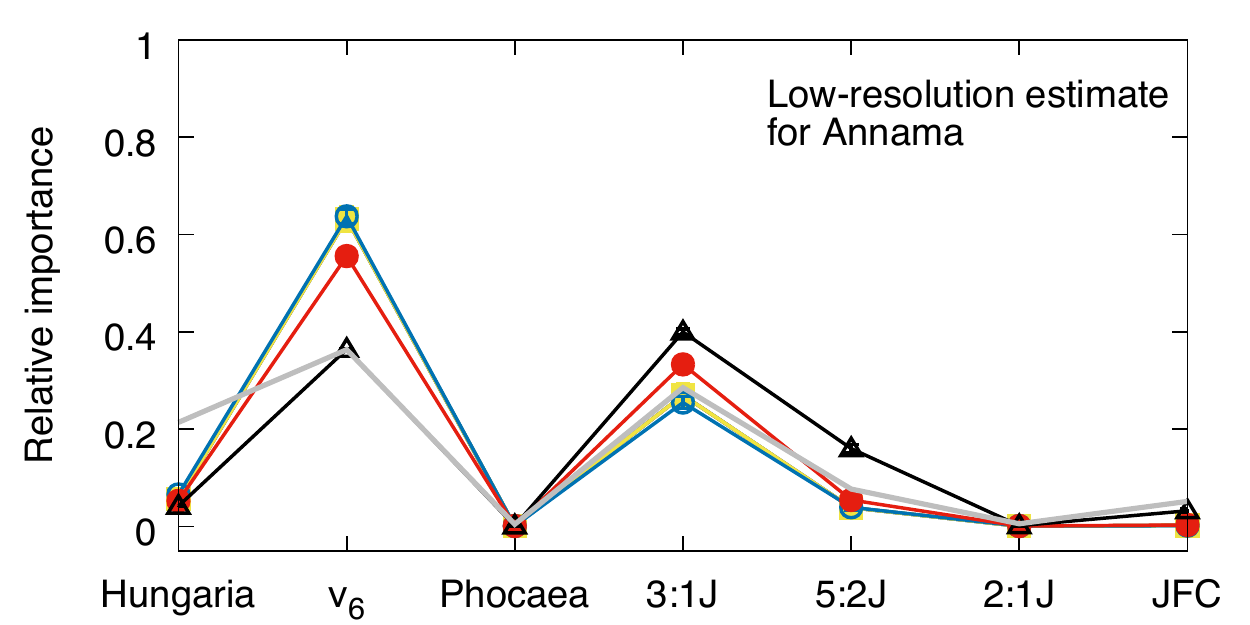}
  \includegraphics[width=0.66\columnwidth]{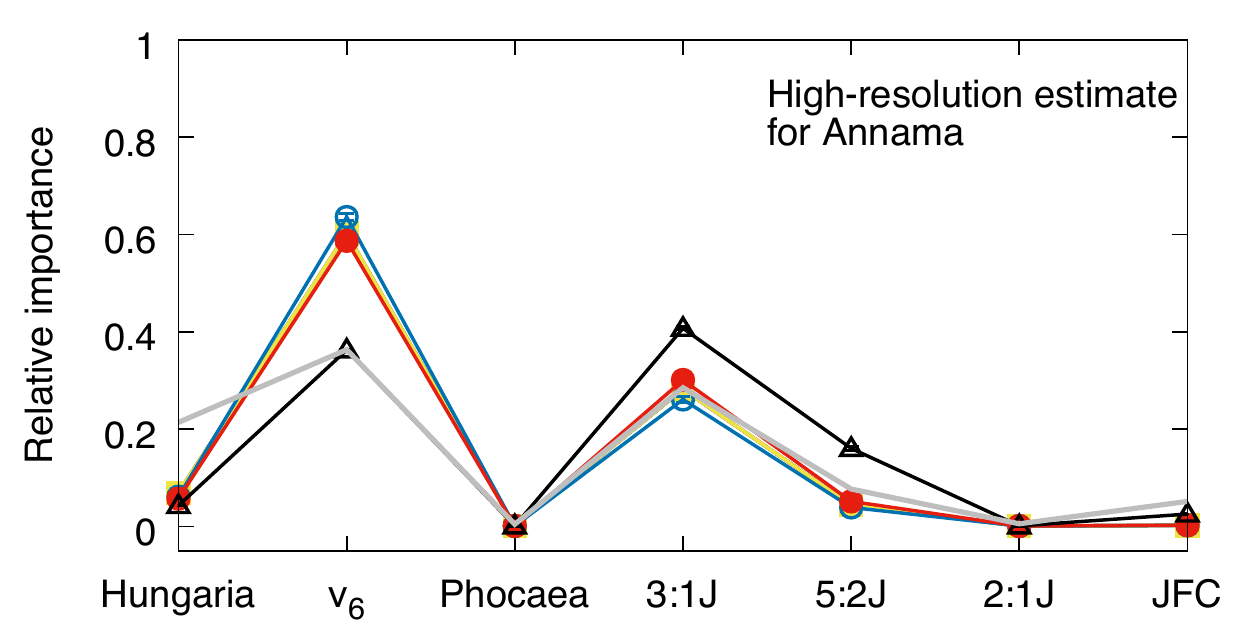}
  \includegraphics[width=0.66\columnwidth]{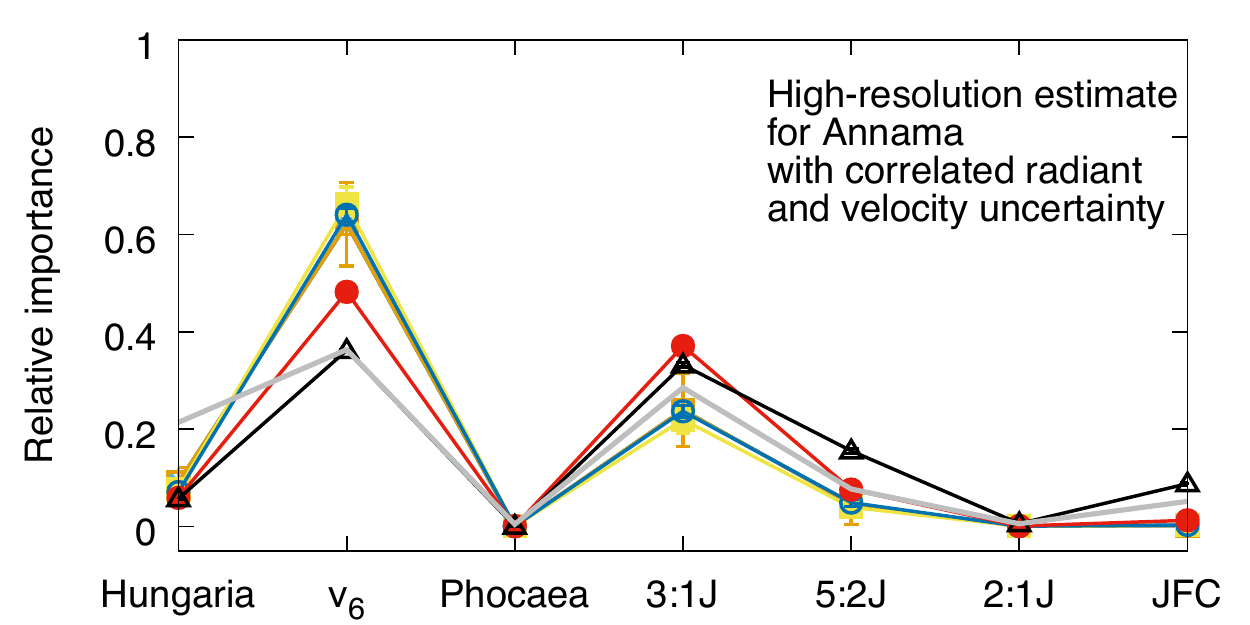}
  \includegraphics[width=0.66\columnwidth]{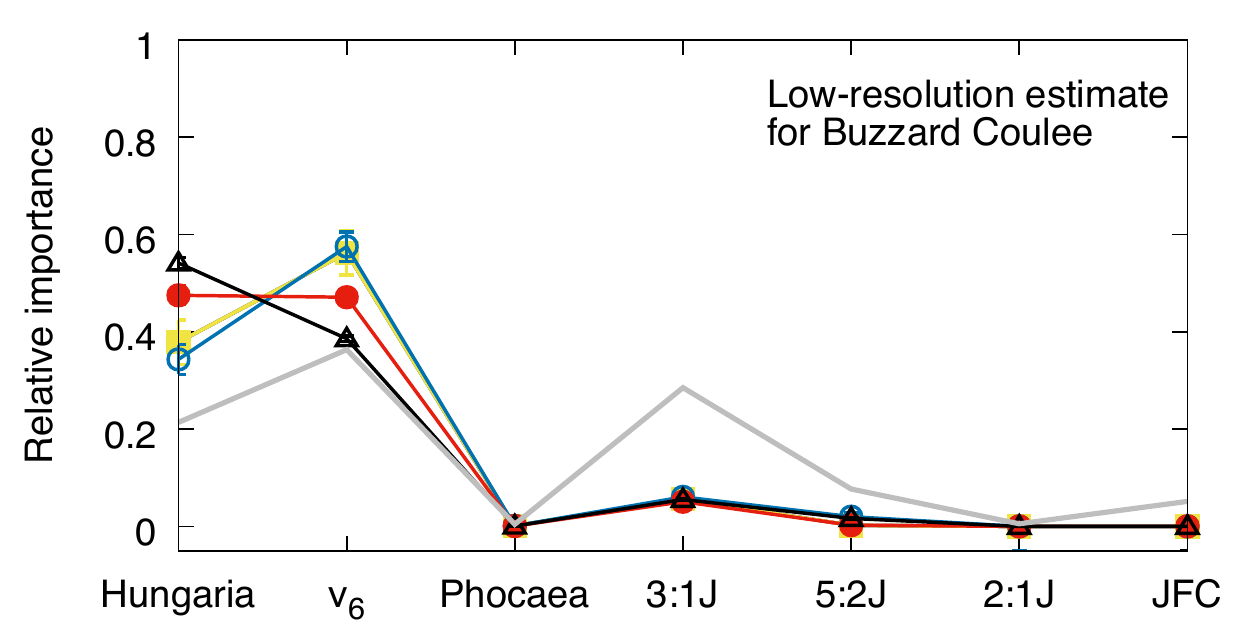}
  \includegraphics[width=0.66\columnwidth]{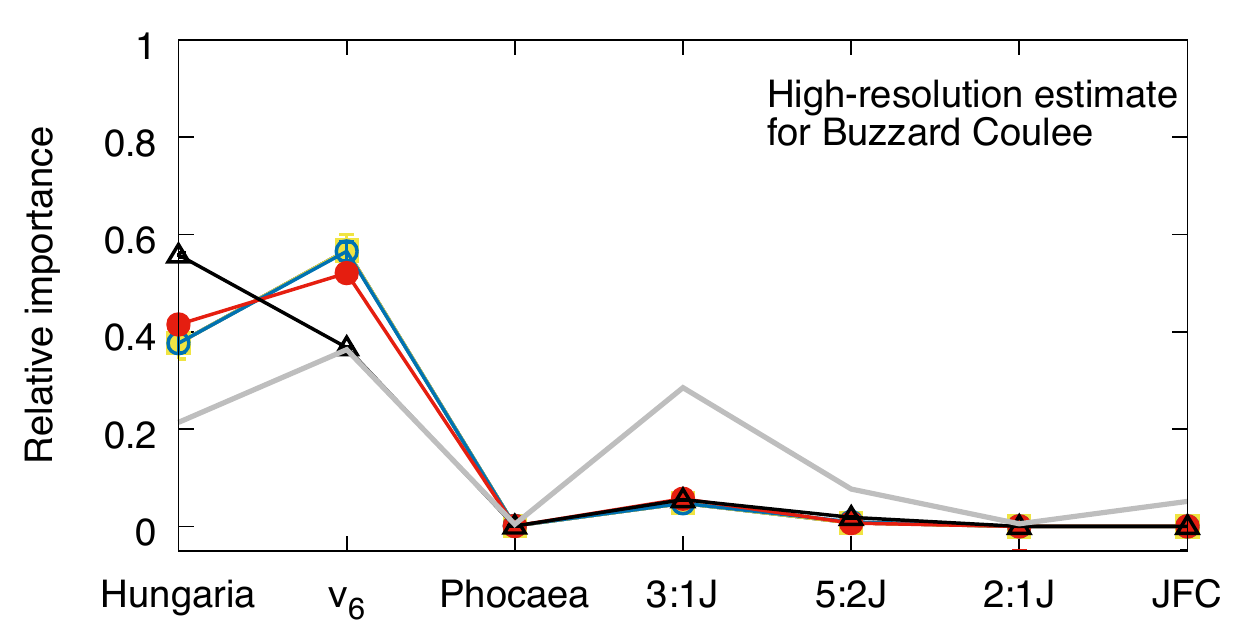}
  \includegraphics[width=0.66\columnwidth]{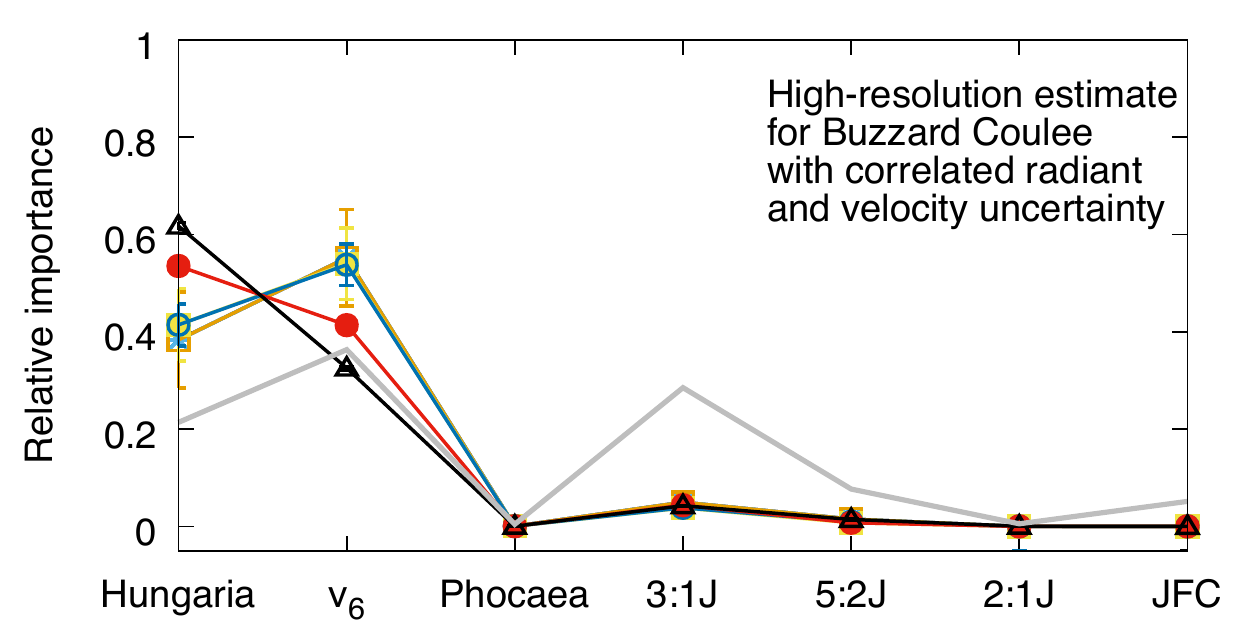}
  \includegraphics[width=0.66\columnwidth]{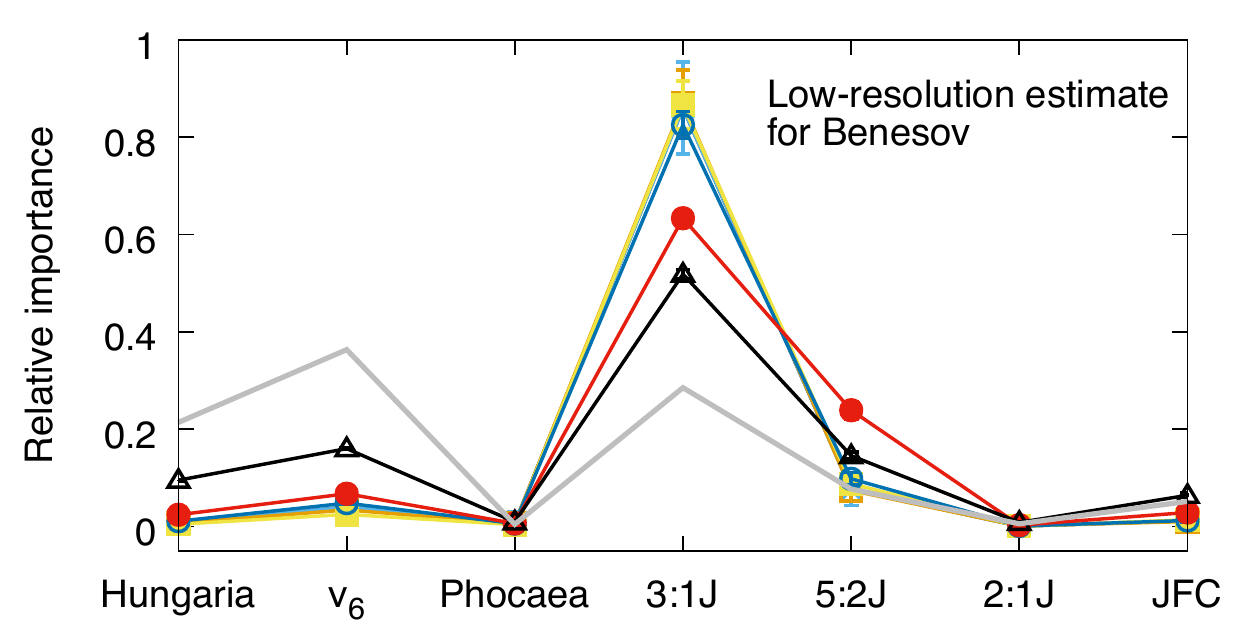}
  \includegraphics[width=0.66\columnwidth]{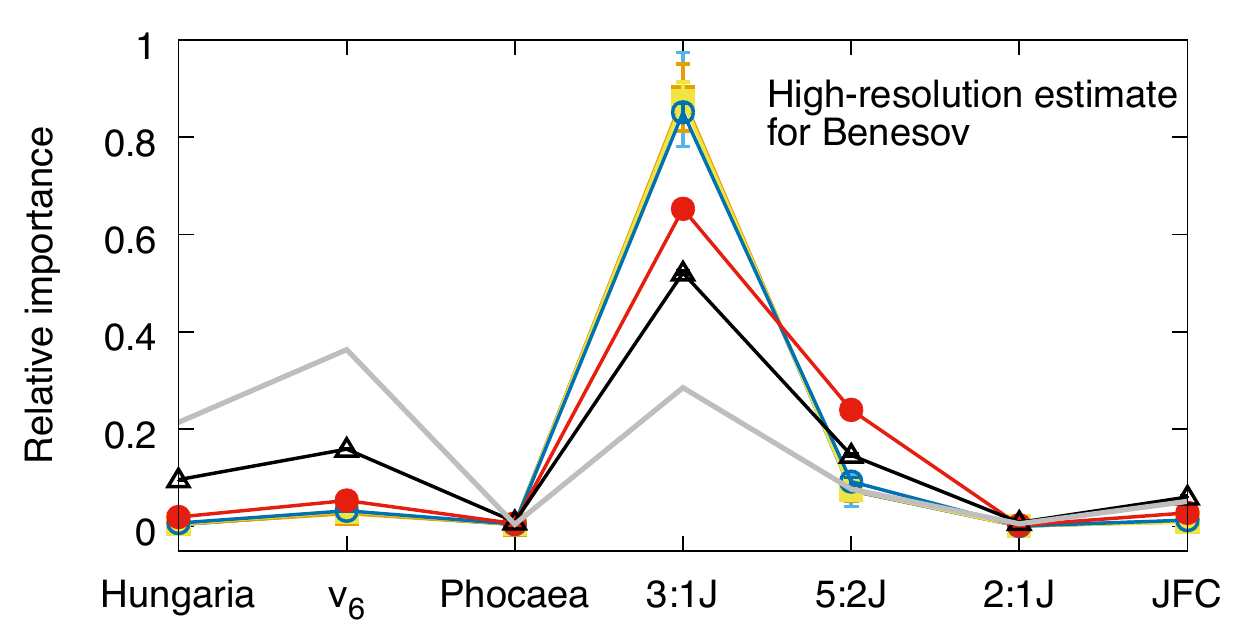}
  \includegraphics[width=0.66\columnwidth]{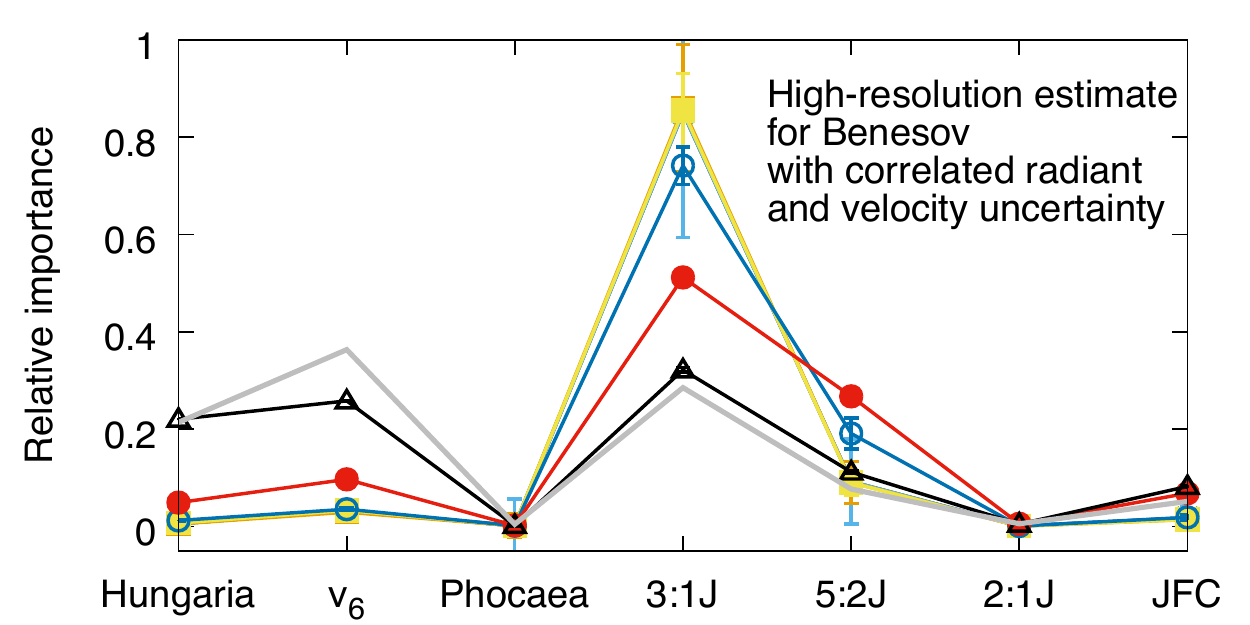}
  \includegraphics[width=0.66\columnwidth]{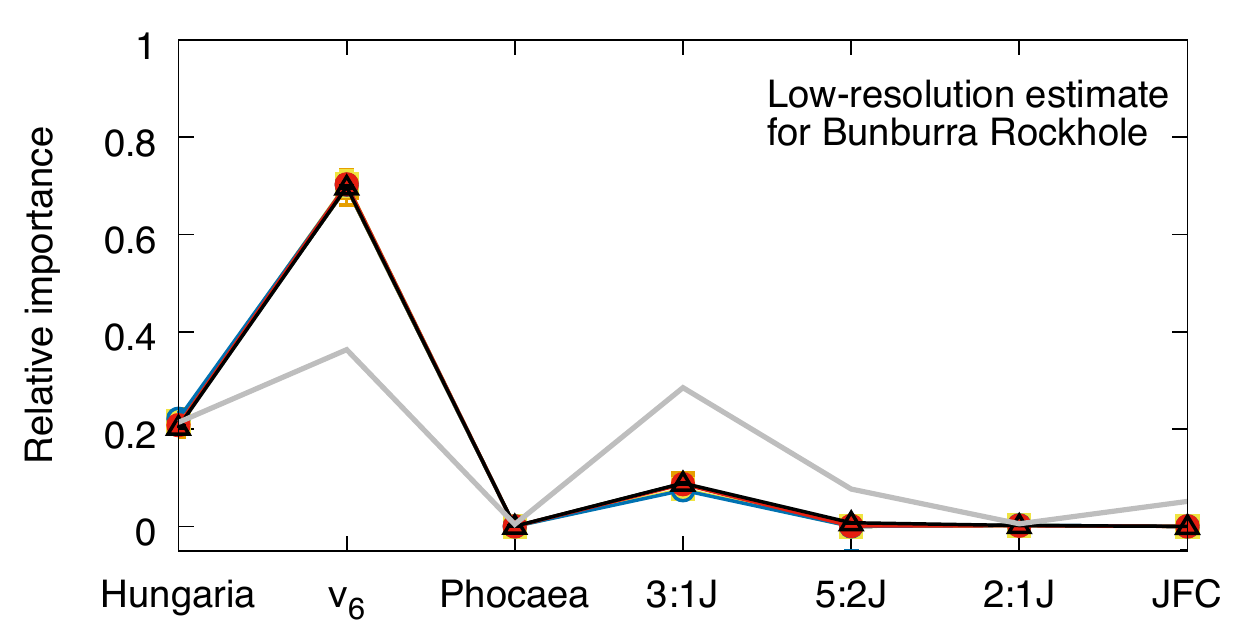}
  \includegraphics[width=0.66\columnwidth]{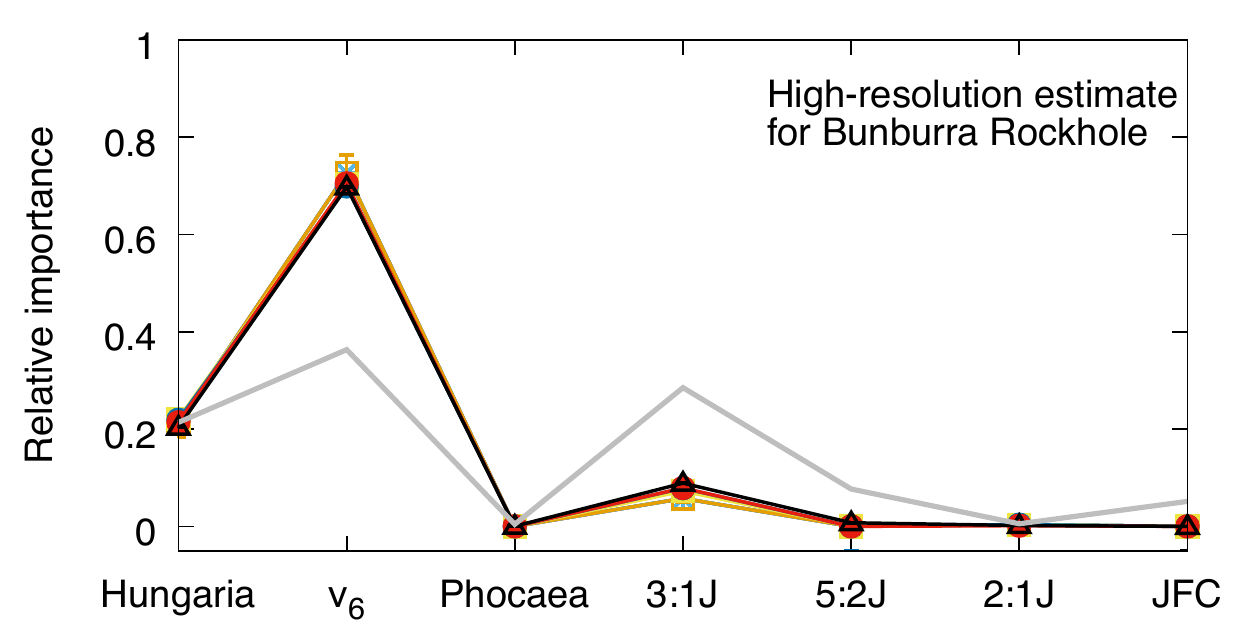}
  \includegraphics[width=0.66\columnwidth]{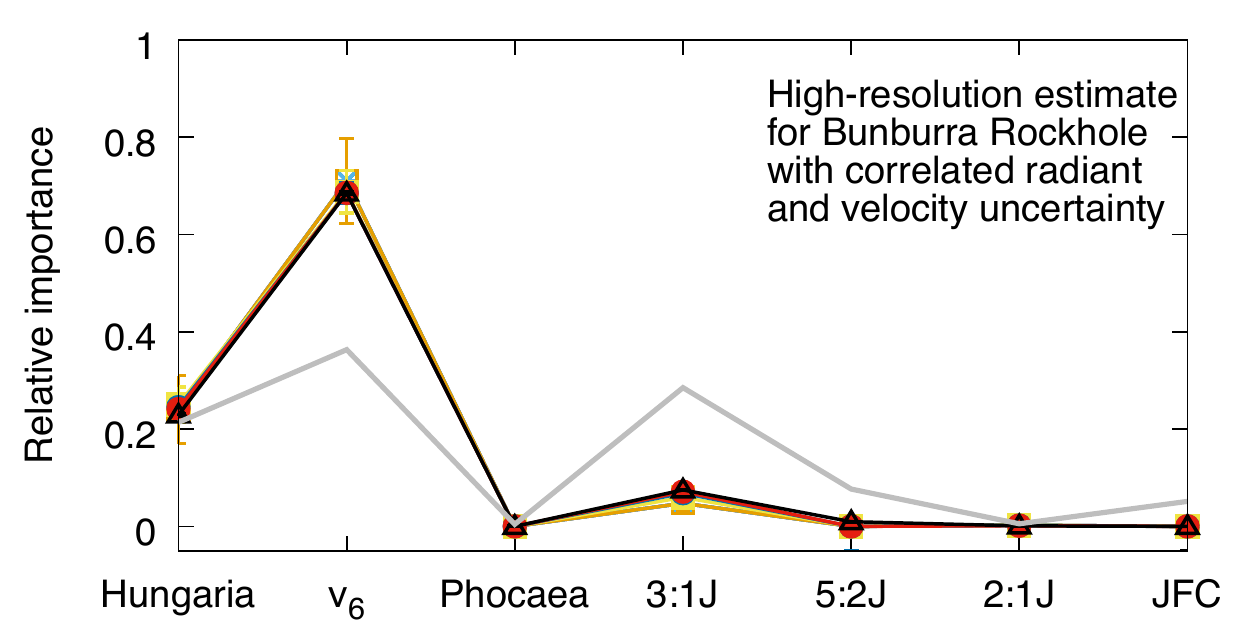}
  \includegraphics[width=0.66\columnwidth]{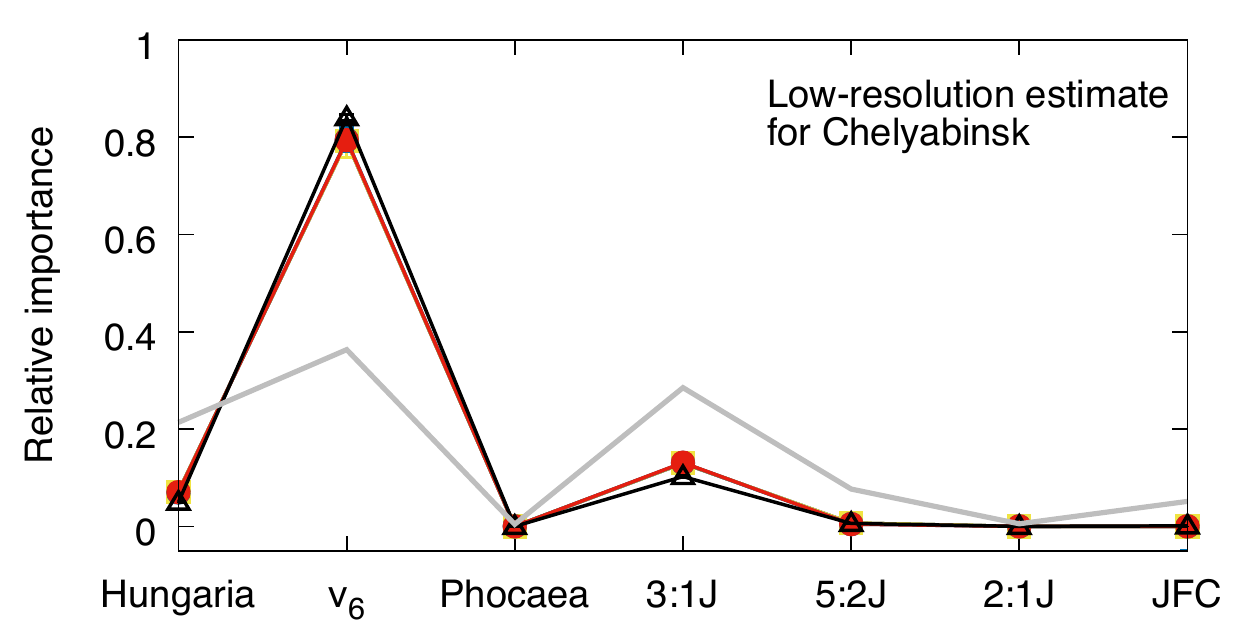}
  \includegraphics[width=0.66\columnwidth]{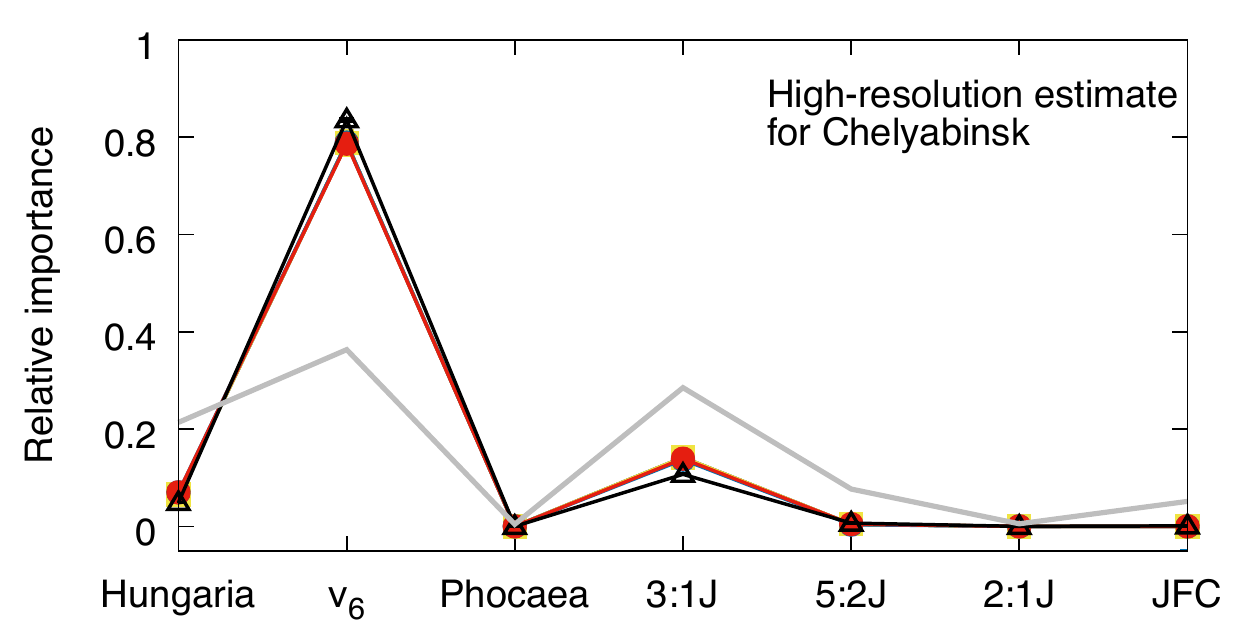}
  \includegraphics[width=0.66\columnwidth]{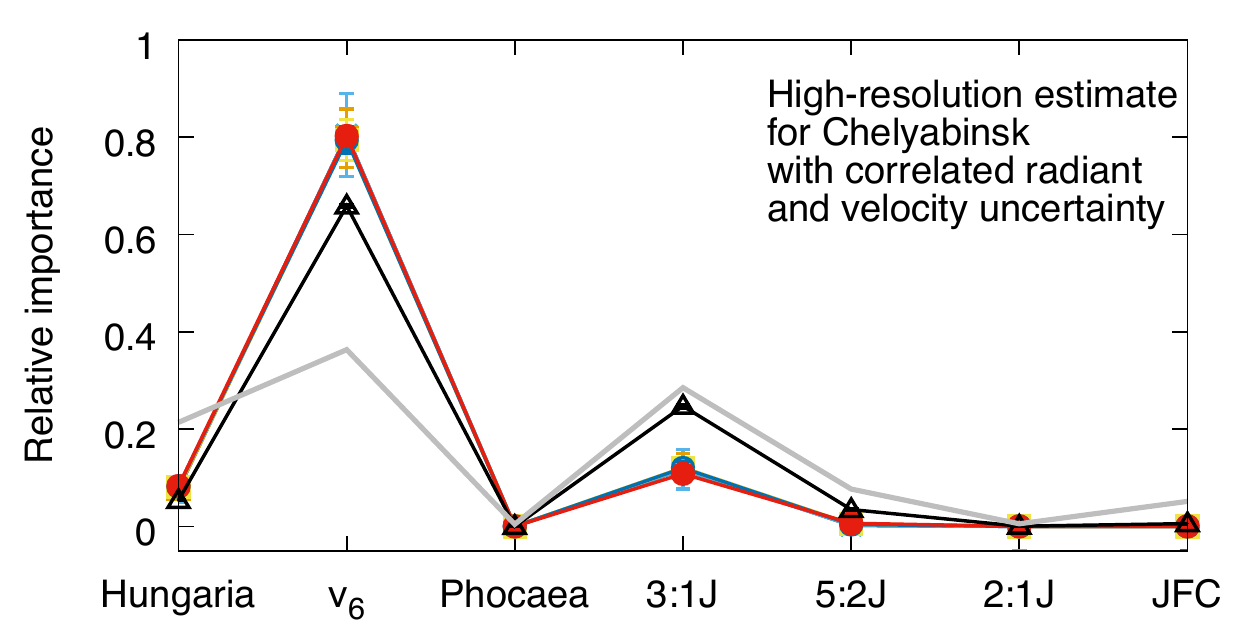}
  \includegraphics[width=0.66\columnwidth]{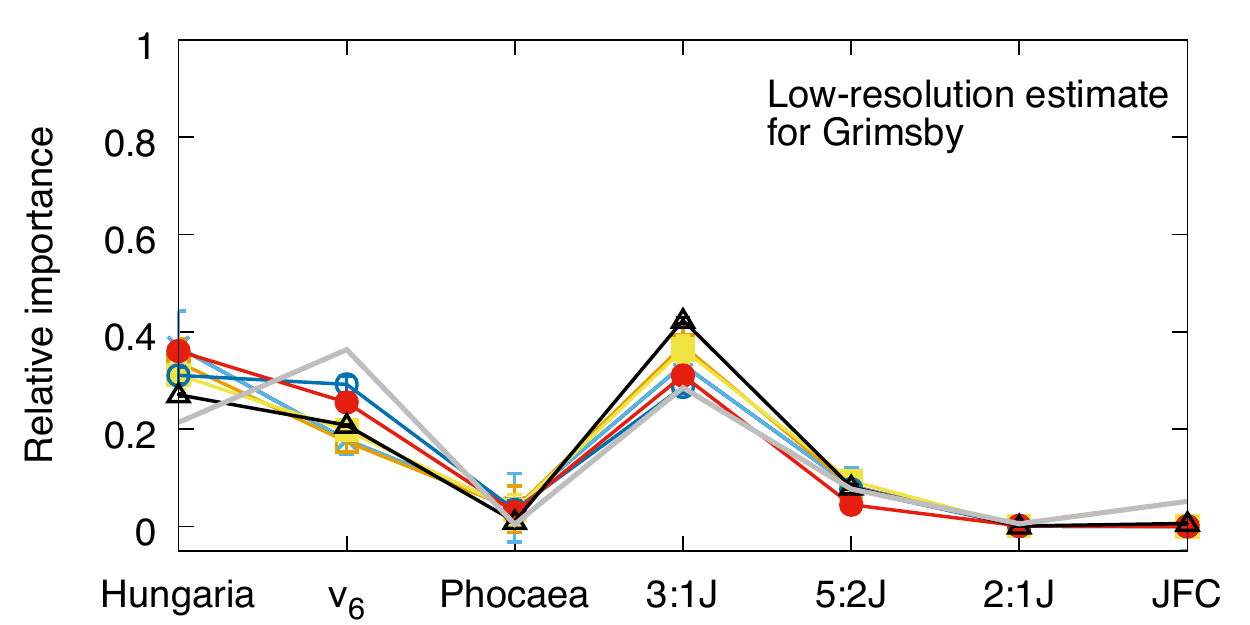}
  \includegraphics[width=0.66\columnwidth]{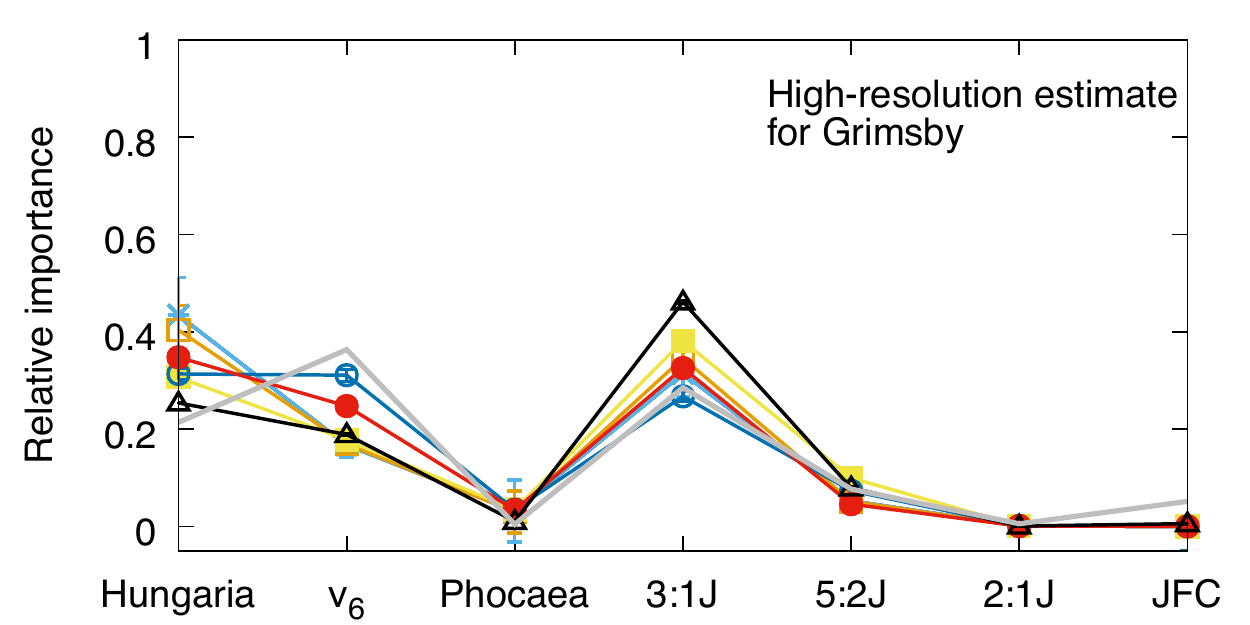}
  \includegraphics[width=0.66\columnwidth]{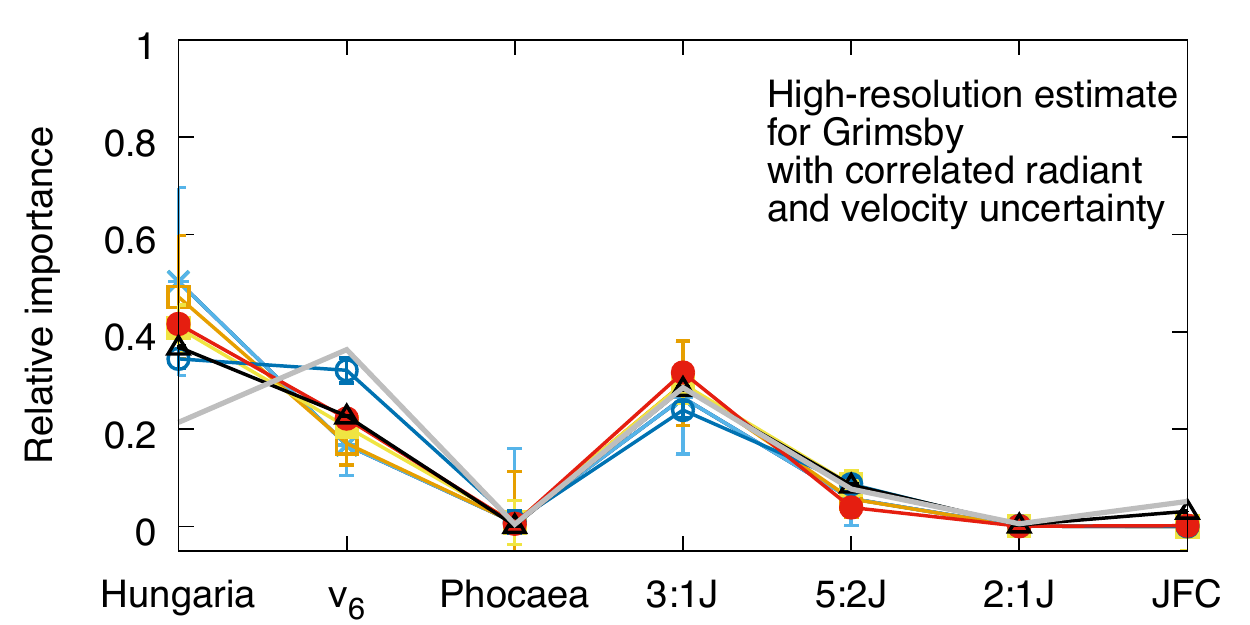}
  \caption{ER prediction corresponding to different speed
    uncertainties (defined in the legend) based on low-resolution
    model with fixed radiant uncertainty for the meteor (left),
    high-resolution model with fixed radiant uncertainty for the
    meteor (middle), and high-resolution model when assuming linear
    correlation between radiant uncertainty and velocity uncertainty
    (right). The gray line refers to the prior distribution, that is,
    the predicted ER ratios in the absence of orbital information or,
    equivalently, infinite orbital uncertainty. See
    Sect.~\ref{sec:velsens} for a detailed discussion of these plots.}
  \label{fig:set1}
\end{figure*}

\begin{figure*}
  \centering
  \includegraphics[width=0.95\textwidth]{key.pdf}
  \includegraphics[width=0.66\columnwidth]{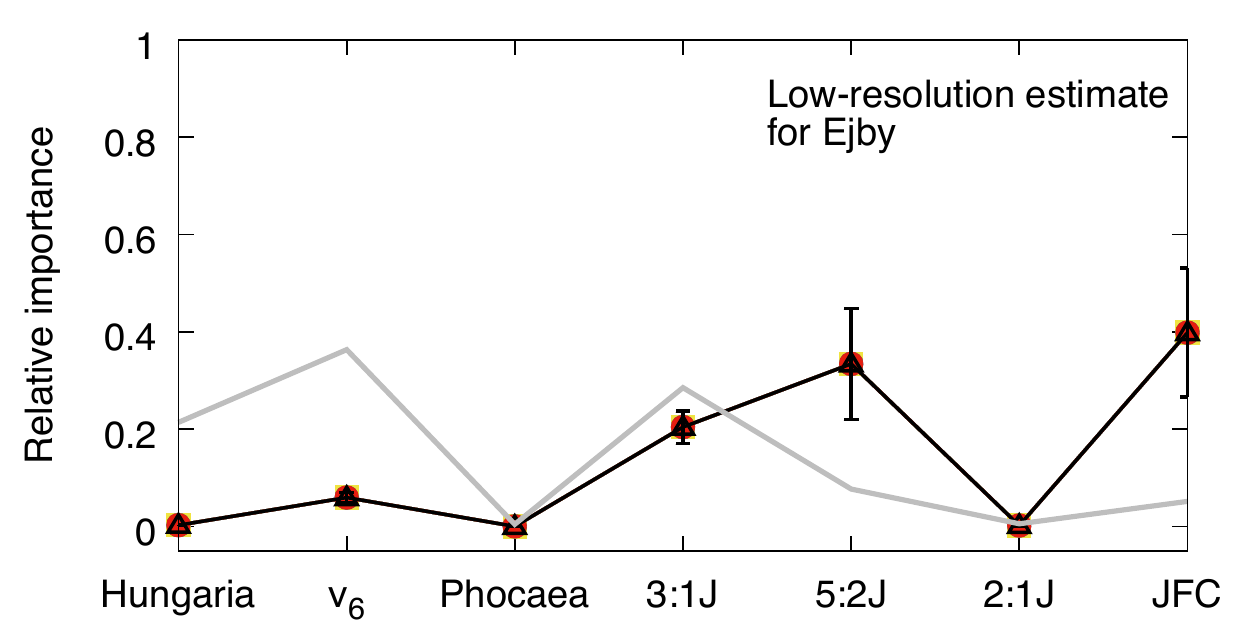}
  \includegraphics[width=0.66\columnwidth]{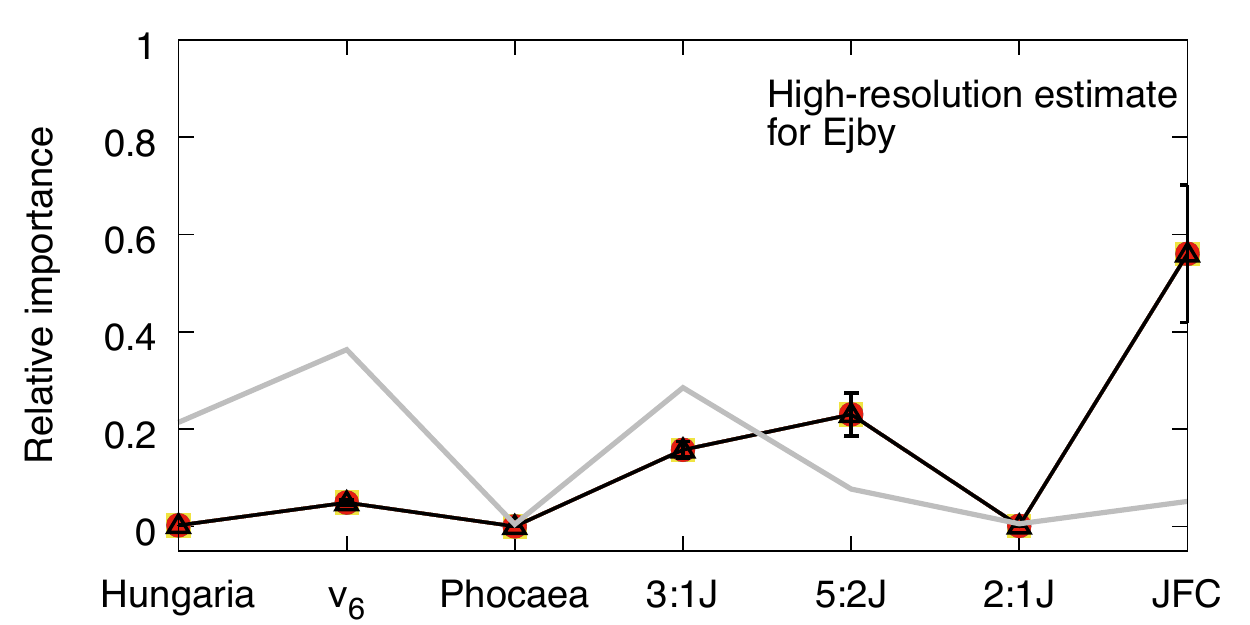}
  \includegraphics[width=0.66\columnwidth]{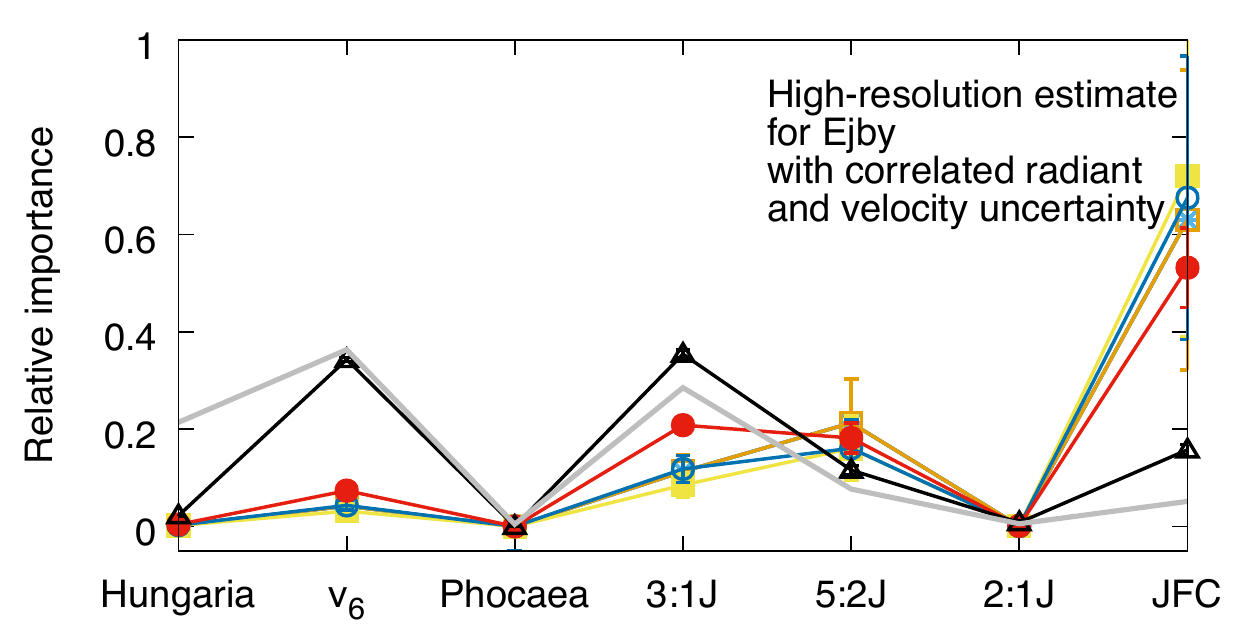}
  \includegraphics[width=0.66\columnwidth]{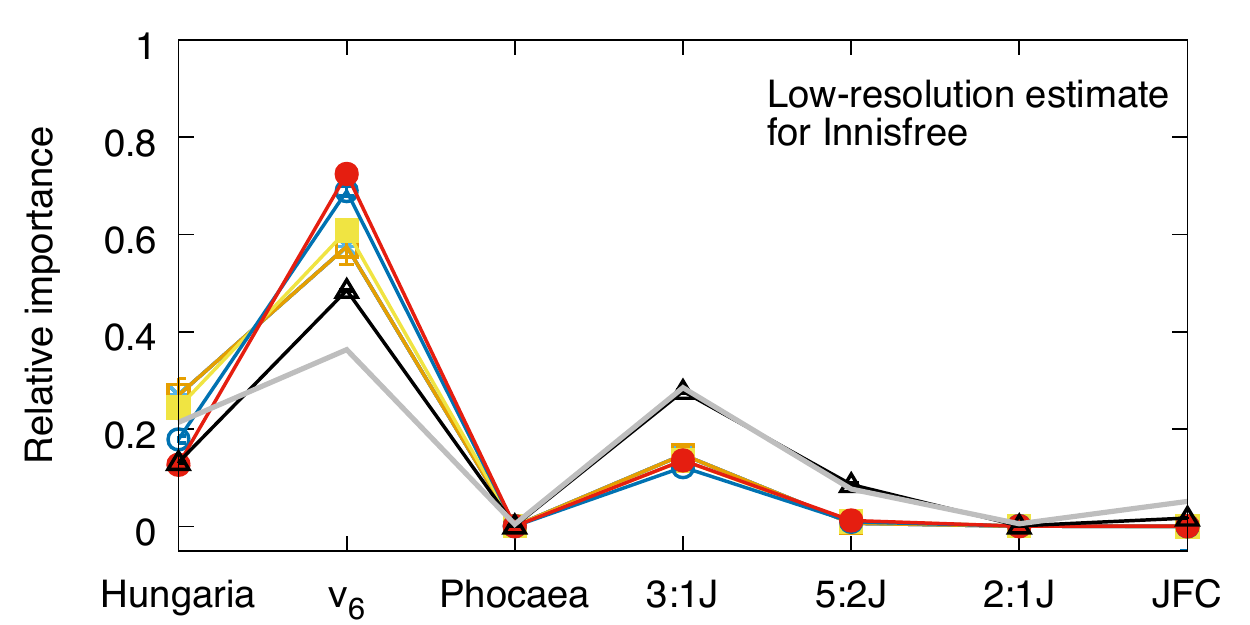}
  \includegraphics[width=0.66\columnwidth]{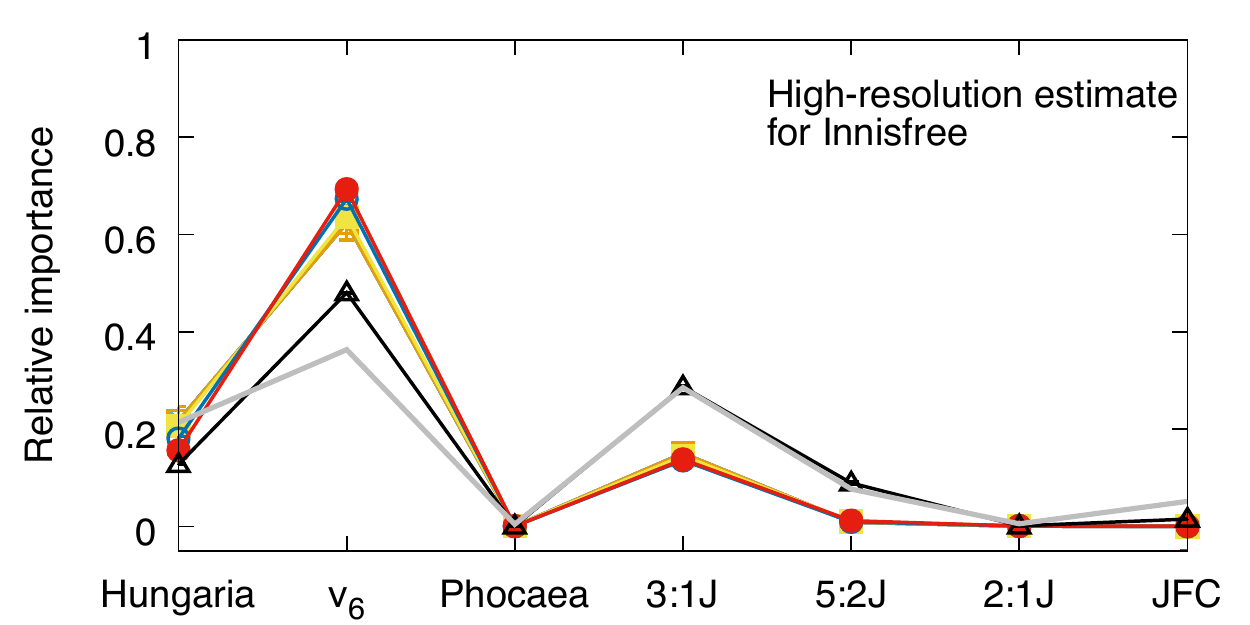}
  \includegraphics[width=0.66\columnwidth]{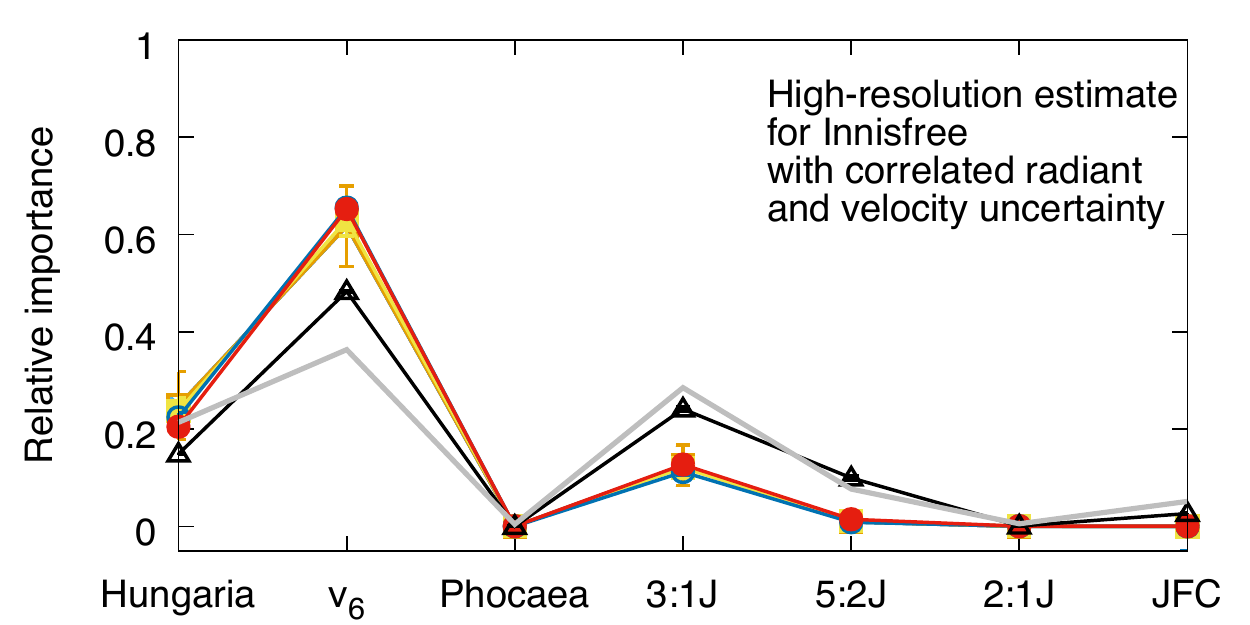}
  \includegraphics[width=0.66\columnwidth]{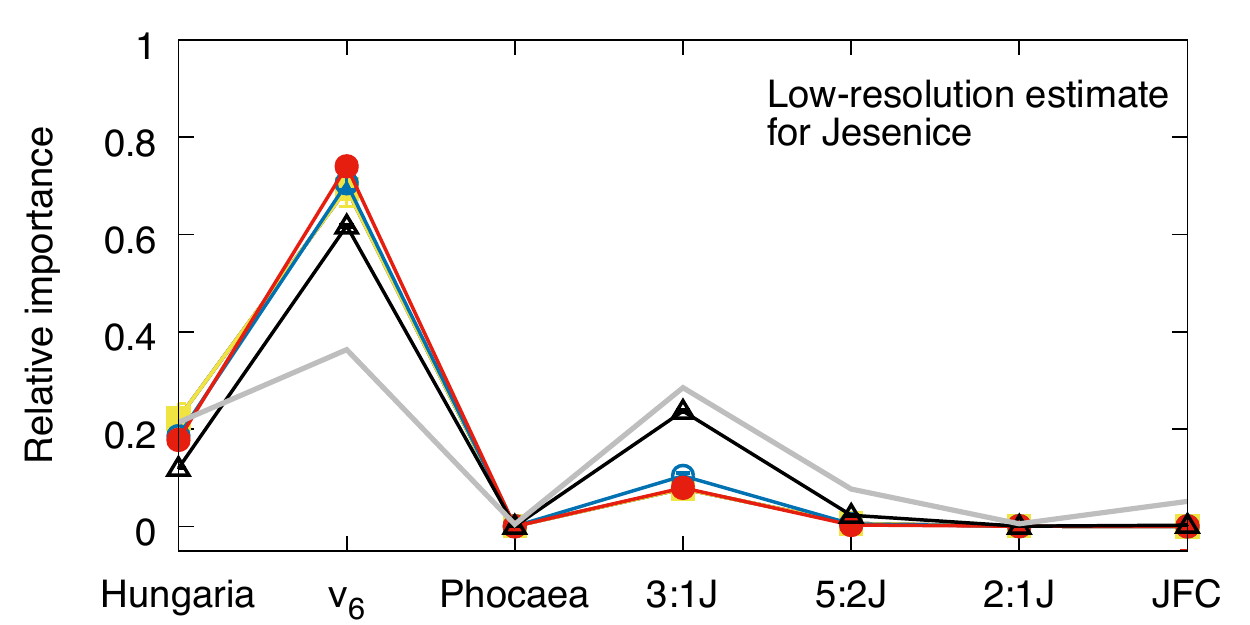}
  \includegraphics[width=0.66\columnwidth]{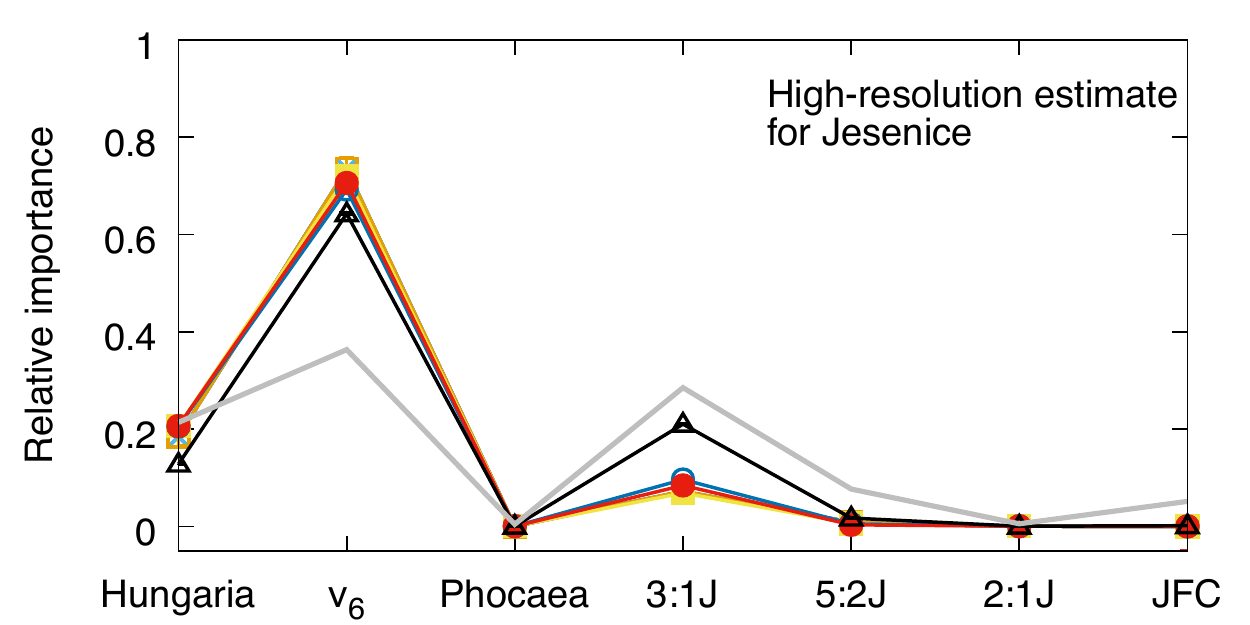}
  \includegraphics[width=0.66\columnwidth]{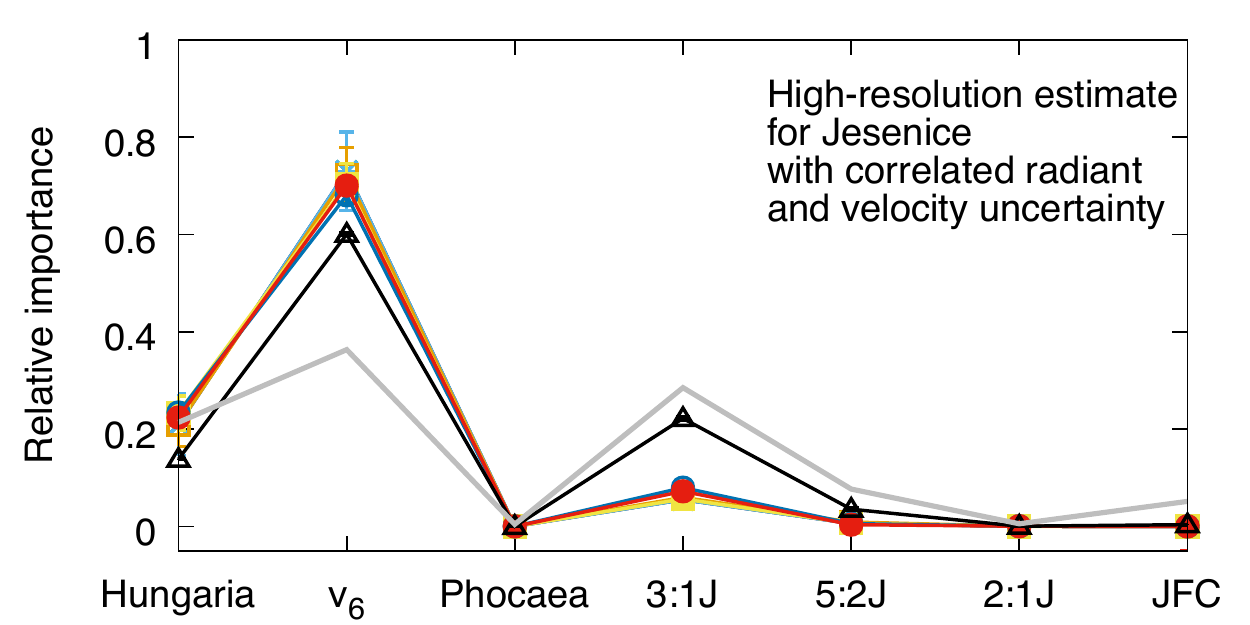}
  \includegraphics[width=0.66\columnwidth]{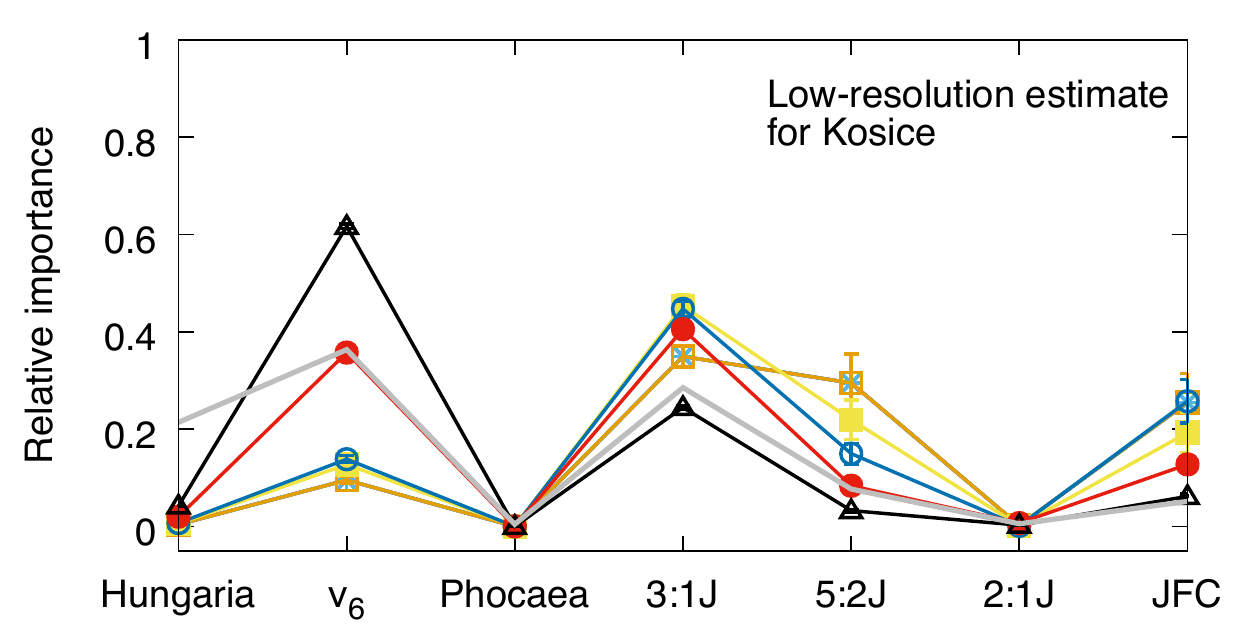}
  \includegraphics[width=0.66\columnwidth]{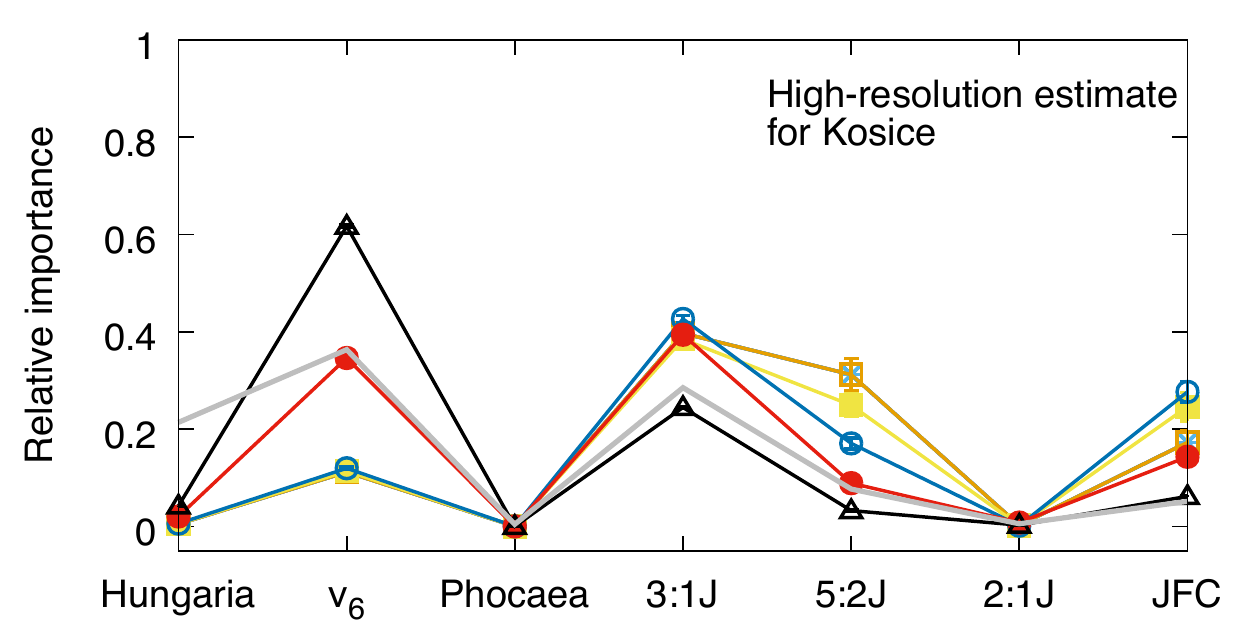}
  \includegraphics[width=0.66\columnwidth]{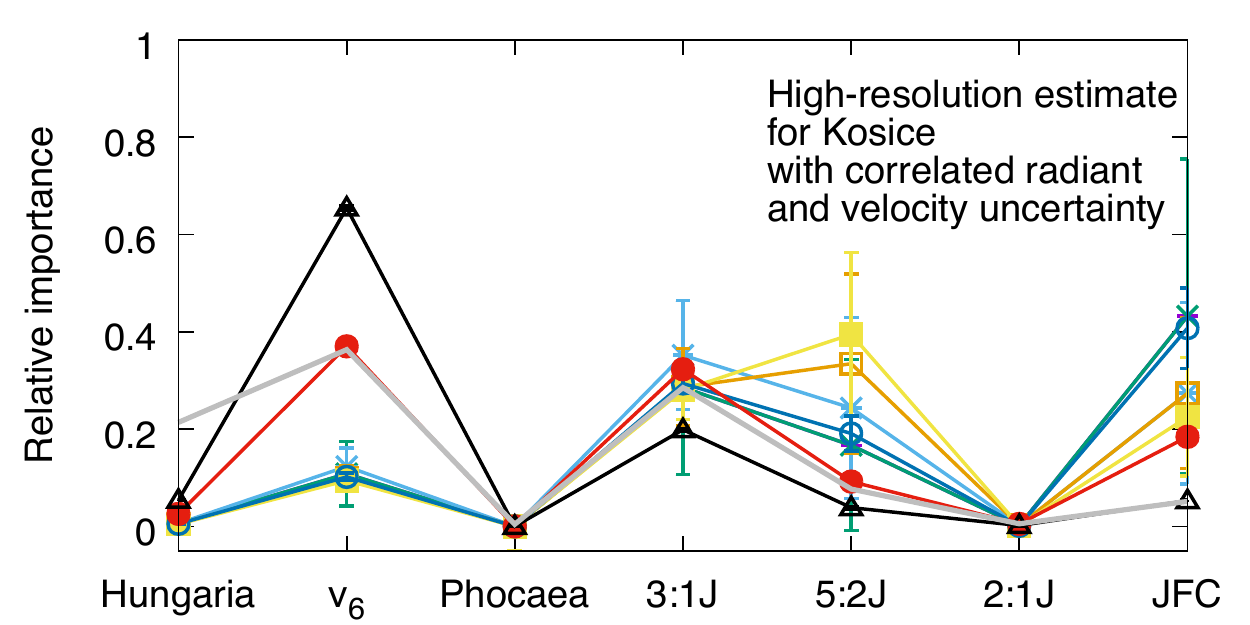}
  \includegraphics[width=0.66\columnwidth]{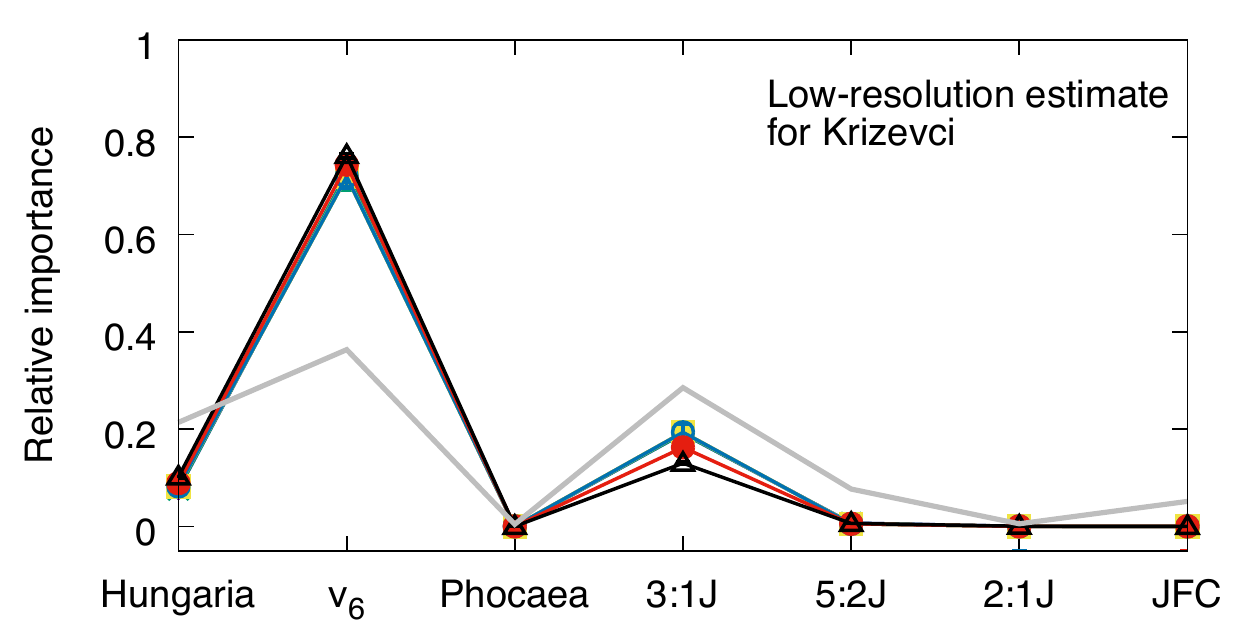}
  \includegraphics[width=0.66\columnwidth]{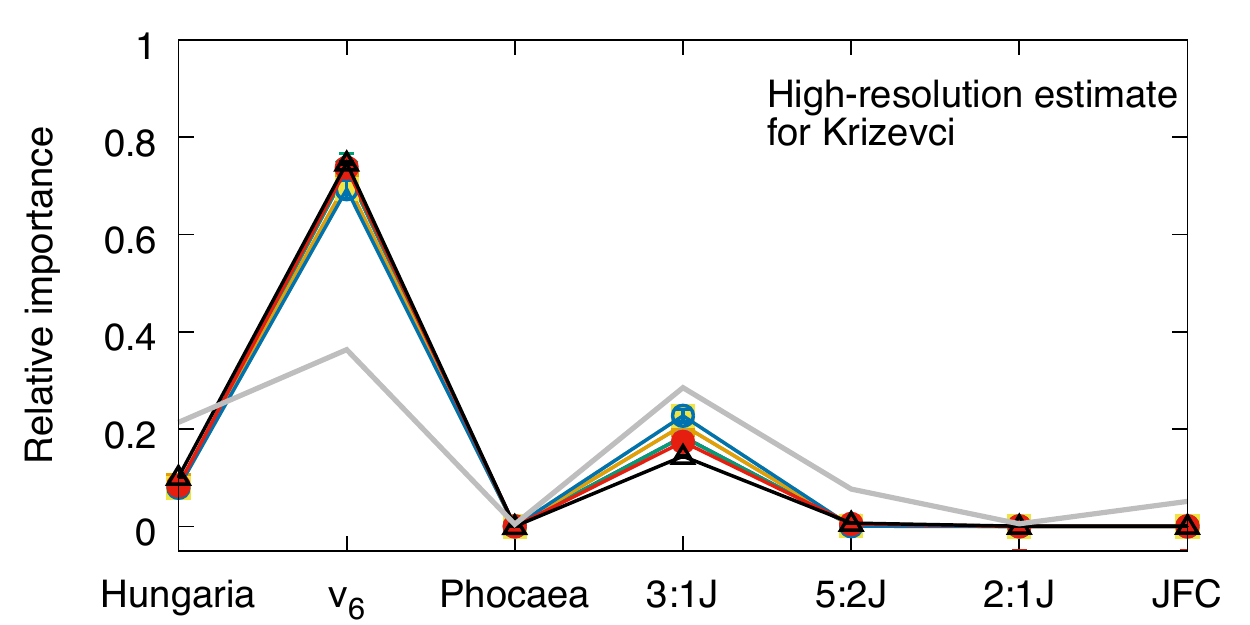}
  \includegraphics[width=0.66\columnwidth]{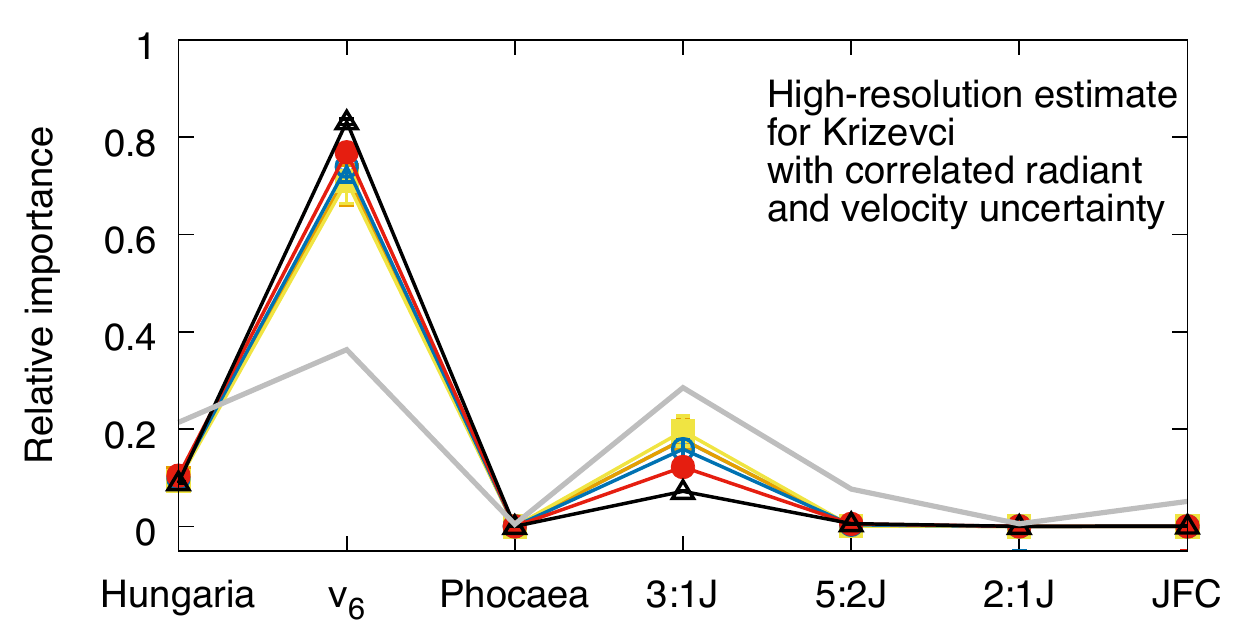}
  \includegraphics[width=0.66\columnwidth]{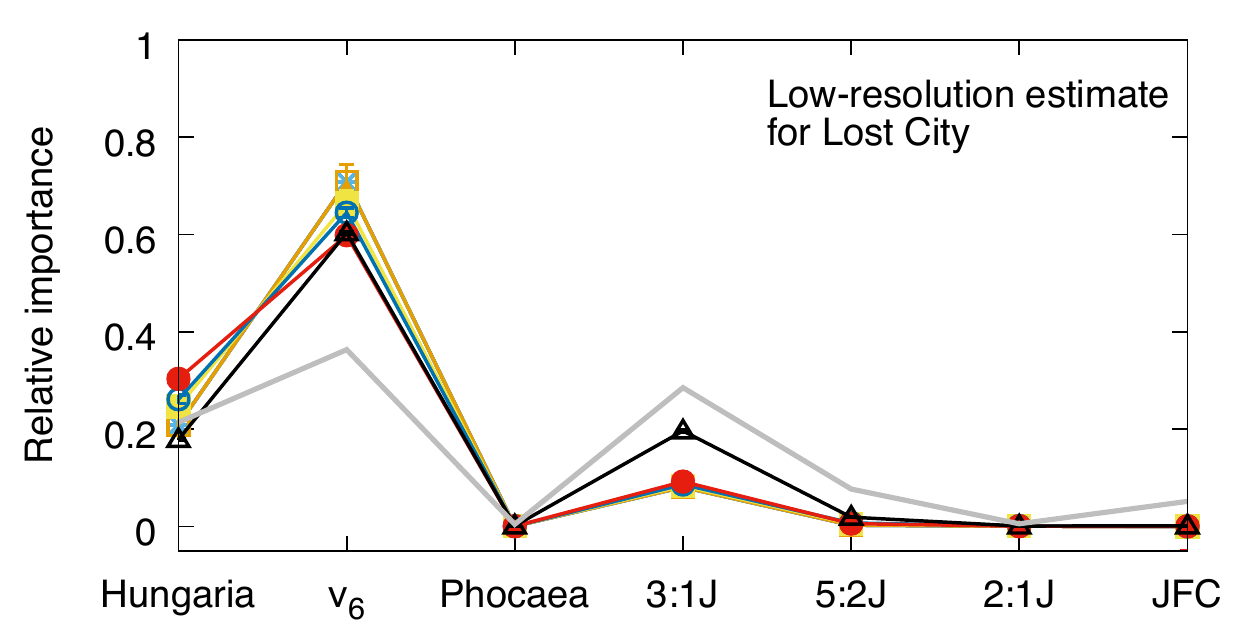}
  \includegraphics[width=0.66\columnwidth]{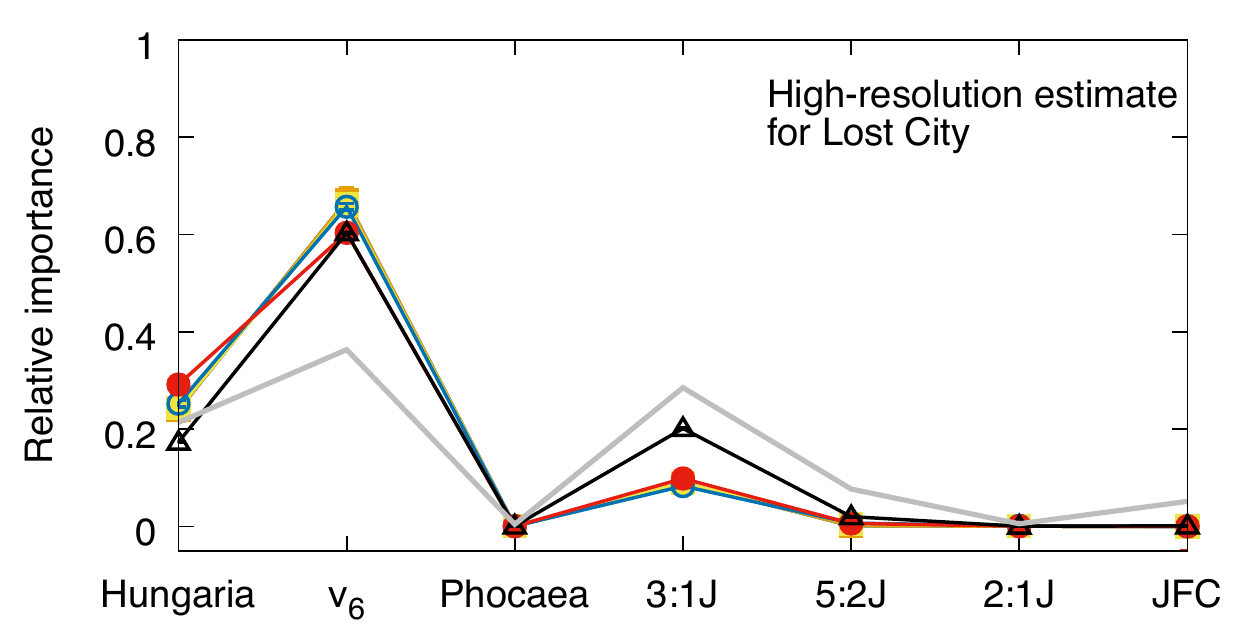}
  \includegraphics[width=0.66\columnwidth]{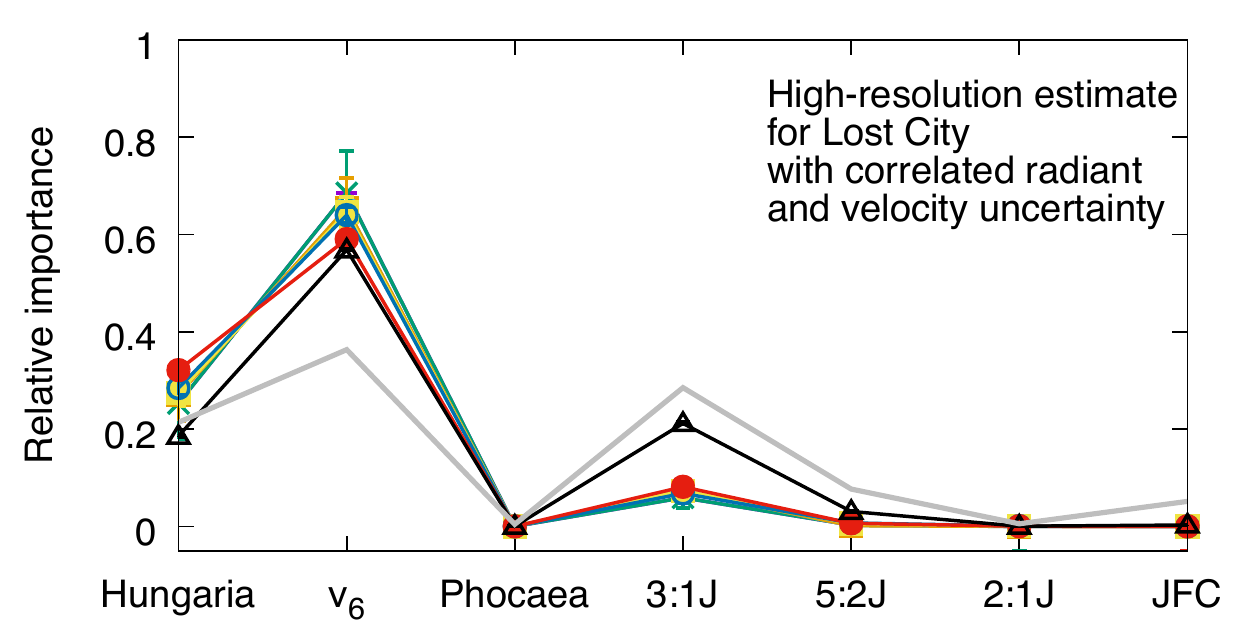}
  \includegraphics[width=0.66\columnwidth]{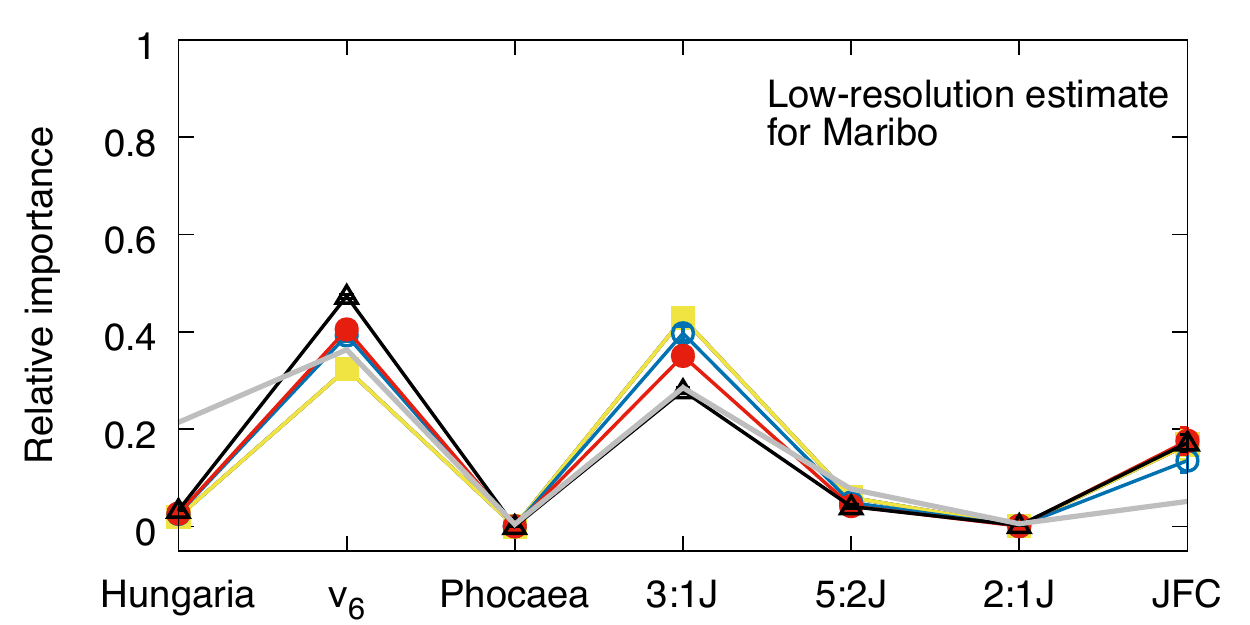}
  \includegraphics[width=0.66\columnwidth]{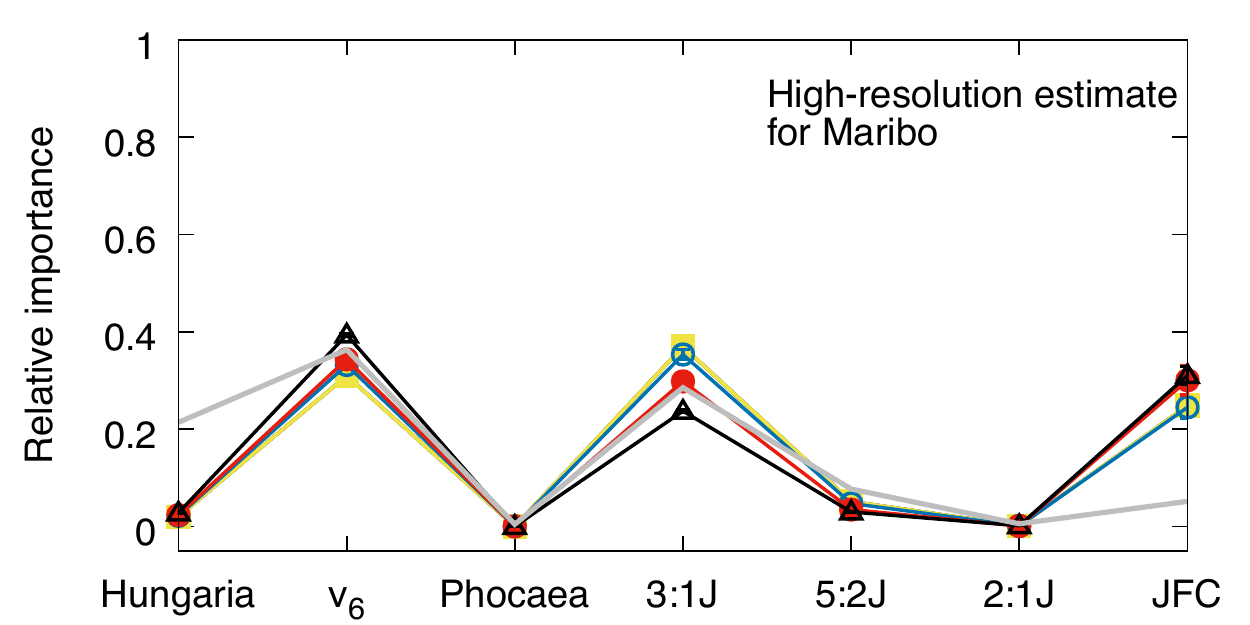}
  \includegraphics[width=0.66\columnwidth]{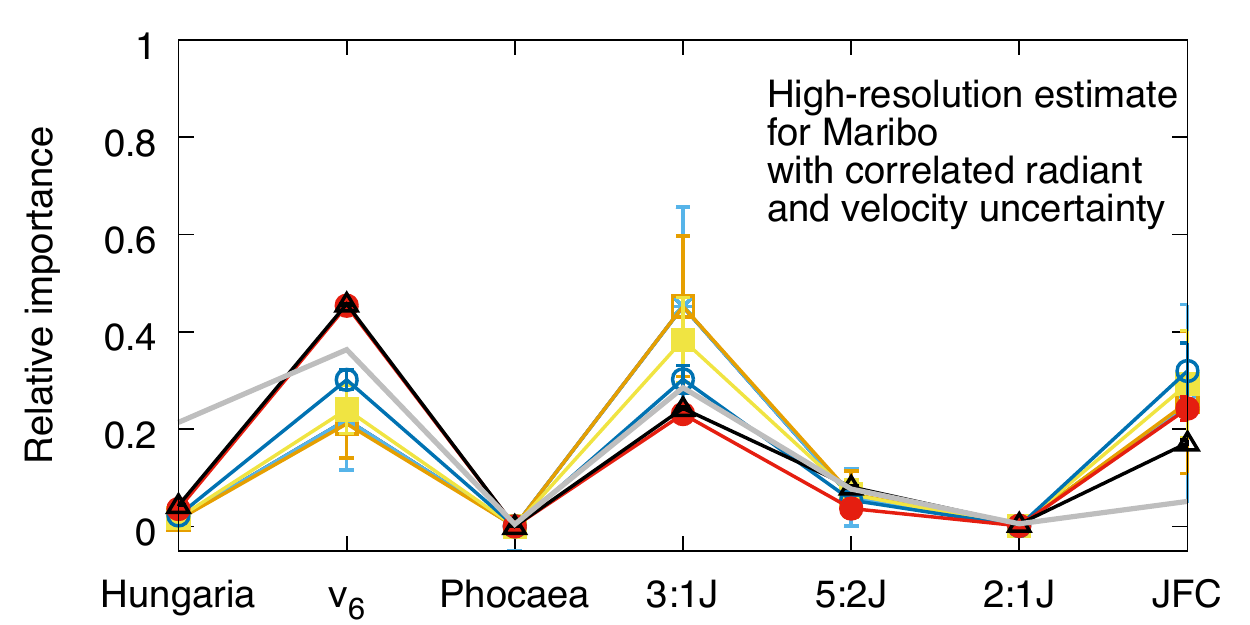}
  \caption{As Fig.~\ref{fig:set1}.}
  \label{fig:set2}
\end{figure*}

\begin{figure*}
  \centering
  \includegraphics[width=0.95\textwidth]{key.pdf}
  \includegraphics[width=0.66\columnwidth]{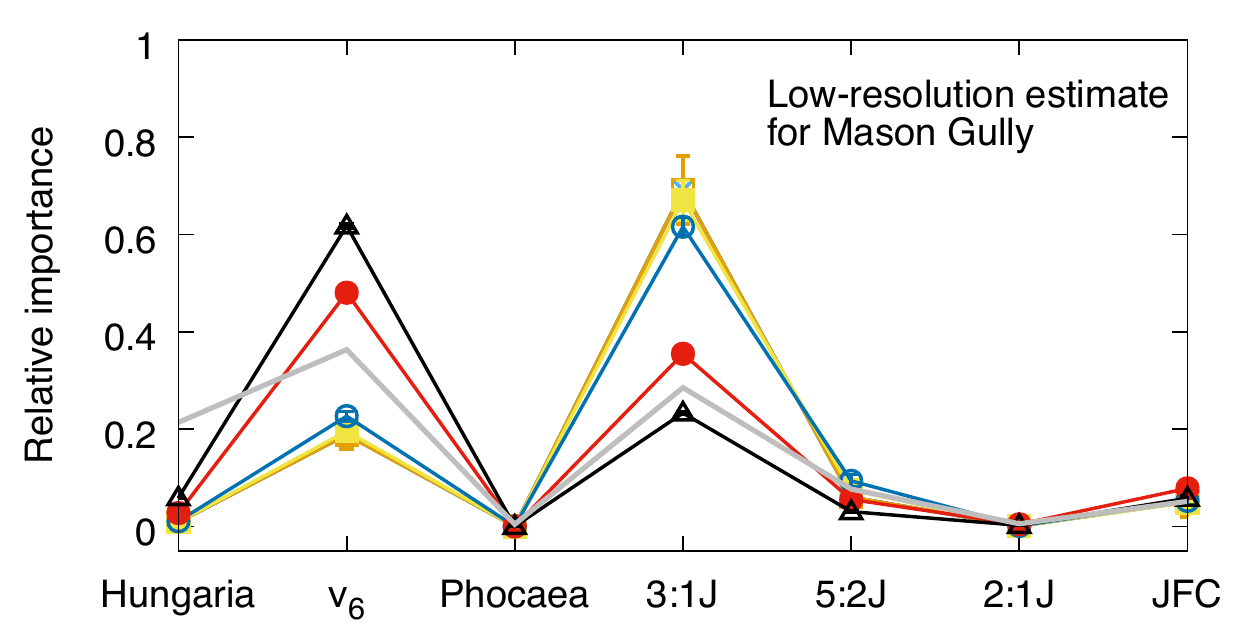}
  \includegraphics[width=0.66\columnwidth]{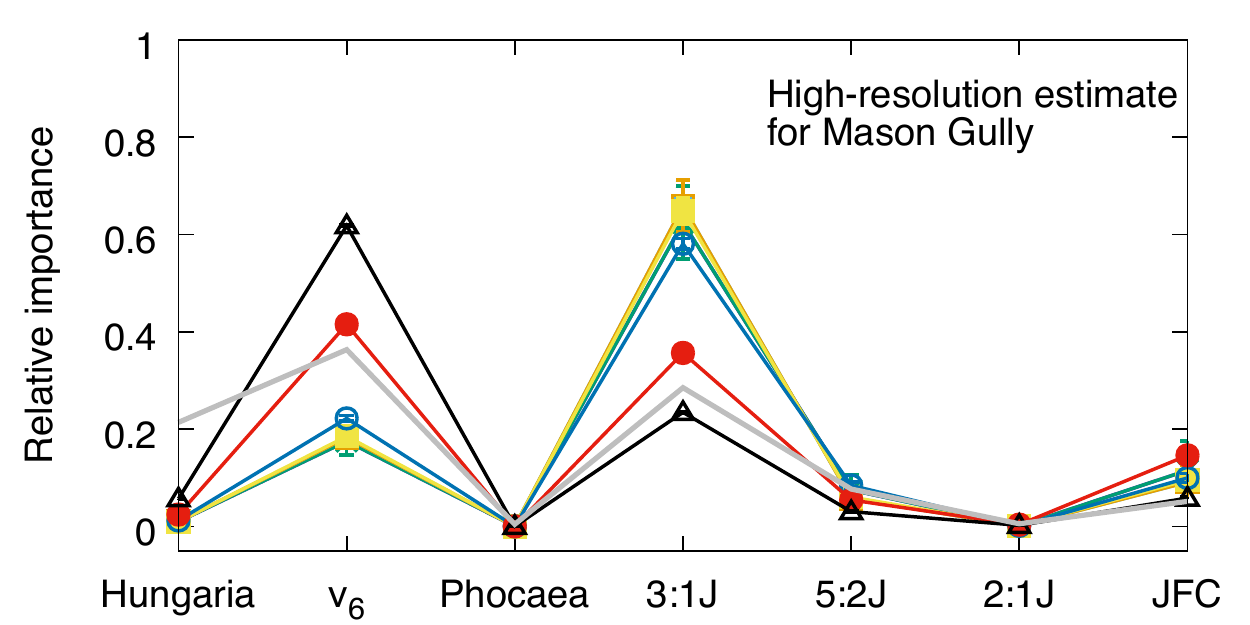}
  \includegraphics[width=0.66\columnwidth]{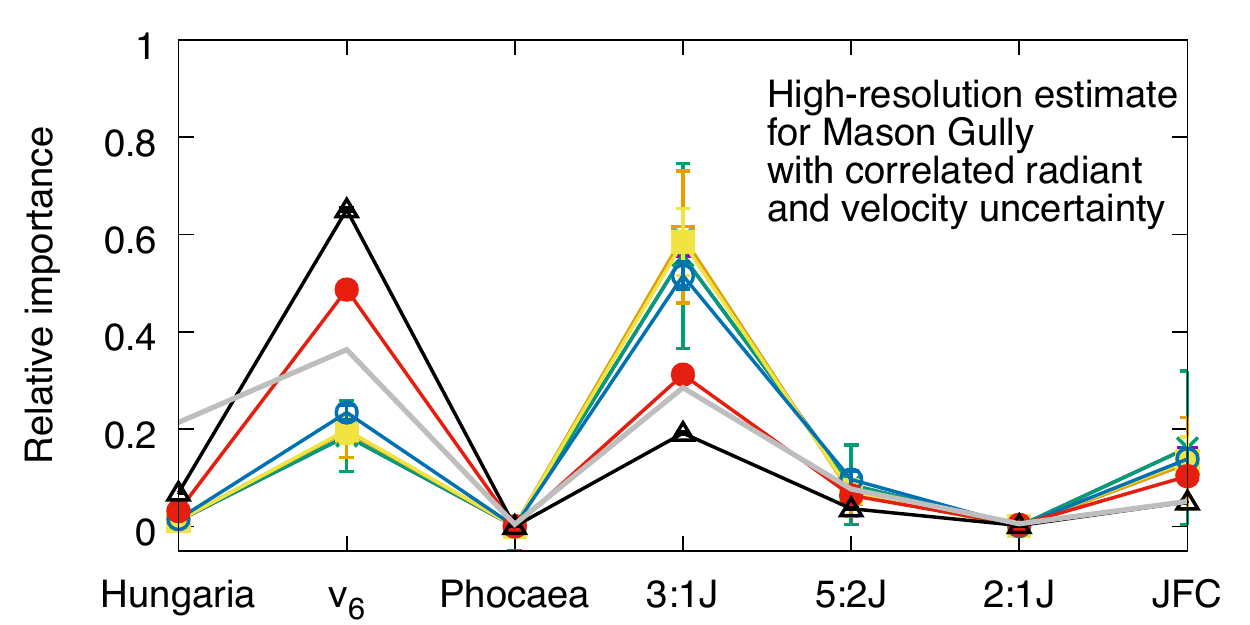}
  \includegraphics[width=0.66\columnwidth]{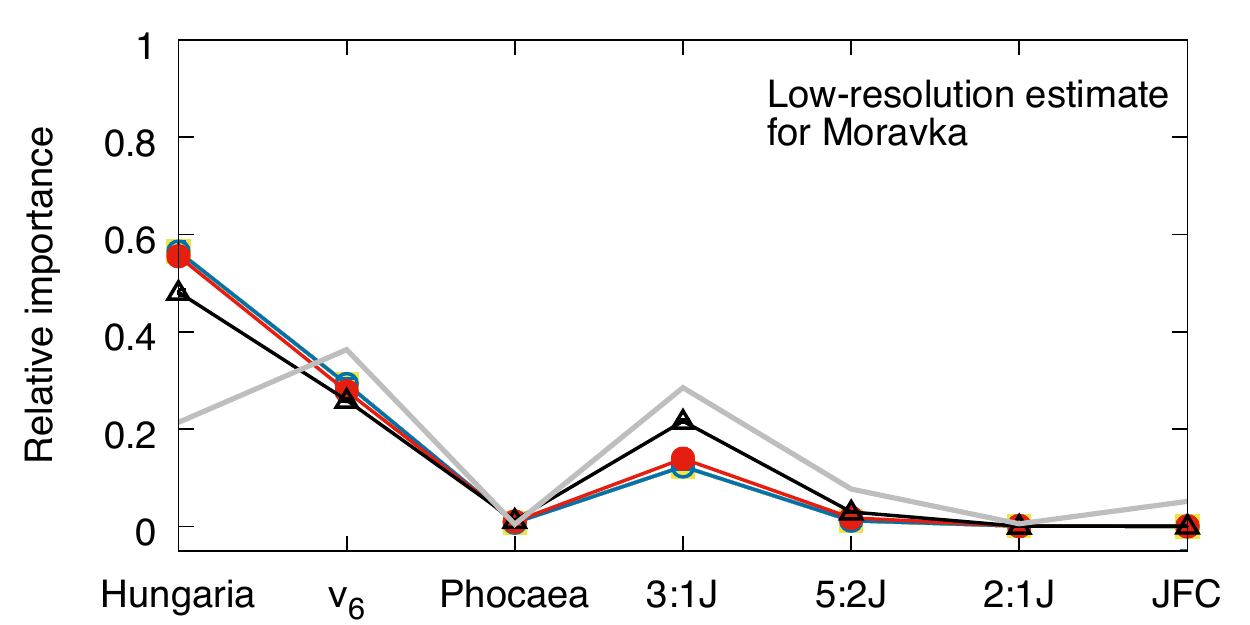}
  \includegraphics[width=0.66\columnwidth]{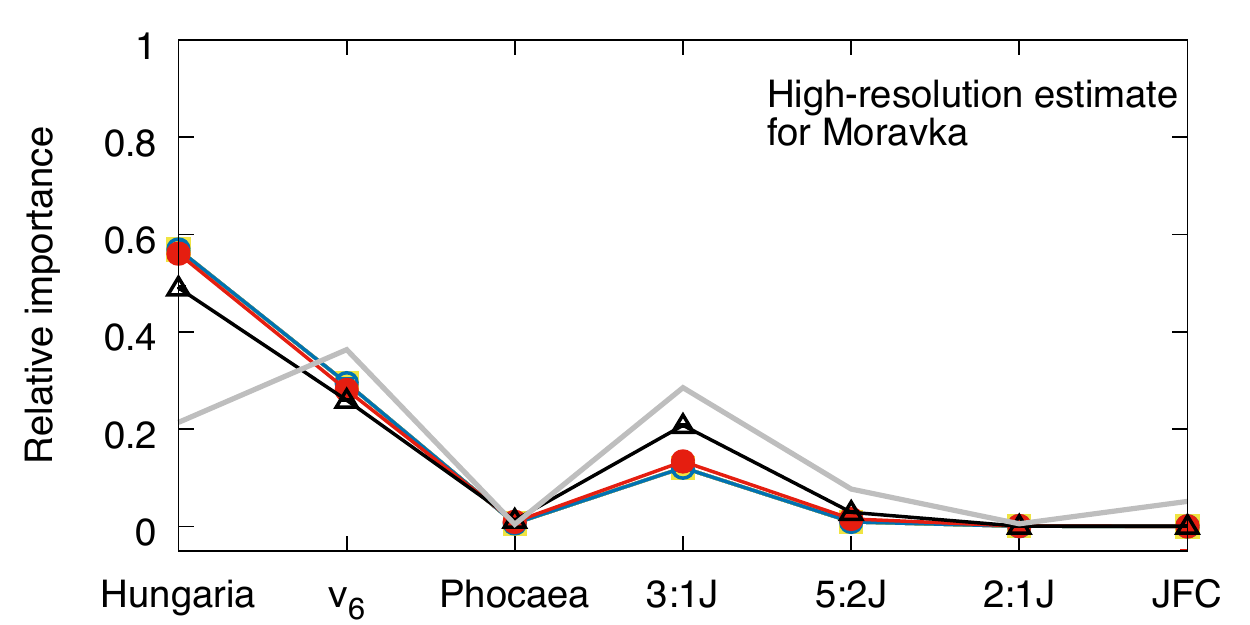}
  \includegraphics[width=0.66\columnwidth]{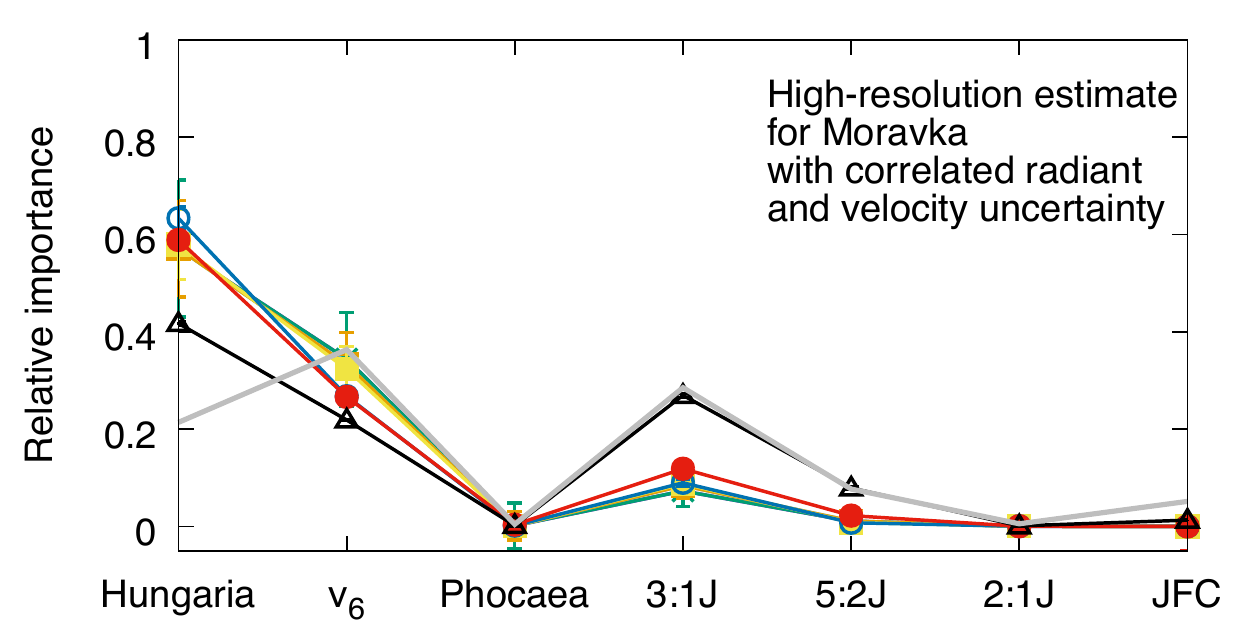}
  \includegraphics[width=0.66\columnwidth]{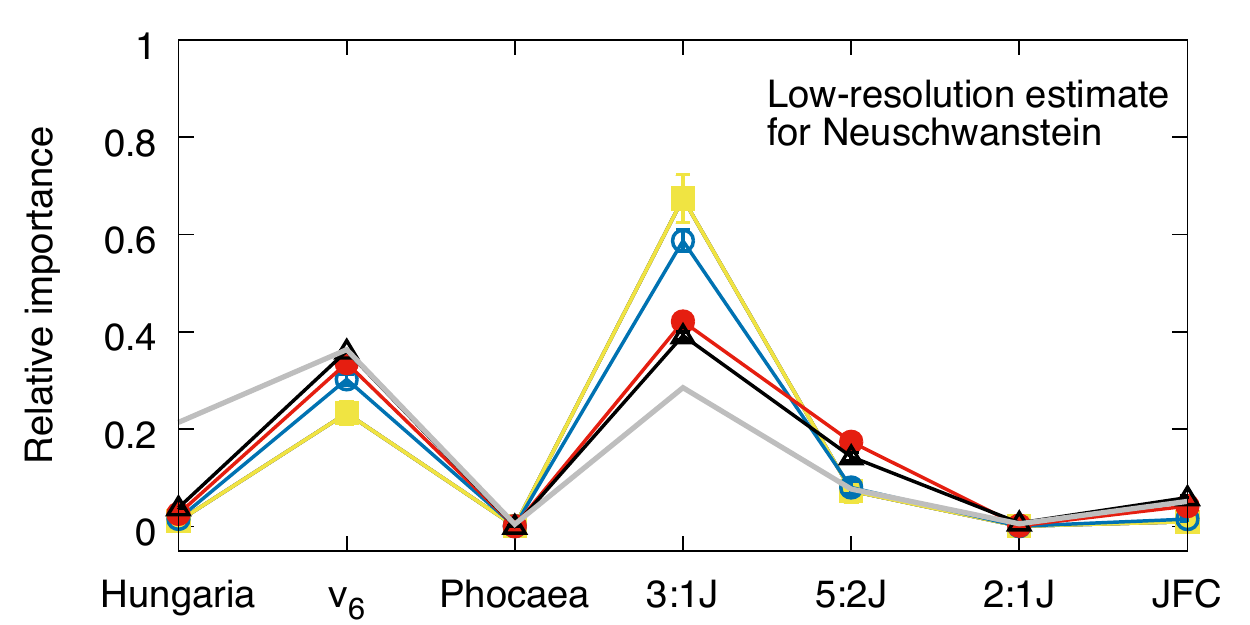}
  \includegraphics[width=0.66\columnwidth]{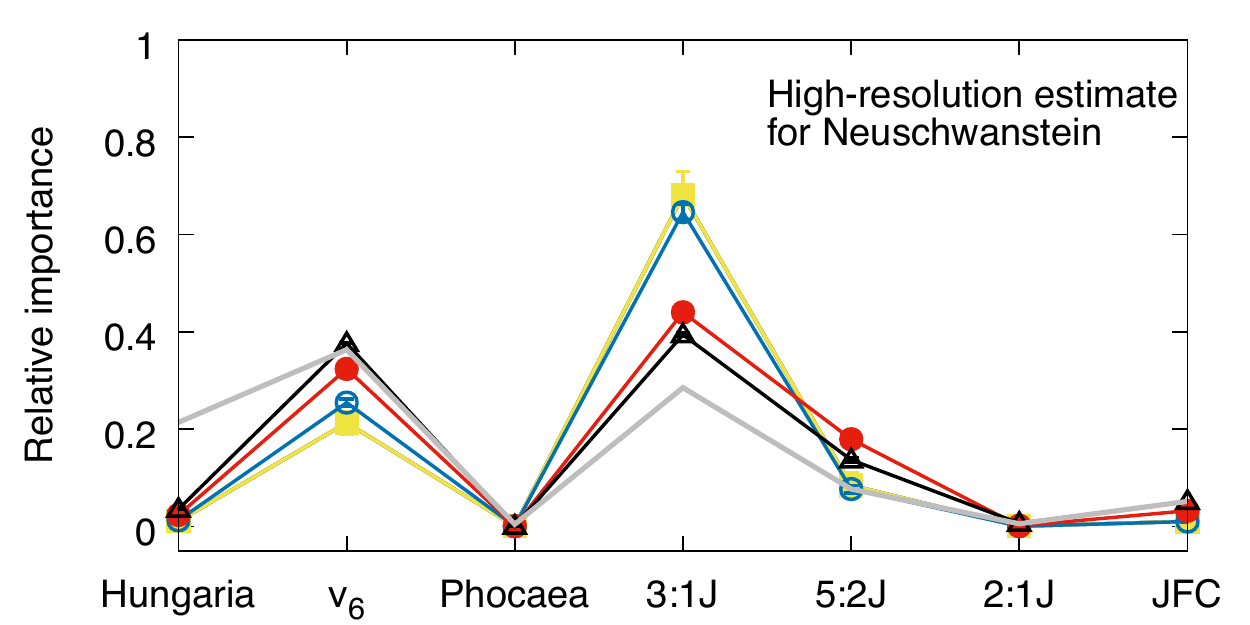}
  \includegraphics[width=0.66\columnwidth]{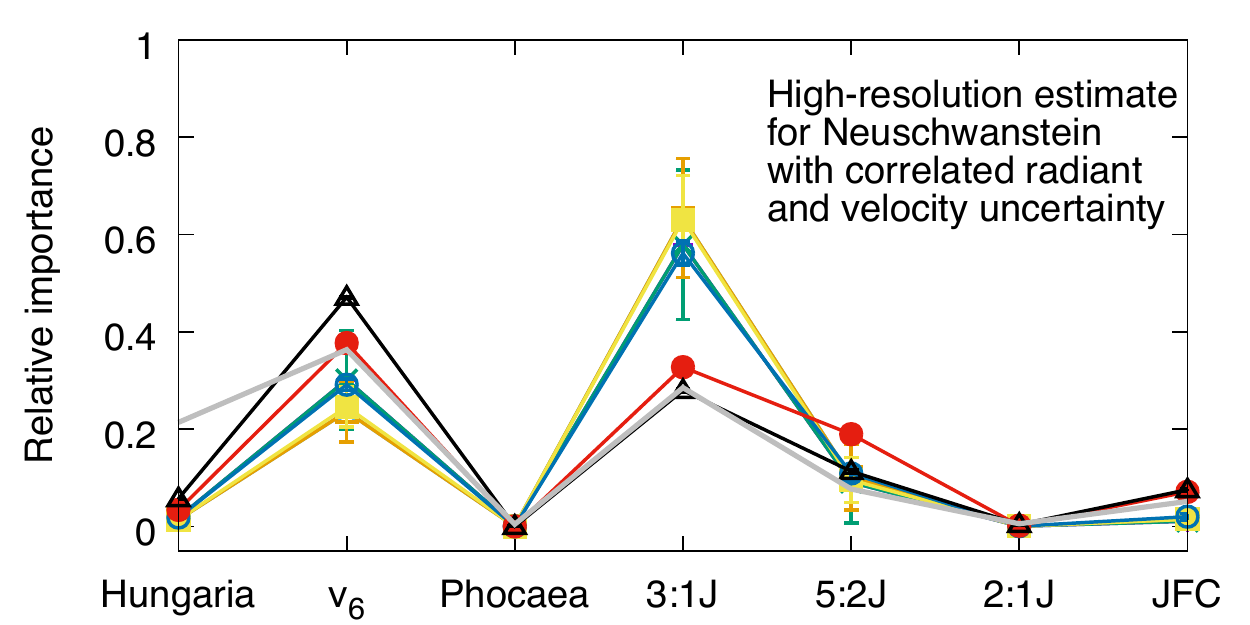}
  \includegraphics[width=0.66\columnwidth]{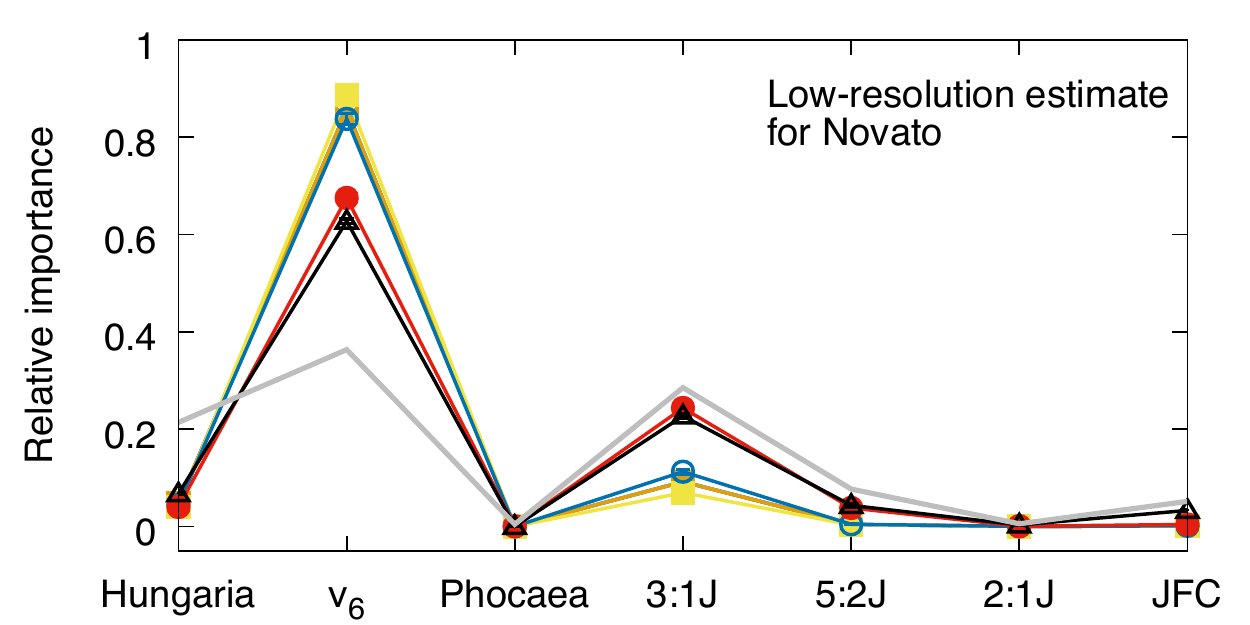}
  \includegraphics[width=0.66\columnwidth]{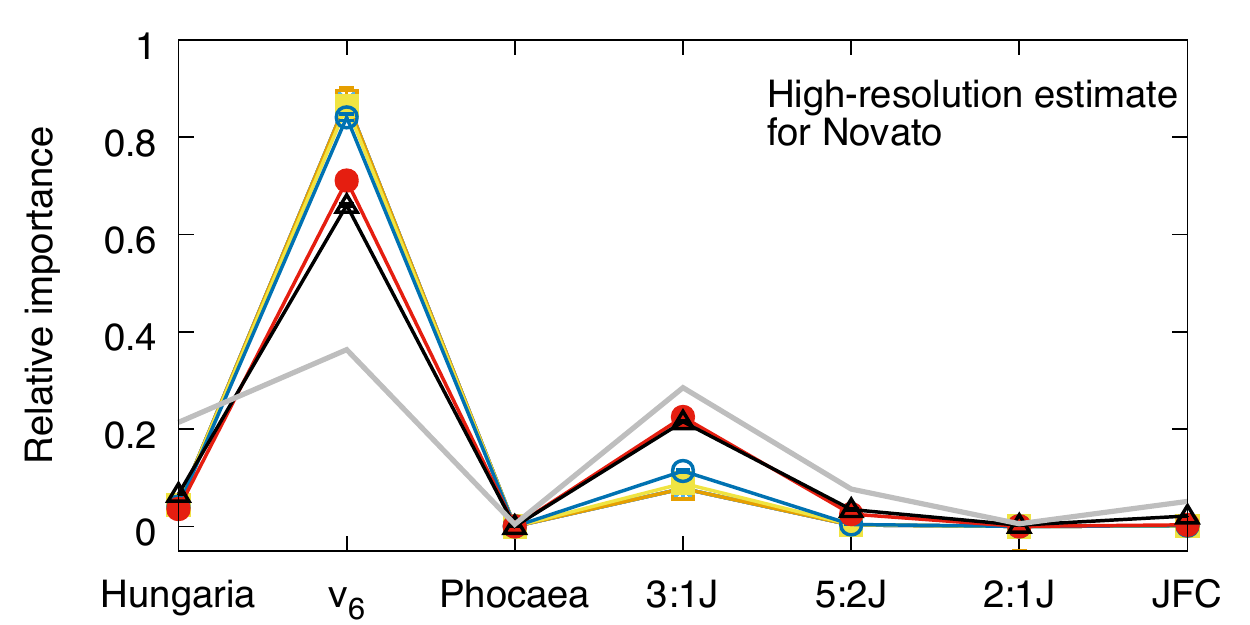}
  \includegraphics[width=0.66\columnwidth]{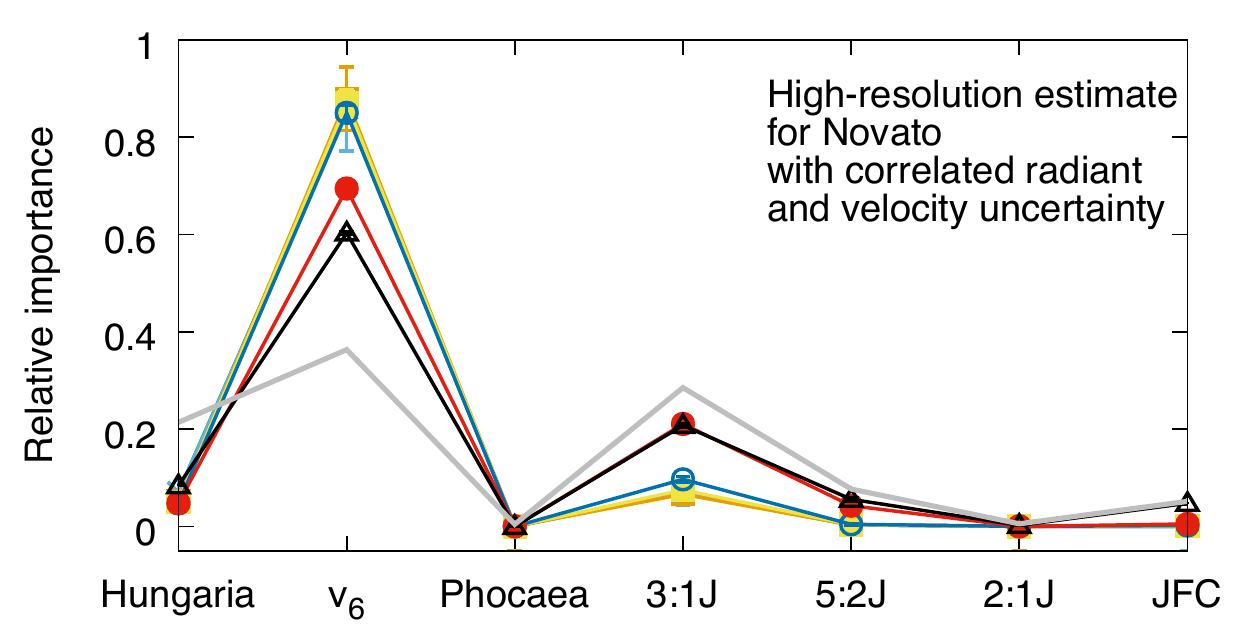}
  \includegraphics[width=0.66\columnwidth]{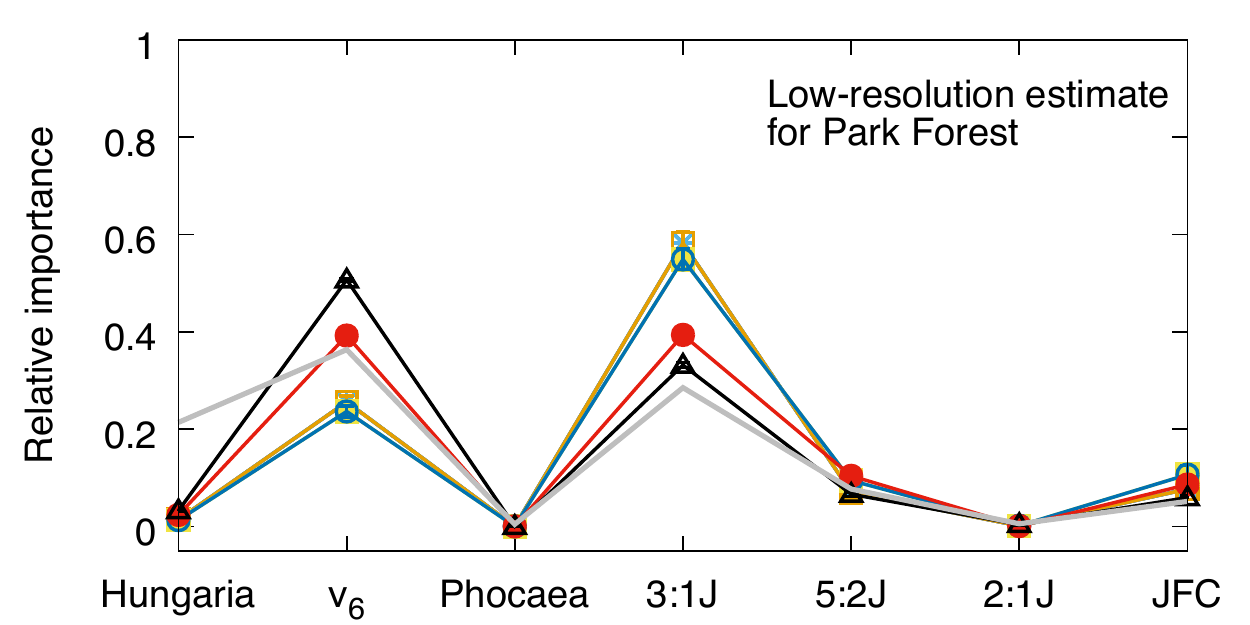}
  \includegraphics[width=0.66\columnwidth]{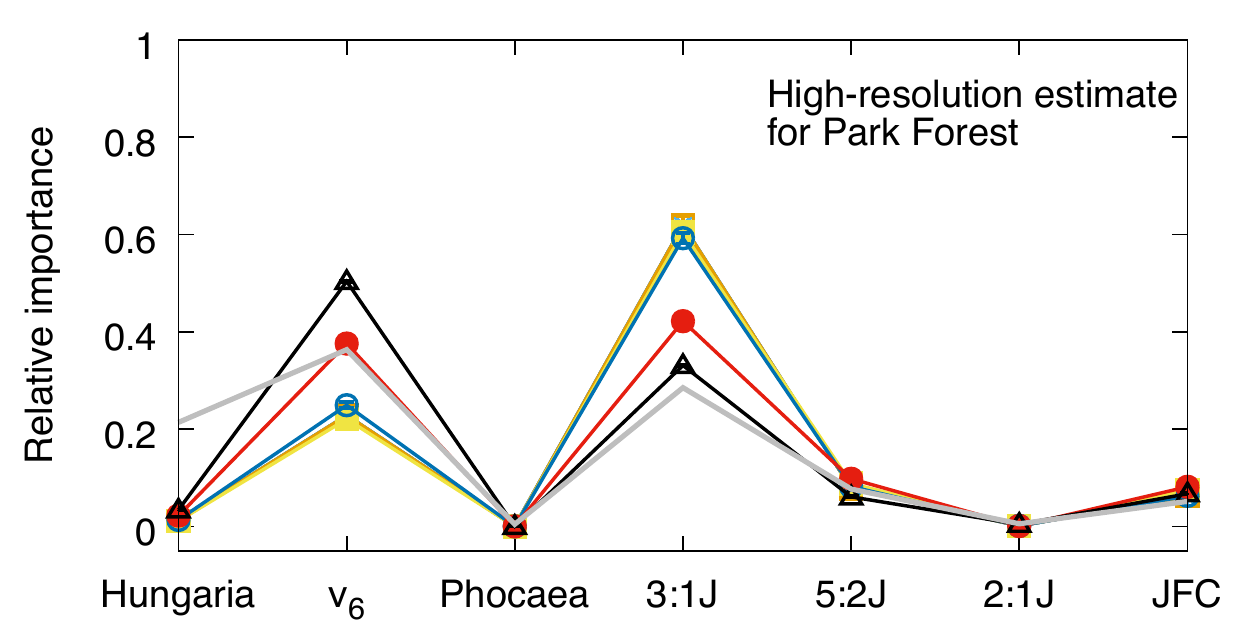}
  \includegraphics[width=0.66\columnwidth]{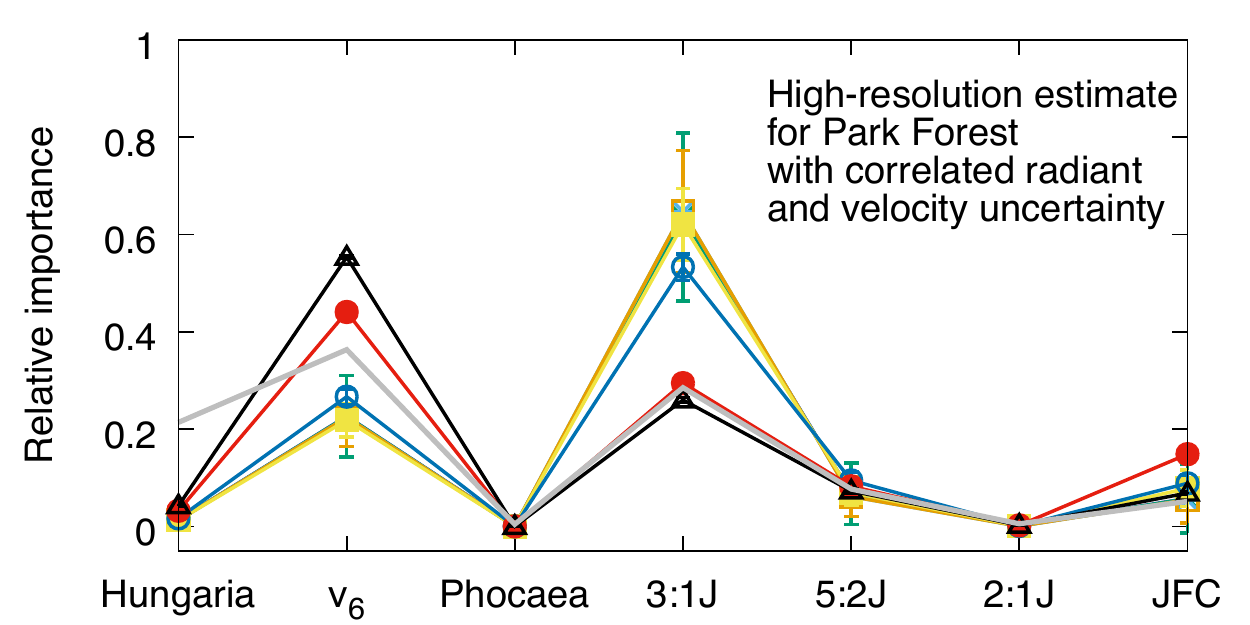}
  \includegraphics[width=0.66\columnwidth]{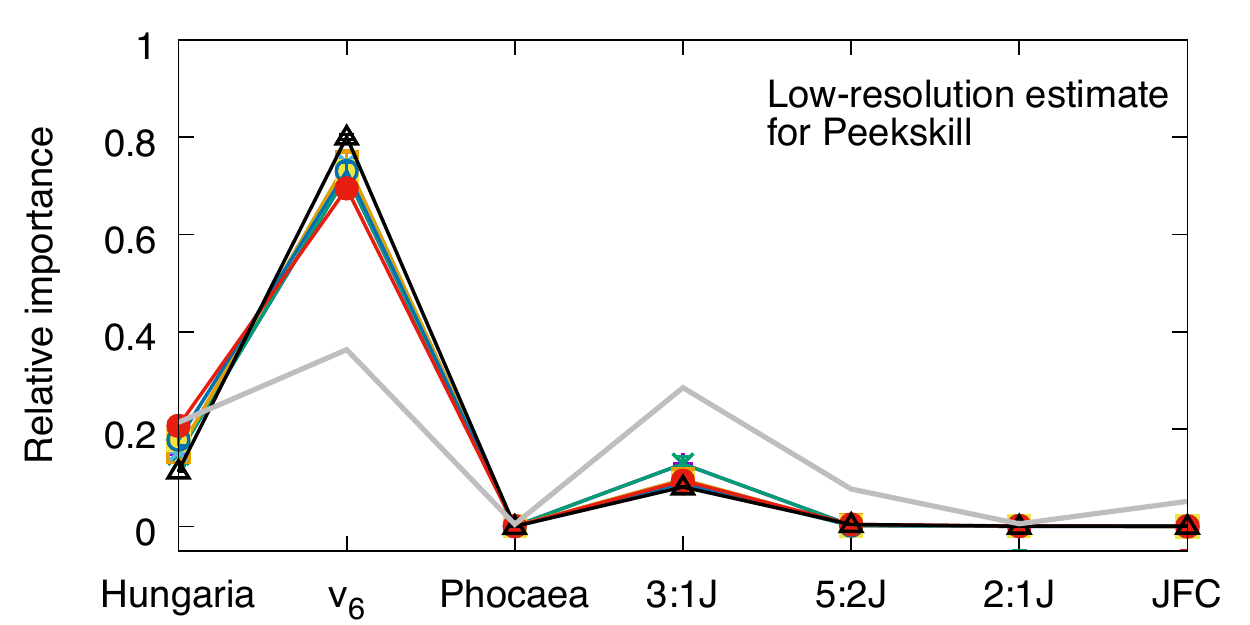}
  \includegraphics[width=0.66\columnwidth]{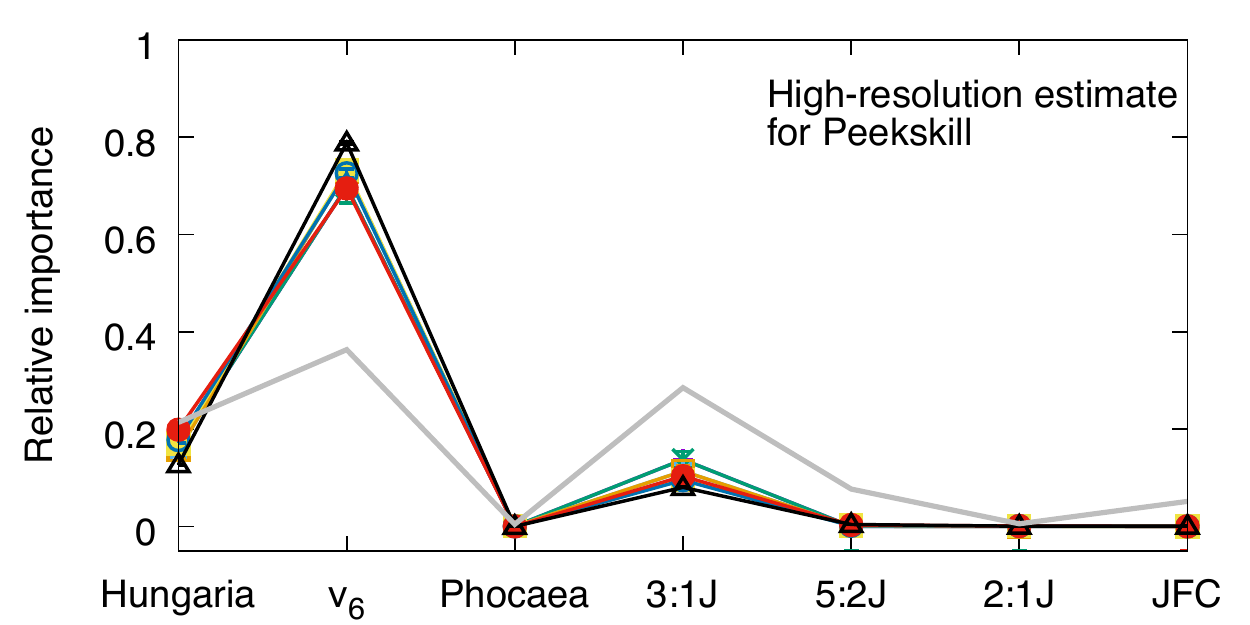}
  \includegraphics[width=0.66\columnwidth]{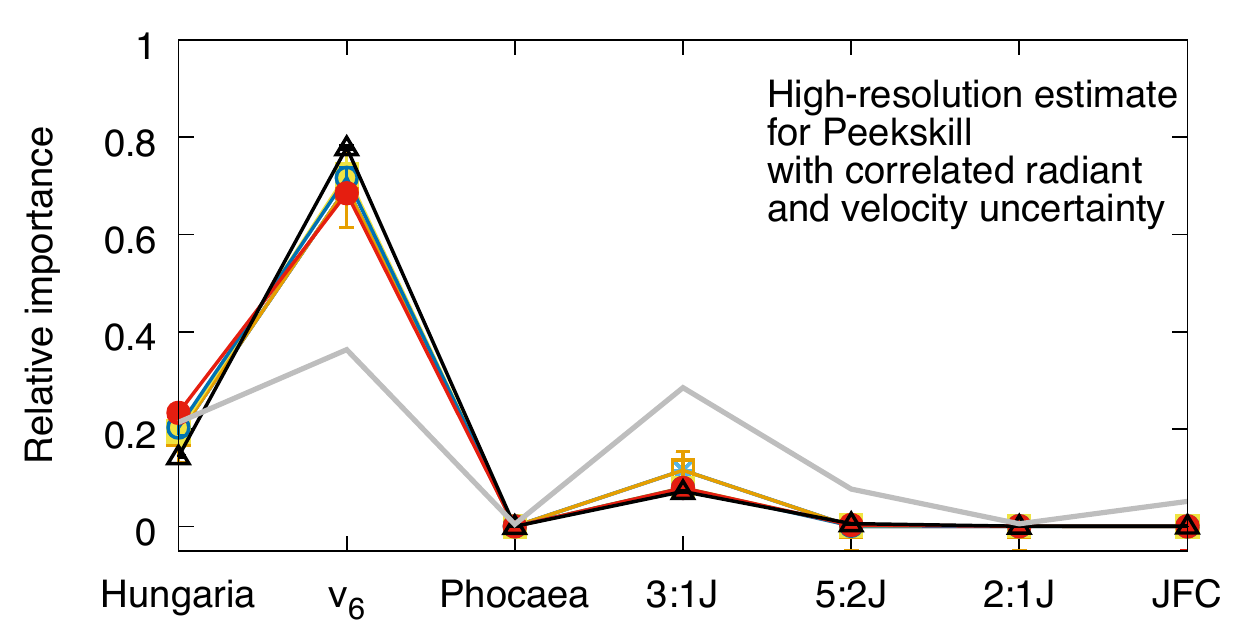}
  \includegraphics[width=0.66\columnwidth]{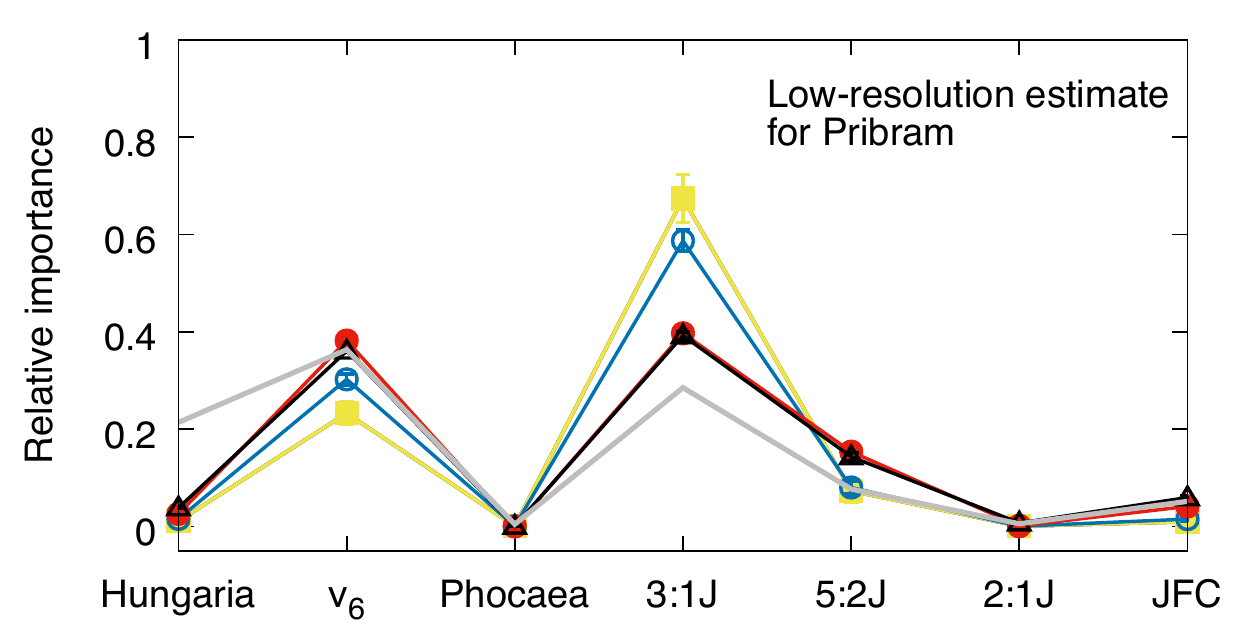}
  \includegraphics[width=0.66\columnwidth]{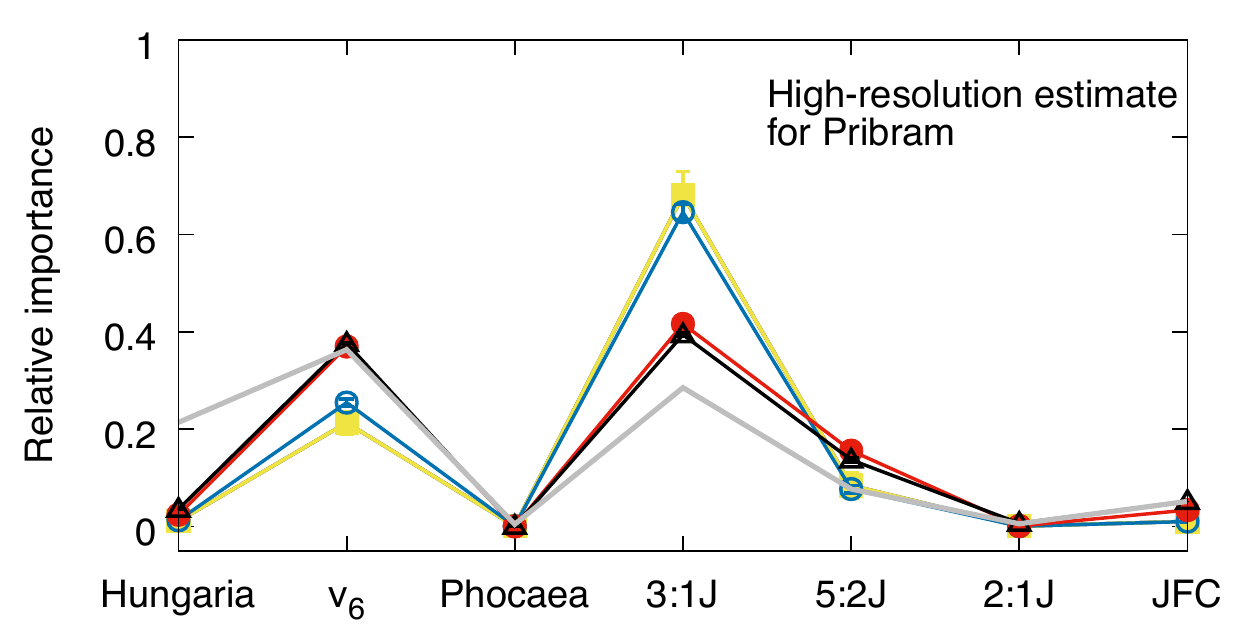}
  \includegraphics[width=0.66\columnwidth]{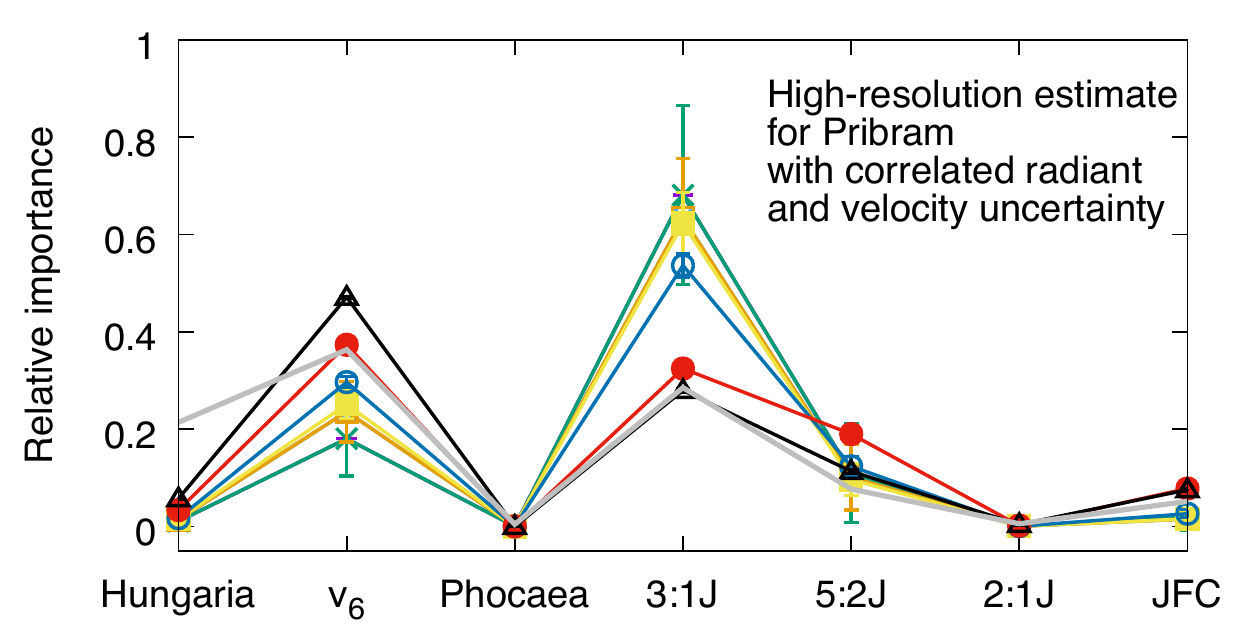}
  \caption{As Fig.~\ref{fig:set1}.}
  \label{fig:set3}
\end{figure*}

\begin{figure*}
  \centering
  \includegraphics[width=0.95\textwidth]{key.pdf}
  \includegraphics[width=0.66\columnwidth]{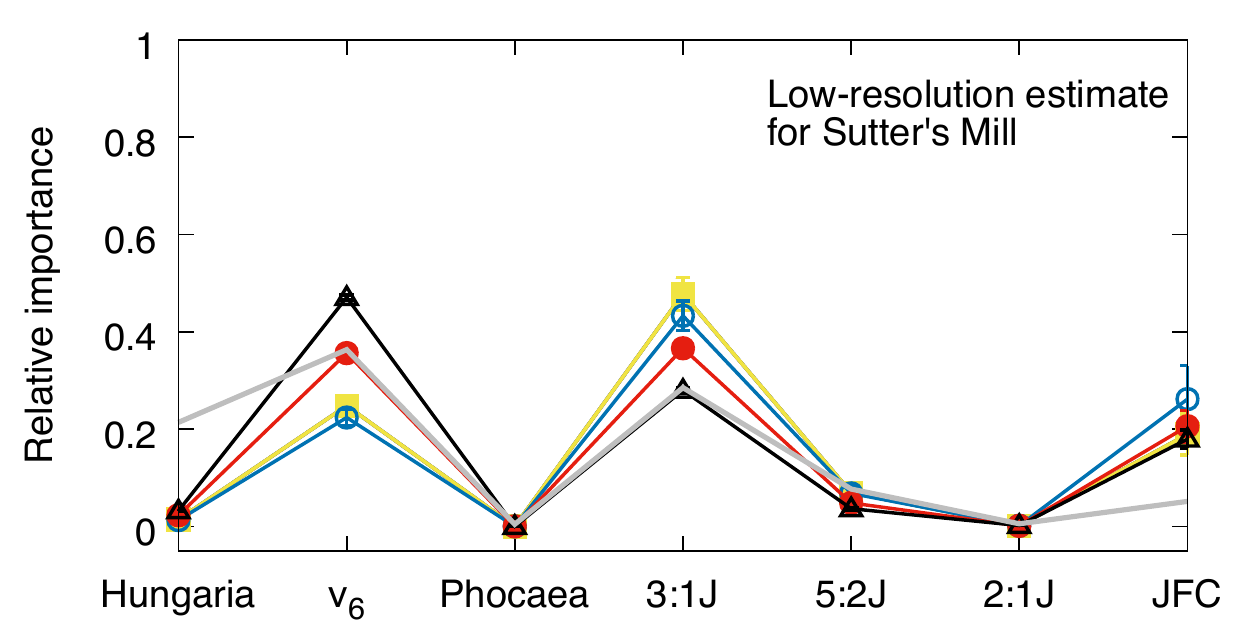}
  \includegraphics[width=0.66\columnwidth]{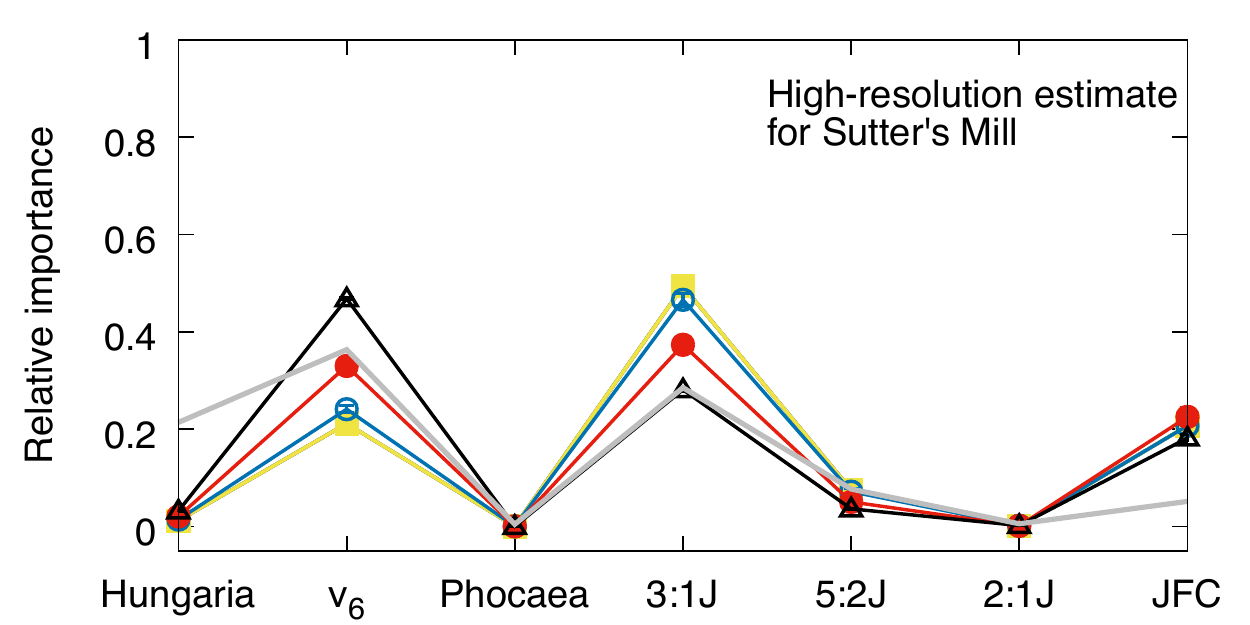}
  \includegraphics[width=0.66\columnwidth]{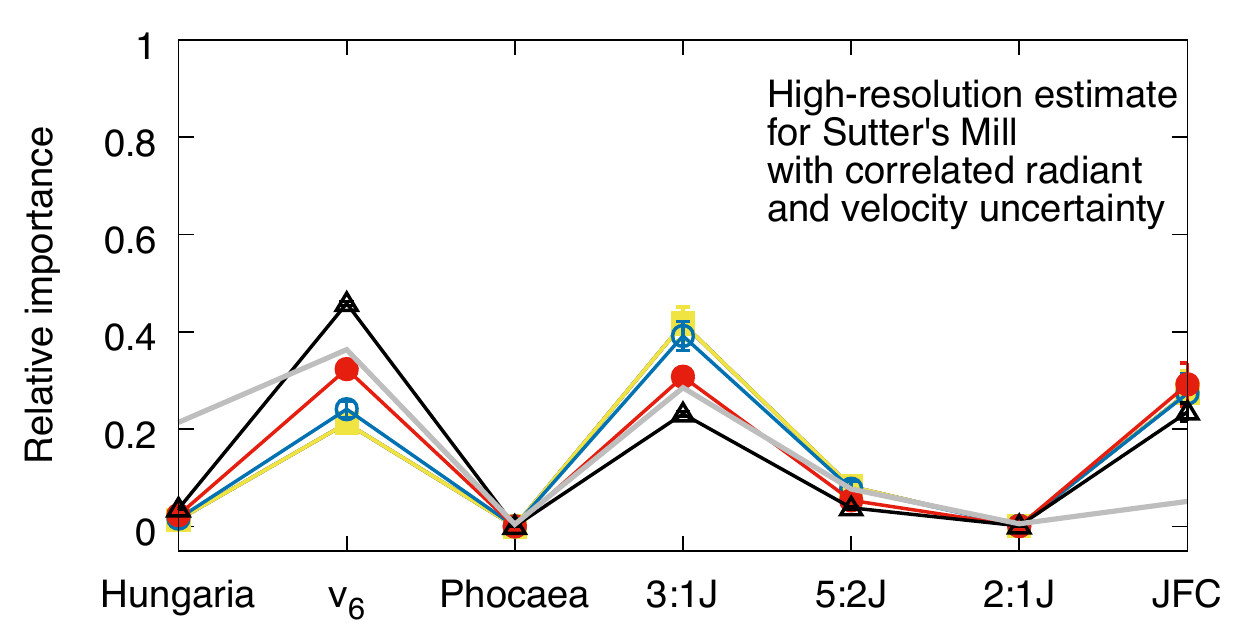}
  \includegraphics[width=0.66\columnwidth]{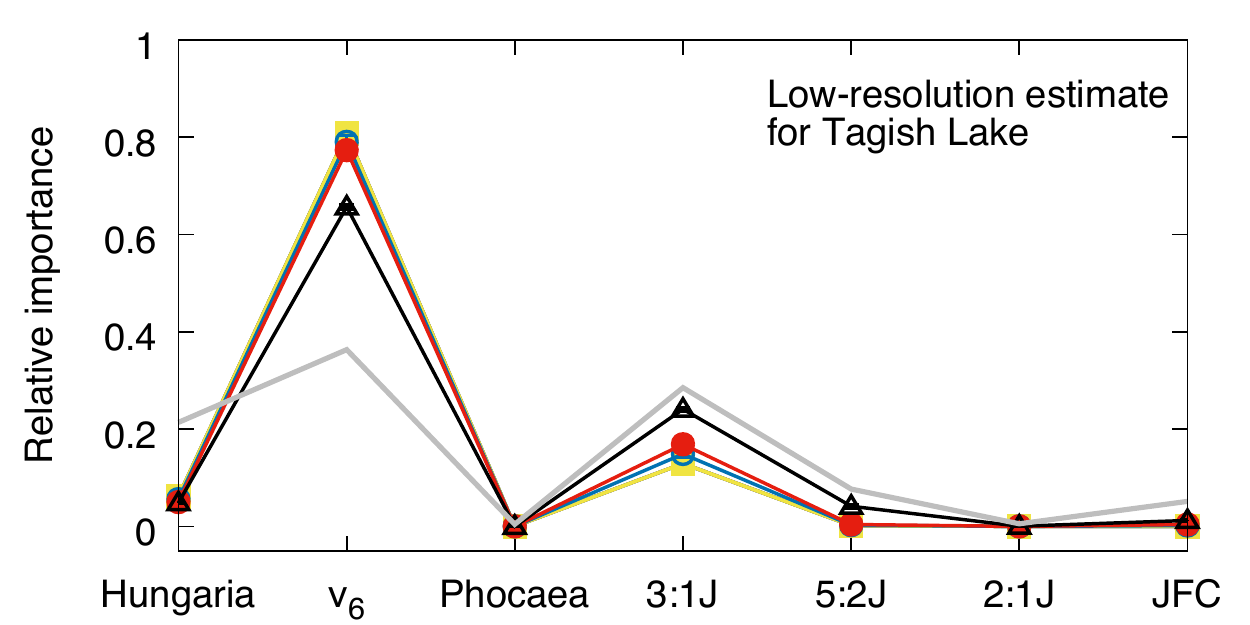}
  \includegraphics[width=0.66\columnwidth]{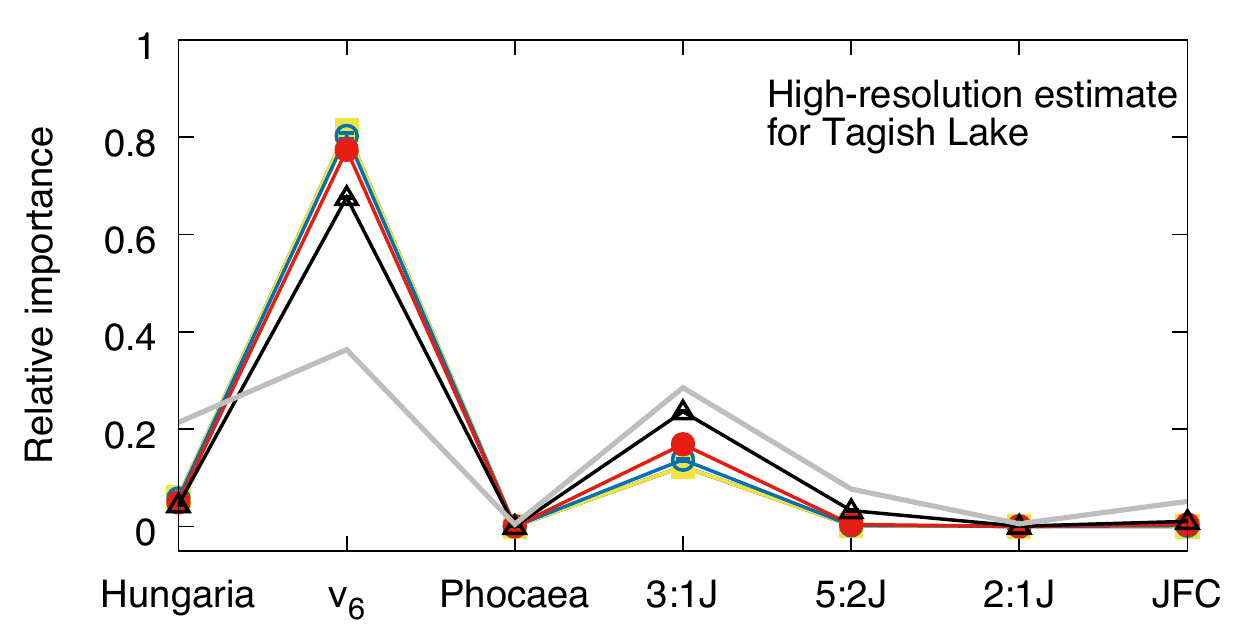}
  \includegraphics[width=0.66\columnwidth]{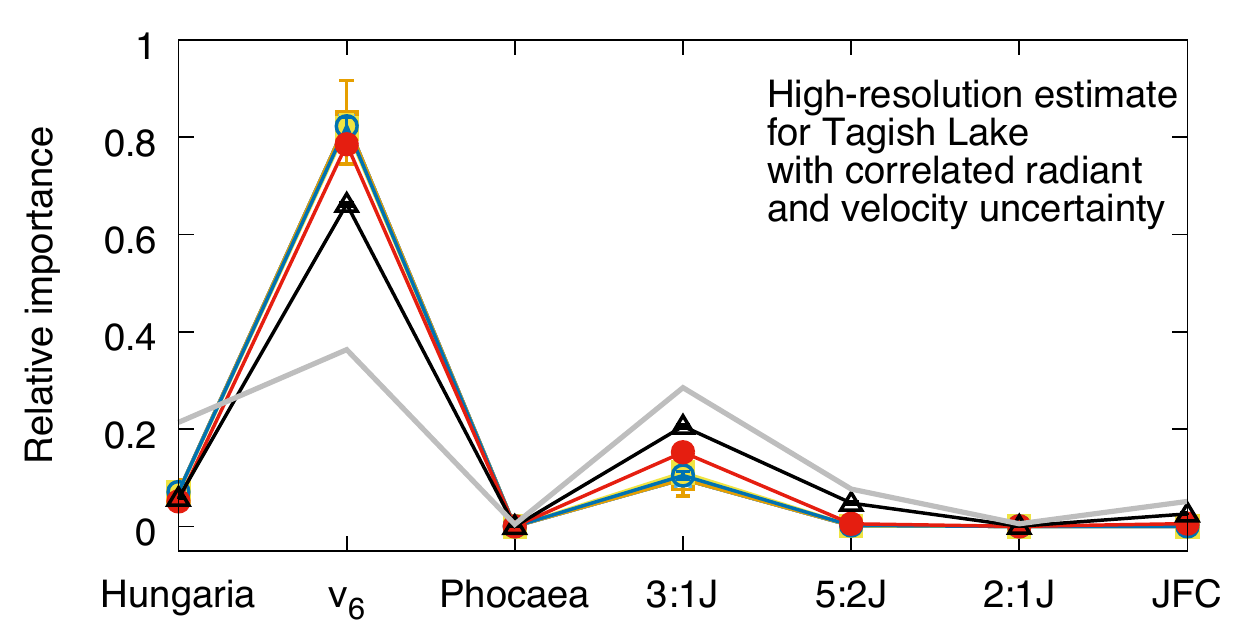}
  \includegraphics[width=0.66\columnwidth]{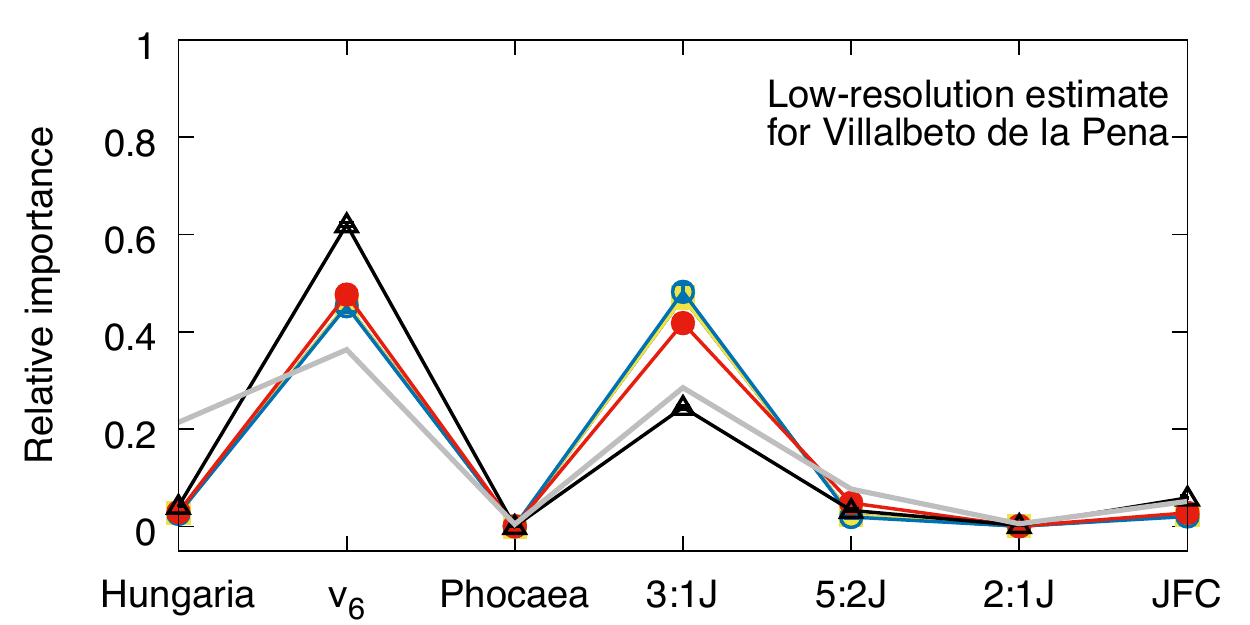}
  \includegraphics[width=0.66\columnwidth]{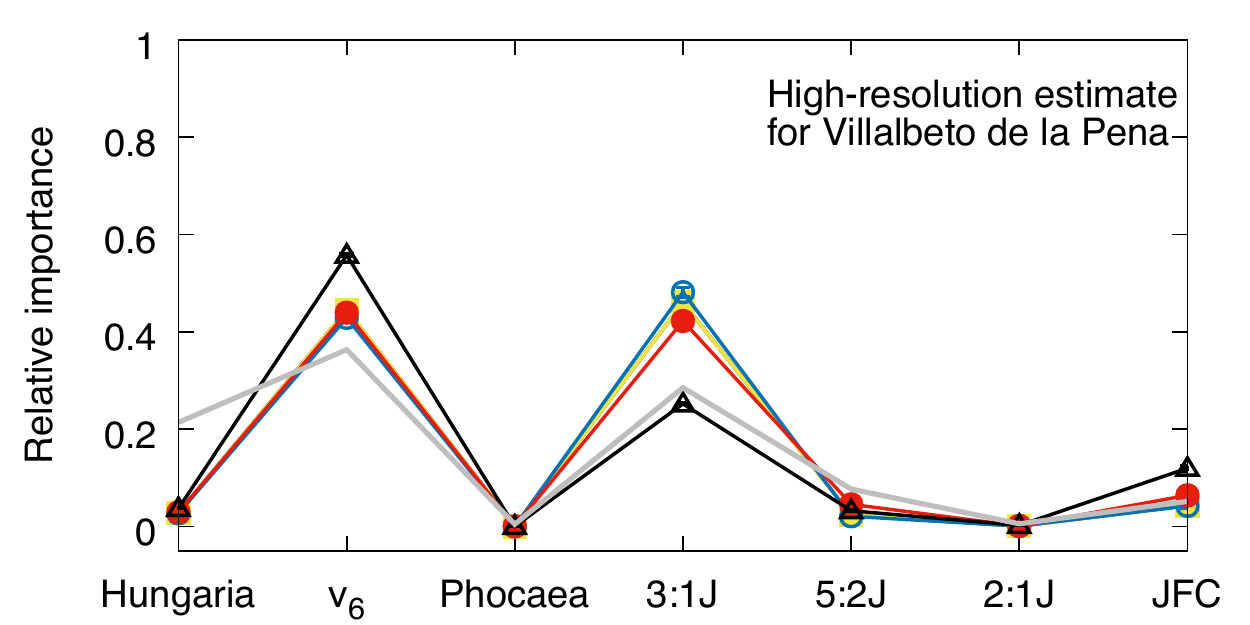}
  \includegraphics[width=0.66\columnwidth]{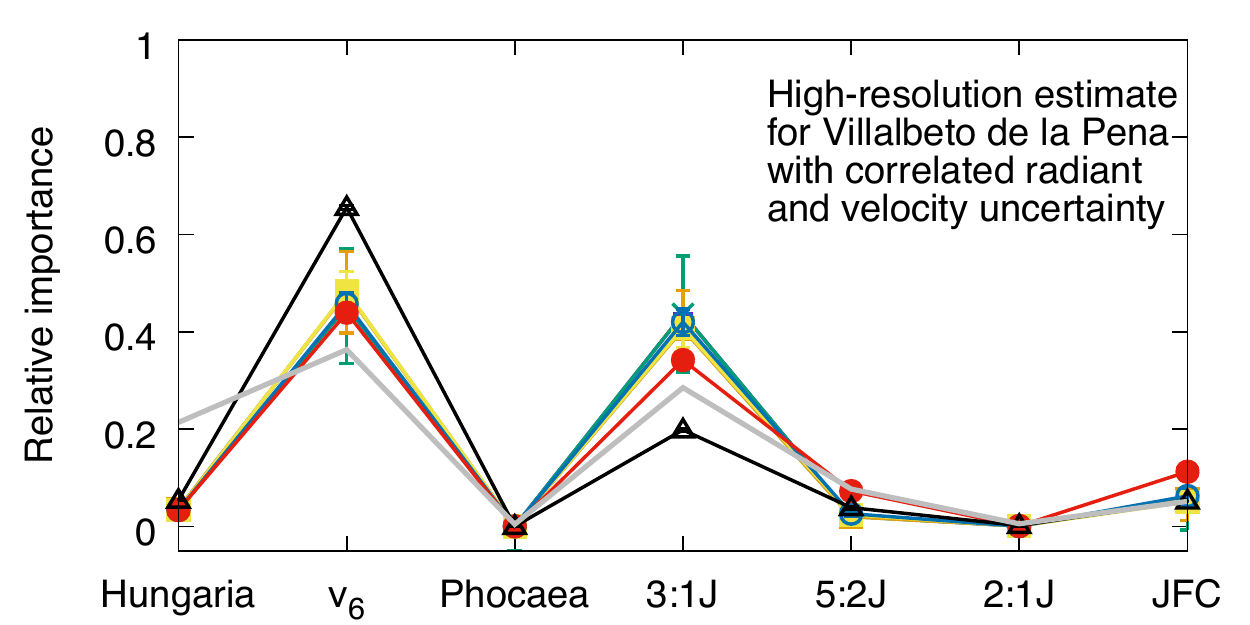}
  \includegraphics[width=0.66\columnwidth]{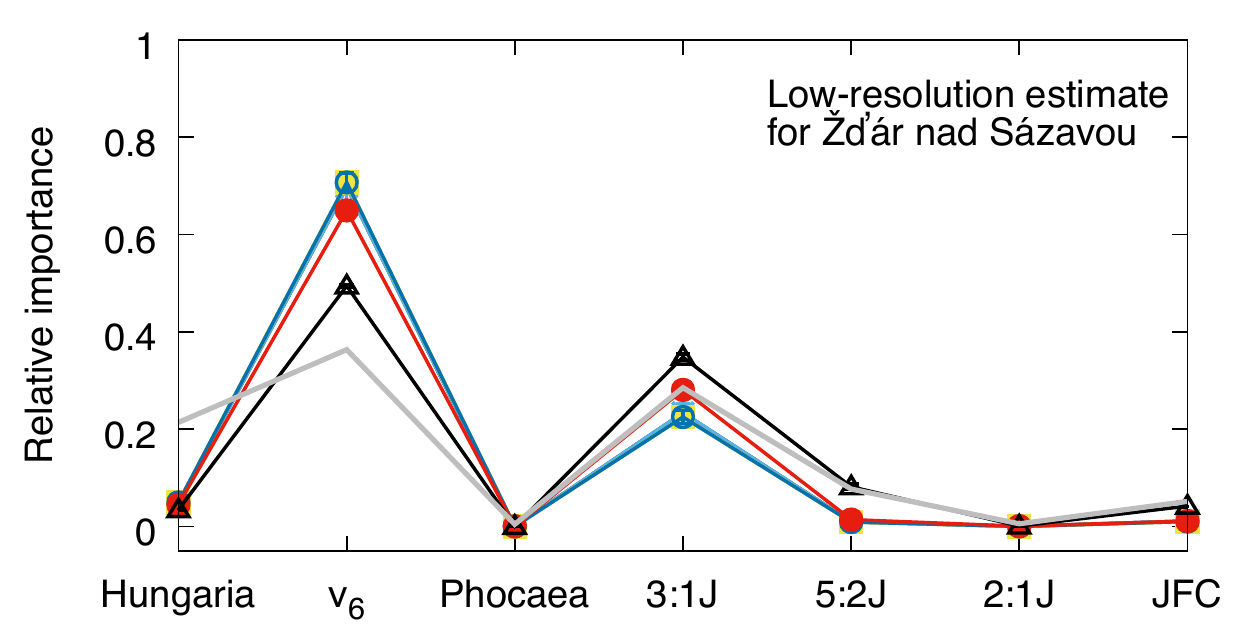}
  \includegraphics[width=0.66\columnwidth]{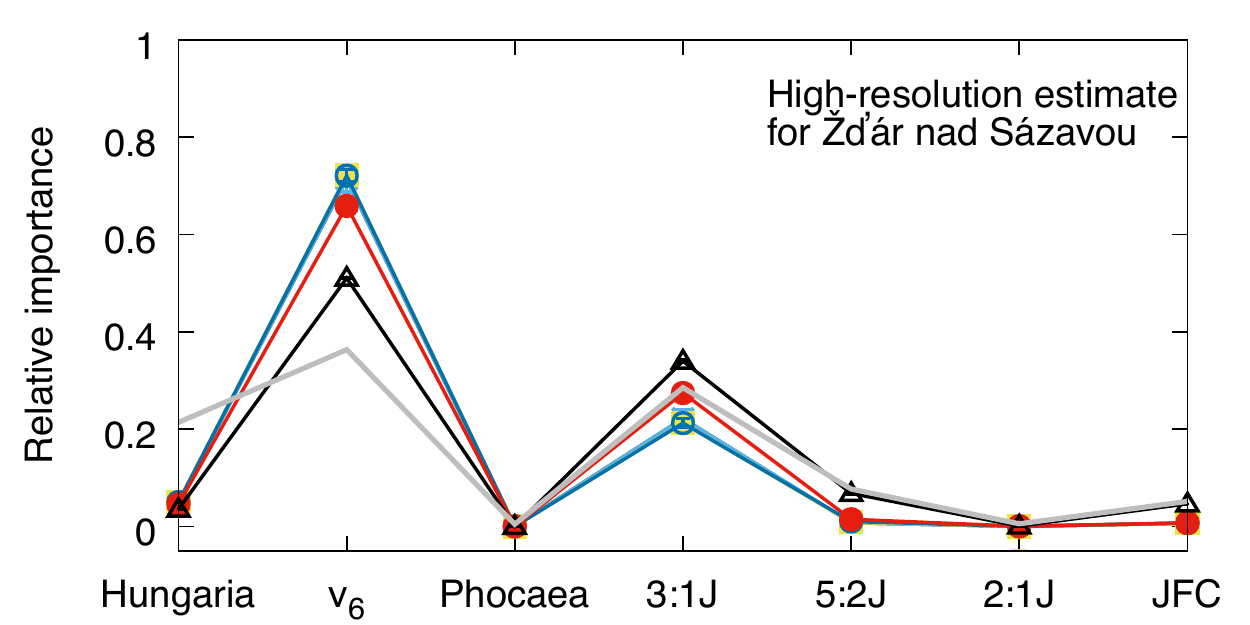}
  \includegraphics[width=0.66\columnwidth]{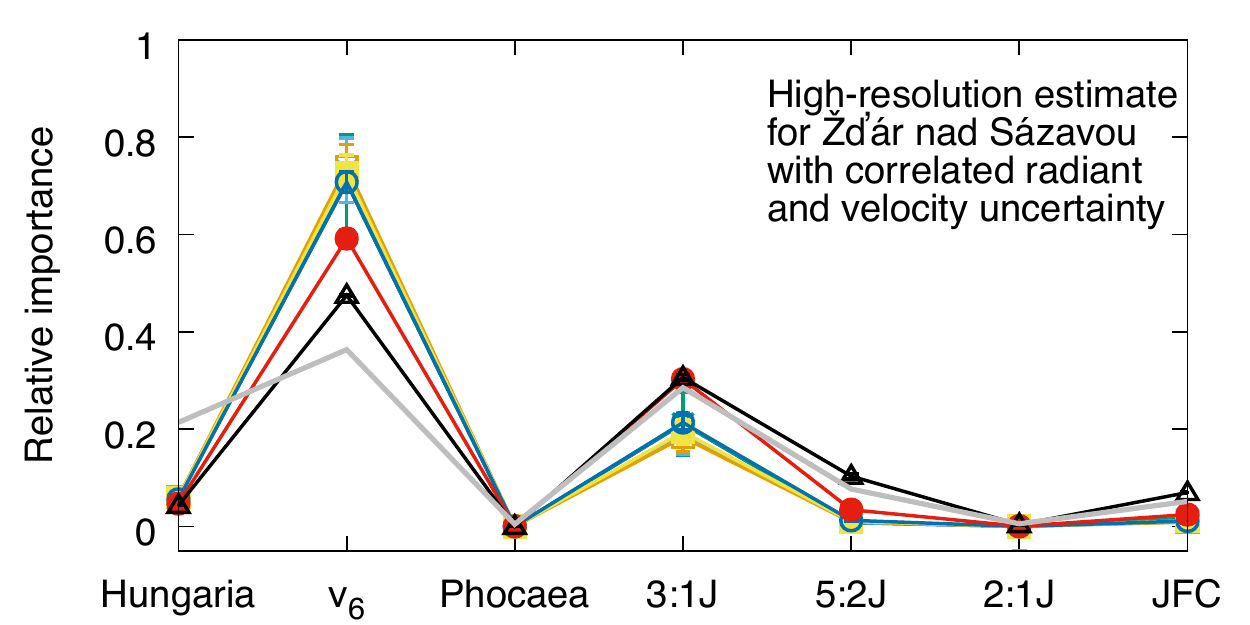}
  \caption{As Fig.~\ref{fig:set1}.}
  \label{fig:set4}
\end{figure*}

Considering the substantial reduction in orbital uncertainty when
reducing the velocity uncertainty, it is somewhat surprising that the
determination of meteorite ERs---the primary scientific motivation for
setting up fireball networks---is much less improved (left column in
Figs.~\ref{fig:set1}--\ref{fig:set4}). In all but a few cases the
largest improvement takes place when the velocity uncertainty is
pushed down to about $1\km\second^{-1}$. It is reassuring to see that
in most cases the primary ERs coincide with $\nu_6$ and 3:1J, the
most important NEO ERs. Note also that in all cases the predicted
probabilities are substantially different from the prior distribution,
i.e., the limiting case in which the observations would provide no
constraints on the meteoroid orbit and, further, the ER.

The results change only marginally when the resolution of the NEO
model is increased by a factor of 8 (left versus middle column in
Figs.~\ref{fig:set1}--\ref{fig:set4}). The explanation for the
insensitivity to model resolution is that the model itself is
probabilistic. Increasing the model resolution can never result in a
unique determination of ER in the volumes of orbital space that
harbour most NEOs, because those volumes are fed by multiple ERs. As
explained in the Introduction, the physical reason for the
non-determinism is the chaotic long-term orbital evolution of NEOs.
The chaotic orbital evolution leads to overlapping but statistically
distinct steady-state orbit distributions for NEOs originating in
different ERs. A comparison of the left and middle columns in
Figs.~\ref{fig:set1}--\ref{fig:set4} suggests that the low-resolution
model accurately captures the overall picture and the high-resolution
model provides little added value.

\begin{figure}[h!]
  \centering
  \includegraphics[width=1.0\columnwidth]{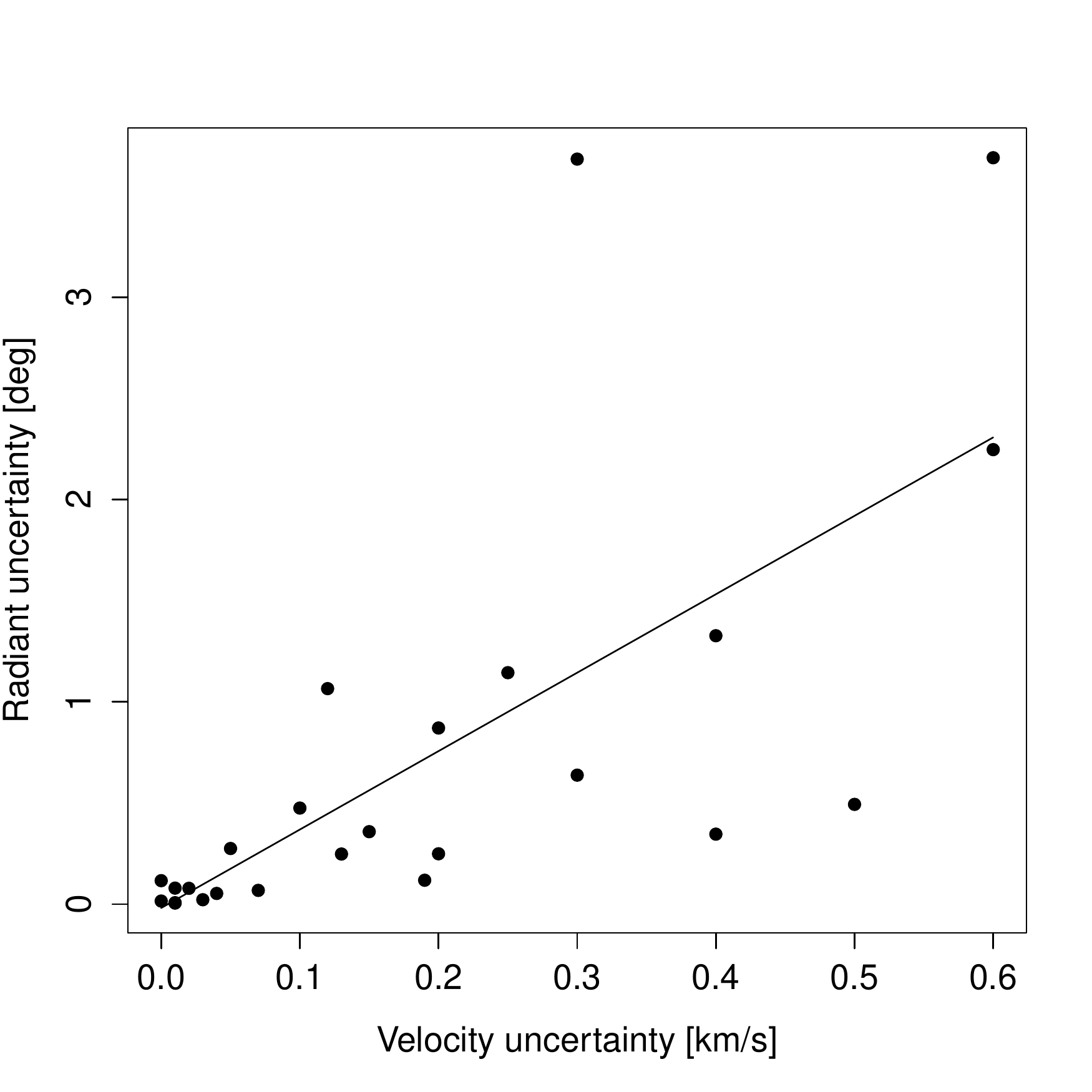}
  \caption{The observed correlation between velocity uncertainty and
    radiant uncertainty ($\sqrt{(\Delta{\rm RA}\cos({\rm Dec}))^2 +
      (\Delta{\rm Dec})^2}$) for meteorite-producing fireballs. The
    black line is a linear fit ($y=ax+b$) to the data with
    $a=(3.9\pm0.8)\deg\second\km^{-1}$ and $b=(-0.02\pm0.22)\deg$.}
  \label{fig:obsuncscosd}
\end{figure}

The assumption that the radiant uncertainty would not be reduced when
the velocity uncertainty is reduced is not entirely correct
(Fig.~\ref{fig:obsuncscosd}). An accurate treatment of the correlation
would require detailed consideration of the hardware and
trajectory-computation approach for each case; this is not within the
scope of this paper, nor is it possible in most cases given the lack
of primary published data for each fireball producing meteorite fall.

Instead we take a simplistic approach and assume a linear correlation
between the radiant uncertainty and the velocity uncertainty. Based on
a linear regression we find that the radiant uncertainty (in degrees)
can be approximated as $3.9\delta v_\infty$ where the constant $b$ is
statistically indistinguishable from zero and hence omitted, and
$\delta v_\infty$ is provided in $\km\second^{-1}$
(Fig.~\ref{fig:obsuncscosd}). Given the velocity uncertainties used
above the radiant uncertainties are approximately $0.004\deg$,
$0.012\deg$, $0.04\deg$, $0.12\deg$, $0.4\deg$, $1.2\deg$, $4\deg$,
$12\deg$, respectively. Again, note that the two last radiant
uncertainties are larger than reported for any of these events.

In this simple model, the reduced radiant uncertainties that
correspond to small velocity uncertainties, lead to orbital
uncertainties which are also much reduced (right column in
Fig.~\ref{fig:velunc_vs_orb}). The reduction of both velocity and
radiant uncertainty leads to a corresponding reduction in orbital
uncertainty, and for most cases there is no indication across the
range of uncertainties considered here that the orbital uncertainty
would reach a plateau. There are only two events that have clearly
different behaviour compared to the rest: Ejby and Sutter's Mill. The
orbital uncertainty for these two reaches a plateau at some hundreds
of meters per second.

The ER probabilities estimated with the correlated radiant
uncertainties hardly show variation when compared to the ER
probabilities estimated using fixed radiant uncertainties,
particularly for $\delta v_\infty \lesssim 0.03\km\second^{-1}$
(middle versus right column in
Figs.~\ref{fig:set1}--\ref{fig:set4}). This is explained by the fact
that the more accurate trajectories typically correspond to
orbital-element uncertainties smaller than the resolution of the NEO
model. Hence an improvement in the orbital accuracy does not directly
translate to an improvement in the source probabilities.

\begin{figure}[h!]
  \centering
  \includegraphics[width=1.0\columnwidth]{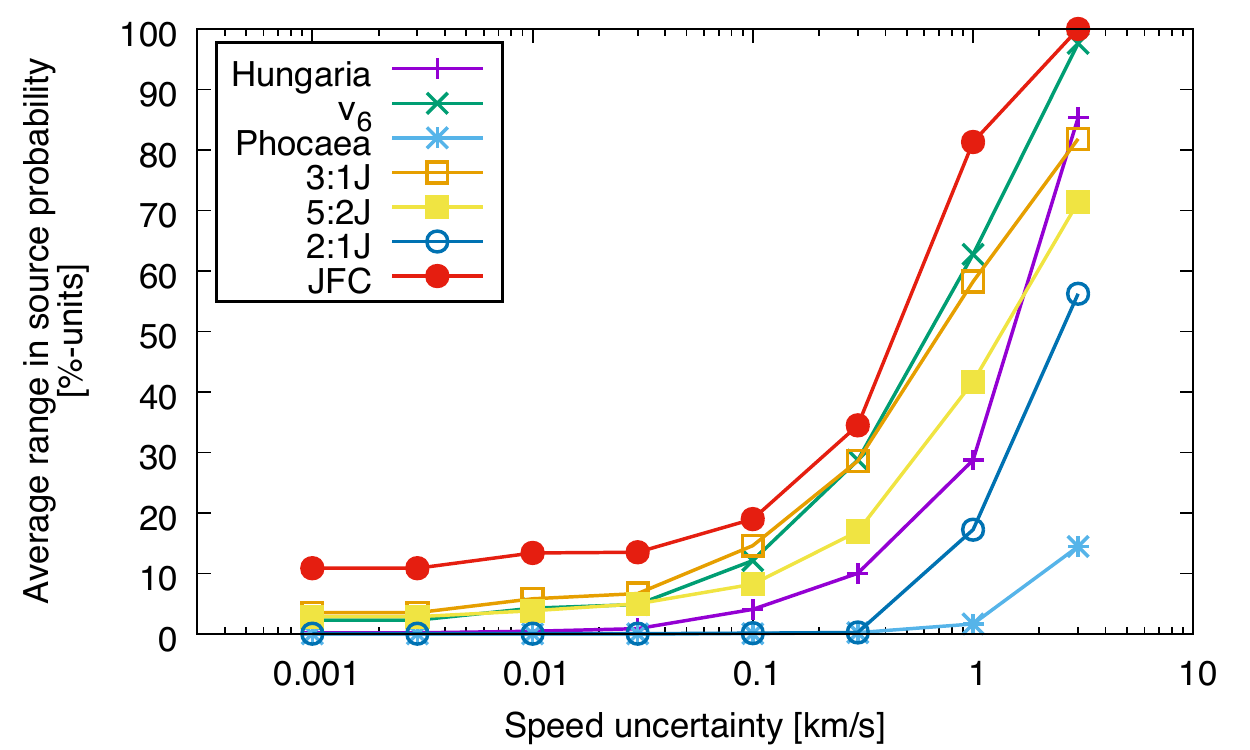}
  \caption{The average range in source probability as a function of
    uncertainty on measured meteor speed.}
  \label{fig:summary}
\end{figure}

To summarize the effect of the correlated radiant and speed
uncertainty on the ER prediction uncertainty, we consider the sample
average of the maximum range of ER probability (a proxy for the
uncertainty) for each ER, and speed and radiant uncertainty value
(Fig.~\ref{fig:summary}). The maximum range of ER probabilities for a
particular orbit solution is computed by finding the minimum and
maximum ER probabilities among the cells in the high-resolution model
that are within the orbital-uncertainty volume. The sample averages
show a steep reduction in ER-probability uncertainty when reducing the
speed uncertainty (and the correlated radiant uncertainty) from 3~km/s
to about 0.1--0.3~km/s. Reducing the speed uncertainty to below
0.1~km/s will only have a marginal impact on the uncertainty of the ER
probability. We also note that the JFC source will always have an
uncertainty greater than 10\%.

\section{Conclusions}

In this work, we first computed heliocentric orbits for 25 meteorite
falls based on the trajectory information reported in the
literature. For several meteorite falls the trajectory information has
only been reported in conference abstracts. We urge members of the
community to have their trajectory analysis peer-reviewed, and
published with enough supporting data to allow colleagues to reproduce
their results.

We then used the orbits (including the orbital
uncertainties) and the NEO model by \citet{2016Natur.530..303G} to
estimate the most likely entrance routes (ERs) for the falls. Both for
orbits and ERs our results broadly agree with previously published
results in that most meteorite falls originate in the inner main
asteroid belt and escape through the 3:1J mean-motion resonance or the
$\nu_6$ secular resonance.

We then proceeded to test how the velocity uncertainty affects the
results for both orbits and ER probabilities. We found that pushing
the velocity uncertainty below some tens of meters per second does not
lead to a substantial reduction of uncertainties related to ER
probabilities.  Similarly, increasing the resolution of the
steady-state orbit distribution of the NEO model also had a limited
effect on the uncertainties related to ER probabilities.

For the purpose of more accurately identifying meteorite ERs it would
thus be more useful to increase the number of ERs that go into the NEO
model rather than determining the meteoroid velocity more accurately
or increasing the resolution of the NEO orbit model. That is, if the
fairly extensive ER complexes used currently would be divided into
their sub-components, it would become possible to provide more
specific (but still probabilistic) estimates for the most likely
ERs. Such an improvement primarily depends on the amount of NEO data
available --- the current choice to use 7 ERs in the state-of-the-art
NEO model is dictated by the need to ensure that the model is not
degenerate \citep{2016Natur.530..303G}.

Finally, we may apply our results in terms of the optimal approaches
to designing and operating fireball camera networks. In particular,
our results suggest that a larger number of moderately precise
meteorite recoveries, with speed precision of order a few hundred
meters/sec will provide the most information regarding meteorite ERs
and, ultimately, source regions.

\section*{Acknowledgements}

The authors wish to thank the two anonymous reviewers for their
constructive suggestions that helped improve the paper. MG is 
grateful for the kind hospitality and financial support when
visiting the University of Western Ontario's Meteor Physics Group. MG
acknowledges funding from the Academy of Finland (grant \#299543) and
the Ruth and Nils-Erik Stenb\"ack foundation. PGB was supported by
funding from the Meteoroid Environment Office through NASA
co-operative agreement NNX15AC94A, the Natural Sciences Research
Council and the Canada Research Chairs program.


\bibliography{asteroid}

\end{document}